\documentclass[twocolumn,aps,prb,superscriptaddress,nofootinbib]{revtex4-2}
\usepackage{graphicx}
\usepackage{times}
\usepackage{bm}
\usepackage[linktocpage=true,colorlinks=true,urlcolor=blue,citecolor=red]{hyperref}
\usepackage{pgfplots}
\usepackage{amsfonts}
\usepackage{amsmath}
\usepackage{xcolor}
\usepackage{braket}
\usepackage{relsize}
\usepackage{placeins}
\usetikzlibrary{arrows}
\usepackage{enumerate}
\usetikzlibrary[patterns] 
\usetikzlibrary[snakes] 
\usetikzlibrary[shapes] 
\usepackage{pgfplots}
\usepackage{newfloat}

\DeclareFloatingEnvironment[name={Supplementary Figure}]{suppfigure}

\graphicspath{ {Figures/} }
\begin{document}

\title{Charge impurity effects in hybrid Majorana nanowires}

\author{Benjamin D. Woods}
\affiliation{Department of Physics and Astronomy, West Virginia University, Morgantown, WV 26506, USA}
\author{Sankar Das Sarma}
\affiliation{Condensed Matter Theory Center and Joint Quantum Institute, Department of Physics, University of Maryland, College Park, Maryland, 20742-4111, USA}
\author{Tudor D. Stanescu}
\affiliation{Department of Physics and Astronomy, West Virginia University, Morgantown, WV 26506, USA}

\begin{abstract}
We address an outstanding problem that represents a critical roadblock in the development of the Majorana-based topological qubit using semiconductor-superconductor hybrid structures: the quantitative characterization of disorder effects generated by the unintentional presence of charge impurities within the hybrid device. Given that disorder can have far-reaching consequences for the Majorana physics, but is intrinsically difficult to probe experimentally in a hybrid structure, providing a quantitative theoretical description of disorder effects becomes essential. To accomplish this task, we develop a microscopic theory that (i) provides a quantitative characterization of the effective potential generated by a charge impurity embedded inside a semiconductor wire proximity-coupled to a superconductor layer by solving self-consistently the associated three-dimensional Schr{\"o}dinger-Poisson problem, (ii) describes the low-energy physics of the hybrid structure in the presence of s-wave superconductivity, spin-orbit coupling, Zeeman splitting, and disorder arising from multiple charge impurities by using the results of (i) within a standard free fermion approach, and (iii) links the microscopic results to experimentally observable features by generating tunneling differential conductance maps as function of the control parameters (e.g., Zeeman field and chemical potential). We find that charge impurities lead to serious complications regarding the realization and observation of Majorana zero modes, which have direct implications for the development of Majorana-based qubits. More importantly, our work provides  a clear direction regarding what needs to be done for progress in the field, including specific materials quality and semiconductor purity targets that must be achieved to create a topological qubit.
\end{abstract}

\maketitle

\section{Introduction} 

Majorana nanowires have been among the most intensively studied topics in physics since 2010, when it was theoretically proposed that semiconductor-superconductor (SM-SC) hybrid platforms could host non-Abelian anyonic Majorana zero modes (MZMs) \cite{Lutchyn2010,Sau2010a,Sau2010,Oreg2010} in the combined presence of s-wave superconductivity, spin-orbit coupling, and Zeeman spin splitting.  The subject has also attracted serious technological attention, way beyond its physics context, since Microsoft  Corporation chose this system as its preferred platform for creating a fault-tolerant topological quantum computer \cite{Kitaev2001,Kitaev2003,DSarma2015, Nayak2008,Stanescu2017}. A large number of experiments followed up on the theoretical predictions using InSb or InAs nanowires and Al or Nb superconductors, generating a lot of excitement with reported observations of zero bias conductance peaks in tunneling spectroscopy \cite{Mourik2012,Deng2012,Das2012,Churchill2013,Chen2017,Suominen2017,Grivnin2019}, which were interpreted as possible signatures of the putative MZMs.  It has, however, become clear by now that most of the experimental samples are likely to contain potential disorder, which strongly affects the interpretation of the tunneling experiments and opens the possibility that the ubiquitous  zero bias peaks showing up in the experiments may actually be generated by disorder-induced non-topological fermionic  low-energy Andreev bound states \cite{Kells2012,Prada2012,Liu2012,Sau2013a,Liu2017,Reeg2018b,Stanescu2019a,Pan2020,Prada2020,Yu2021}.  The subject is very much in flux and, in the absence of a clear understanding and characterization of disorder effects, much of what is going on experimentally remains problematic, in spite of high-profile experimental publications with claims of Majorana discovery appearing regularly.
 
This is the background and the context of the current theoretical work, in which we take a step back and ask a fundamental question: What happens if the nanowire, instead of being pristine, has disorder arising from unintentional charge impurities residing in it?  The scenario considered in this question is not hypothetical, since unintentional charge impurities (``low doping'') constitute the commonest type of disorder in high-quality semiconductor materials \cite{Pantelides1978}.  There is evidence that the experimental nanowires do, in fact, have substantial disorder.  Given the considerable confusion about the situation surrounding the Majorana nanowire experiments and the intrinsic difficulty of directly measuring disorder in hybrid nanostructures, 
we think it is appropriate to take a quantitative, microscopic approach to the problem by first solving exactly the single impurity problem within a self-consistent numerical scheme, then using the results to study topological superconductivity and Majorana physics in the presence of impurity disorder within the standard free fermion theory.  

More specifically, this is what we do in this paper. First, we provide a quantitative characterization of the effective potential generated by a charge impurity embedded inside a semiconductor wire proximity-coupled to a superconductor layer by solving self-consistently the associated three-dimensional Schr{\"o}dinger-Poisson problem. Next, using the single impurity effective potential obtained self-consistently, we construct disorder  potentials associated with the presence of multiple charge impurities and solve numerically the Bogoliubov-de Gennes (BdG) equations that describe the hybrid system in the presence of s-wave superconductivity, spin-orbit coupling, Zeeman splitting, and disorder arising from charge impurities. We also carry out first principles charge transport calculations and determine the tunneling differential conductance as a function of various systems parameters (e.g., disorder strength, chemical potential, and Zeeman splitting). Along the way, we introduce a number of quantities that facilitate the characterization of the low-energy physics in the presence of disorder (e.g., the Majorana separation length and the edge-to-edge correlation) and describe several protocols that enable a more efficient extraction and use of experimentally accessible information (e.g., construction of zero-bias conductance correlation maps). 
Given that the work presented in this paper is multifaceted, with many independent results of importance in their specific contexts, we first provide a  summary of our key findings, with references to the relevant equations and figures (see Sec. \ref{KeyR}), so that the reader uninterested in the technical details can simply learn about our main results without going through the rest of the paper, with all its technical complexity. 

We emphasize that our work is of considerable importance to the development of a Majorana-based topological quantum computer (TQC), as it addresses a critical outstanding problem facing the realization of  topological qubits using hybrid  nanostructures, which is the platform Microsoft Corporation is working on.  In particular, our finding that charge impurities in the environment lead to serious complications regarding the realization and observation
of Majorana zero modes has obvious direct implications for the development of Majorana-based qubits and TQC.  
 Our work provides a full microscopic-based description of how experimentally available Majorana nanowire behave in the presence of charge impurity disorder of varying strength.  
 More importantly, our work provides  a clear future direction regarding what needs to be done for progress in the field, as well as quantitative measures of the maximum allowed impurity concentrations consistent with the full manifestation of  topological MZMs in hybrid nanostructures. 
In particular,  based on our extensive realistic calculations, we provide specific materials quality and semiconductor purity targets which must be achieved to create a topological qubit, providing a clear blueprint for future progress towards building a TQC.  Our work establishes a clear goal of using nanowires with impurity concentrations around $10^{15}$ per cm$^3$ or lower for TQC hardware to be feasible using Majorana qubits.  This is a challenging target, but by no means an impossible one. 

The remainder of this paper is organized as follows. In Sec. \ref{KeyR} we provide a summary of our key results and discuss their significance in the context of the ongoing experimental effort to realize topological superconductivity and Majorana zero modes using semiconductor-superconductor hybrid structures. The case of a single charge impurity embedded within a proximity-coupled nanowire is investigated in Sec. \ref{Single}. The model used in our analysis is described in Sec. \ref{Model1}, the details of the self-consistent Schr{\"o}dinger-Poisson scheme for calculating the effective impurity-induced potential are presented in Sec. \ref{Calc1}, and the results of the numerical calculations are discussed in Sec. \ref{Results1}. Section \ref{Mult} is dedicated to the multi-impurity case, with Sec. \ref{Model2} describing the effective single-band model used in our analysis and Sec. \ref{Results2} discussing the results of the numerical calculations and their implications for the low energy physics of hybrid nanostructures with charge impurities.
Our concluding remarks are presented in Sec. \ref{Conclusion}. 

\section{Summary of key results} \label{KeyR} 

In this section we provide a brief summary of our key results and indicate the  relevant equations and/or figures. For technical details and in-depth discussion of the results the reader should consult the corresponding paragraphs in sections \ref{Single} and \ref{Mult}. 

\begin{itemize}
\item We provide a quantitative description of the effective potential [see Eq. (\ref{effPot}) and Fig. \ref{FIG3}] generated by a charge impurity embedded into a semiconductor wire-superconductor nanostructure (Fig. \ref{FIG1}) by solving self-consistently the corresponding three-dimensional Schr{\"o}dinger-Poisson problem [Eqs. (\ref{SM_Ham}) and (\ref{Pois})]. 
\item We show that the position dependence of the effective impurity potential has a simple functional form [see Eq. (\ref{Vfit})], 
with two controlling parameters: the amplitude and decay length of the impurity potential in the absence of redistribution of free charge. This can help future device modeling in the presence of disorder, by circumventing the need to explicitly address a numerically demanding three-dimensional Schr{\"o}dinger-Poisson problem.
\item We determine the distribution of the effective impurity potential parameters by sampling 169 possible  impurity locations evenly distributed over the hexagonal cross-section of the semiconductor wire and show that the typical values of the amplitude are on the order of $1.5-2~$meV, while the typical decay lengths are about $8-12$ nm (Fig. \ref{FIG5}). 
\item We demonstrate that the screening by the superconductor has a limited effect on reducing the magnitude and characteristic length scale of the effective impurity potential inside the semiconductor (Fig. \ref{FIG6}).  On the other hand, screening by the free charge in the wire has considerable effects (Fig. \ref{FIG6bis}) and has to be incorporated self-consistently to obtain a quantitative description of the low-energy physics in the presence of charge impurities.
\item We show that the presence of multiple charge impurities embedded inside the wire generates a correlated disorder potential (Fig. \ref{FIGM2bis}) characterized by a correlation function having a central peak of height on the order $1~$meV$^2$ and width at half maximum in the range $20-40~$nm (Fig. \ref{FIGM3}). The  correlation function scales with the impurity concentration.
\item We introduce the precisely defined concepts of Majorana separation length [Eqs. (\ref{lsepn}-\ref{lsep})] and the edge-to-edge correlation [Eqs. (\ref{Cn}-\ref{Wn})] as useful theoretical tools for characterizing the effects of impurity-induced disorder and we connect them to the differential tunnel conductance  [Eq. (\ref{CG})].  
\item We show that generating comprehensive maps that cover large ranges of control parameters (Figs. \ref{FIGM5}, \ref{FIGM11}, \ref{FIGM12}, and \ref{FIGM14}), rather than focusing on specific post-selected traces, constitutes a productive approach to understanding disorder effects in hybrid devices. We suggest that this should be the standard protocol for the experimental characterization of these devices, instead of the current focus on post-selected fine-tuned features, which is potentially prone to serious confirmation bias problems and provides no relevant information on the effects of disorder.
\item We find that in the low impurity density regime the system is characterized by well separated Majorana modes and finite edge-to-edge correlations within large areas inside the nominally topological region,  demonstrating topological immunity to weak disorder (Figs. \ref{FIGM5}, \ref{FIGM6}, \ref{FIGM8}, and \ref{FIGM11}).
\item  In the intermediate  impurity density regime, the parameter regions corresponding to significant edge-to-edge correlations reduce to relatively small, isolated islands located both inside and outside the nominally topological region (Figs. \ref{FIGM12} and \ref{FIGM14}). There is still a significant region corresponding to well separated  Majorana modes (Fig. \ref{FIGM12}), but, typically, these modes are localized away from the edges of the system and remain ``invisible''  to local probes applied to these edges  (e.g., tunneling spectroscopy at the wire ends).  
\item We show that the zero-bias conductance maps (in the tuning parameter space) are characterized by qualitatively different features inside and outside the nominally topological regime (Figs. \ref{FIGM11} and \ref{FIGM14}).  This suggests that detailed zero-bias conductance maps could help identify nominally topological regions even when the presence of disorder suppresses the ``standard'' Majorana phenomenology expected in a clean system.
\item We introduce ``global'' parameters that characterize the  properties of the Majorana bound states emerging in the system in the presence of charge impurities [Eqs. (\ref{MDef}-\ref{EgDef})] and we calculate the dependence of the disorder-averaged ``global'' parameters on the impurity concentration (Figs. \ref{FIGM17} and \ref{FIGM18}) and spin-orbit coupling strength (Fig. \ref{FIGM19}).
\item We find that well separated Majorana modes can generically emerge in the presence of charge impurities up to relatively high impurity concentration levels, but, for a given wire length, the presence of these well-separated Majoranas translates into significant edge-to-edge correlations only if the impurity concentration is below a critical threshold (Figs. \ref{FIGM17} and \ref{FIGM18}). The existence of a disorder-dependent characteristic length scale is particularly significant in the context of the exponential protection of Majorana modes, which is necessary for fault tolerant qubit operations. 
\end{itemize}

\section{Single charge impurity} \label{Single} 

In this section, we investigate a single charge impurity embedded  within a semiconductor (SM)  nanowire proximity-coupled to a superconductor (SC). In particular, we address the key question regarding the magnitude and characteristic length scale of the potential inhomogeneity induced by the charge impurity. The screening due to the presence of the superconductor and of a nearby metallic gate, as well as the effects due to the redistribution of free charge within the SM wire are incorporated using a position-dependent self-consistent Schr{\"o}dinger-Poisson scheme. Our model for describing the SM-SC hybrid structure with an embedded charge impurity is introduced in Sec. \ref{Model1},  the self-consistent Schr{\"o}dinger-Poisson method is presented in Sec. \ref{Calc1}, while the results of  our analysis are discussed  in Sec. \ref{Results1}.

\subsection{Model} \label{Model1}

\begin{figure}[t]
\begin{center}
\includegraphics[width=0.4\textwidth]{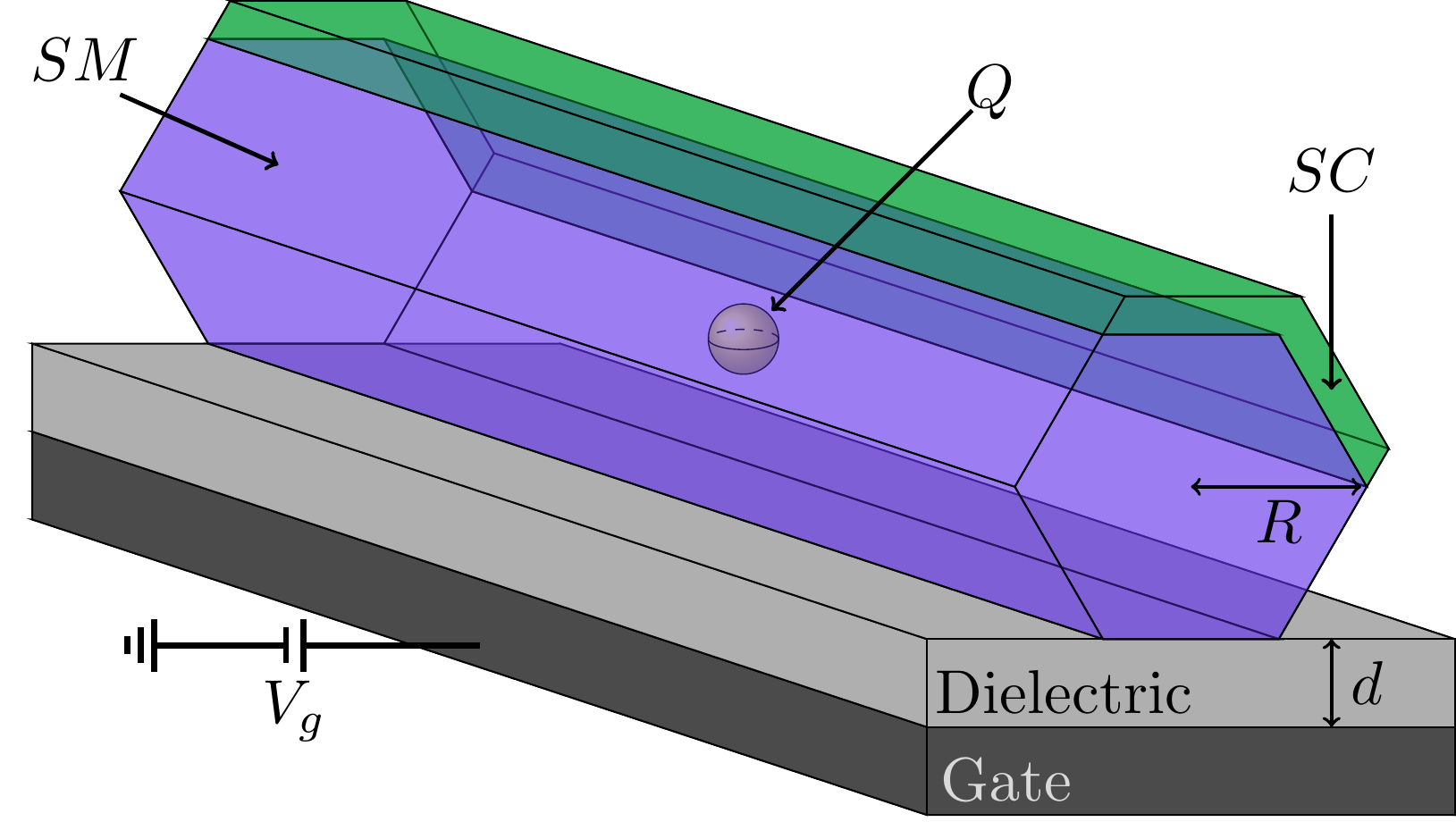}
\end{center}
\vspace{-3mm}
\caption{Schematic representation of the SM-SC hybrid device with an embedded charge impurity. A semiconductor nanowire (purple) of radius $R$ is proximity coupled to a thin superconductor (green).  An impurity (yellow sphere) of charge $Q$ embedded within the semiconductor nanowire will create a potential inhomogeneity.  The band edges of the low-energy SM subbands  can be tuned near the Fermi level using a back gate (black) separated from the wire by a thin dielectric layer (gray).}
\label{FIG1}
\vspace{-1mm}
\end{figure}

We consider the hybrid device represented schematically in Fig. \ref{FIG1}, which consists of a hexagonal semiconductor nanowire of radius $R$ (purple in Fig. \ref{FIG1}) having a thin superconducting layer (green) deposited on two of its facets. A metallic back gate (black) separated from the hybrid nanowire by a thin dielectric layer of thickness $d$  (gray)  is used to tune the band edges of the low-energy SM subbands near the Fermi level. Up to minor modifications of the device geometry, e.g.,  having additional side gates, or depositing the SC on more than two facets, this setup corresponds to the most prevalent type of SM-SC hybrid device used experimentally for exploring  Majorana physics \cite{Mourik2012,Deng2012,Das2012,Chang2015,Albrecht2016,Chen2017,deMoor2018,Lee2019,Bommer2019,Shen2020,Yu2021}. The key additional ingredient, which represents the focus of this study, is a charge impurity $Q$ embedded  inside the SM wire, as is indicated in Fig. \ref{FIG1} by a yellow sphere. In our theory, the effects induced by the presence of the charge impurity are calculated exactly within a  position-dependent self-consistent Schr{\"o}dinger-Poisson formalism.

At this stage, the SM nanowire is modeled using a simple effective mass Hamiltonian given by
\begin{equation}
	H = 
	-\frac{\hbar^2}{2 m^*} \nabla^2 
	- e \phi\left(\vec{r}\right), \label{SM_Ham}
\end{equation}
where $m^*$ is the effective mass, $\nabla^2$ is the Laplacian operator in 3-dimensional space, and $\phi$ is the electrostatic potential inside the wire. We assume that the wire is infinitely long. The potential $\phi$  must satisfy the Poisson equation,
\begin{equation}
   	 \nabla \cdot \left[\epsilon(\vec{r}) \nabla\phi(\vec{r})\right] =
    	  -\rho(\vec{r}), \label{Pois}
\end{equation}
where $\epsilon(\vec{r})$ is a material dependent dielectric constant taking different values inside  the dielectric, the SM wire, and the surrounding vacuum  and $\rho$ is the charge density within the wire. We impose Dirichlet boundary conditions on the bottom gate and the surface of the superconductor with potential values $V_g$ and $V_{SC}$, respectively. Note that the boundary condition on the SC surface accounts for the band-bending of the SM conduction band near the SM-SC interface \cite{Vuik2016,Schuwalow2019}. 
In addition, we impose Neumann boundary conditions on the sides and top of the full simulation region for Eq. (\ref{Pois}), which are a distance $b \gg R$ away from the nanowire. Note that this choice of boundary conditions on the outer boundaries has negligible impact on the potential within the nanowire \cite{Woods2020a}. It is convenient (and physically appealing) to break the total charge density into three components,
\begin{equation}
	\rho\left(\vec{r}\right) = 
	\rho_o\left(x,y\right) 
	+ \rho_{imp}\left(\vec{r}\right)
	+ \rho_{red}\left(\vec{r}\right), \label{chrDen}
\end{equation}
where $\rho_o$ is the free charge density inside the SM wire in the \textit{absence} of a charge impurity, $\rho_{imp}$ is the charge density associated with  the impurity, and $\rho_{red}$ accounts for the redistribution of free charge due to the presence of the impurity, i.e. describes the screening cloud. Note that $\rho_o$ is translation invariant along the direction parallel to the wire, which we take as the $z$ direction.
The charge impurity is modeled as a small sphere of radius $R_{imp}$ and uniform charge density given by
\begin{equation}
    \rho_{imp}(\vec{r}) =
    \begin{cases}
    \frac{3 Q}{4 \pi R_{imp}}, & |\vec{r} - \vec{r}_{imp}| \leq R_{imp} \\
    0, & |\vec{r} - \vec{r}_{imp}| > R_{imp}
    \end{cases},     \label{Eq4}
\end{equation}
where $\vec{r}_{imp}= x_{imp} \hat{e}_x + y_{imp} \hat{e}_y$ is the position vector of the impurity. Note that, without loss of generality, we assume $z_{imp}=0$.
Finally, the free charge density is related to the occupied electronic states,
\begin{equation}
	\rho_f\left(\vec{r}\right) = 	
	\rho_o\left(x,y\right) + \rho_{red}\left(\vec{r}\right) = 
	2 \sum_n \left|\psi_n\left(\vec{r}\right)\right|^2
	f\left(E_n,T\right), \label{chrDen2}
\end{equation}
where $\rho_f$ is the total free charge density, $E_n$ and $\psi_n$ are the $n^\text{th}$ eigenenergy  of the Hamiltonian (\ref{SM_Ham}) and the corresponding eigenstate, respectively,  $f$ is the Fermi function, $T$ the temperature, and the factor of $2$ accounts for spin degeneracy. Note that Eq. (\ref{chrDen2}) couples Eqs. (\ref{SM_Ham}) and (\ref{Pois}), known as the Schr{\"o}dinger-Poisson equations. The free charge density and the electrostatic potential are given by the self-consistent solution of these equations.

Before presenting our method for solving the Schr{\"o}dinger-Poisson problem, a few comments about the model are warranted. First, note that we have neglected the key ingredients responsible for the emergence of Majorana physics in a SM-SC hybrid structure, namely proximity-induced superconductivity, spin-orbit coupling, and Zeeman splitting. These additional contributions to the effective Hamiltonian, which will be included in the finite wire model discussed in Sec. \ref{Mult}, are characterized by energy scales much smaller than the typical inter-band spacing associated with the Hamiltonian in Eq. (\ref{SM_Ham}), the potentials  $V_g$ and $V_{SC}$, and the bare potential of the charge impurity. 
In other words, the spatial profile of eigenstates $\psi_n$ and, implicitly, the charge density $\rho_f\left(\vec{r}\right)$ and the potential $\phi(\vec{r})$ are mainly determined by the terms already included in Eq. (\ref{SM_Ham}) and by the boundary conditions, while the additional terms are expected to generate small perturbations. Also note that we do not explicitly include the SC subsystem in the Hamiltonian, but consider it in the boundary conditions. 
Of course, the coupling between the SM and SC is crucial for inducing superconductivity within the SM wire through proximity effect. Moreover, it is known that the proximity coupling to the superconductor renormalizes the low-energy spectrum of the hybrid system \cite{Stanescu2017} and generates  a shift of the SM subbands \cite{Reeg2018a}. However, these effects  can be accounted for in our model by modifying the effective mass and appropriately shifting $V_{g}$ and $V_{SC}$. Consequently, to avoid the dramatic increase of the computational cost associated with including the SC in the Hamiltonian, we do not explicitly consider the SC degrees of freedom.  We stress, however, that the SC still plays an important role in our model due to the band-bending generated by the Dirichlet boundary condition imposed on $\phi$ at the SC surface.
 
\subsection{Self-consistent Schr{\"o}dinger-Poisson scheme} \label{Calc1}
  
We start by decomposing the electrostatic potential into three components,  similar to Eq. (\ref{chrDen}). Explicitly, we have
\begin{equation}
	\phi\left(\vec{r}\right) = 
	\phi_o\left(x,y\right) 
	+ \phi_{imp}\left(\vec{r}\right)
	+ \phi_{red}\left(\vec{r}\right), \label{phi1}
\end{equation}
where $\phi_o$ is the electrostatic potential in the \textit{absence} of a charge impurity, and $\phi_{imp}$ and $\phi_{red}$ are solutions of the  Poisson equation with $\rho_{imp}$ and $\rho_{red}$ as source terms, respectively. The Dirichlet boundary conditions for non-zero values of $V_g$ and $V_{SC}$ are imposed on $\phi_o$, while  $\phi_{imp}$ and $\phi_{red}$ are subject to trivial boundary conditions. Next, we rewrite the Hamiltonian as 
\begin{equation}
	H = H_o + H^\prime, \label{Ham_terms}
\end{equation}
with
\begin{align}
	H_o =& -\frac{\hbar^2}{2 m^*} \nabla^2 
	- e \phi_o\left(x,y\right), \label{Ham_o} \\
	H^\prime =&  
	- e \phi_{imp}\left(\vec{r}\right) - e \phi_{red}\left(\vec{r}\right). \label{Ham_prime} 
\end{align}
Here, $H_o$ is the Hamiltonian of the clean system (i.e., the wire without a charge impurity) and $H^\prime$ represents the perturbation due to the presence of the impurity.
We first solve the Schr{\"o}dinger-Poisson equations with $H=H_o$. Details regarding the sefl-consistent numerical procedure can be found in Refs. \cite{Woods2018, Woods2020a}. The key output of this initial calculation is a set $\{\left(\varepsilon_{\alpha,o}, \varphi_\alpha \right) |~ \alpha \in \mathbb{N} \}$ of transverse eigenenergies and corresponding eigenmodes. Note that the transverse  wavefunction $\varphi_\alpha$ satisfies the eigenvalue equation
\begin{equation}
	H_o \Big[\varphi_\alpha\left(x,y\right) e^{i k z}\Big] = 
	\left( \varepsilon_{\alpha,o} + \frac{\hbar^2 k^2}{2m^*}\right)
	\Big[ \varphi_\alpha\left(x,y\right) e^{ik z}\Big],
\end{equation}
for arbitrary values of $k$. In other words, $\varphi_\alpha$ represents the $k$-independent transverse profile of the $\alpha$ subband for a  clean system, while $\varepsilon_{\alpha,o}$ is the  energy of the corresponding band edge (i.e., bottom of the band). Since $\{ \varphi_\alpha\}$ is a complete, orthonormal set of transverse functions, we use it as a basis to expand the states of the full Hamiltonian (\ref{Ham_terms}). Explicitly, we have 
\begin{equation}
	\psi_n\left(\vec{r}\right) =
	\sum_\alpha \varphi_{\alpha}\left(x,y\right) g_{n,\alpha}(z), \label{psi_expansion}
\end{equation}
where $\psi_n$ is the $n^\text{th}$ eigenstate of Eq. (\ref{Ham_terms}) and $g_{n,\alpha}(z)$ is a yet-undetermined function of $z$.
In principle, all subbands may contribute to each eigenstate. In practice, however, only a limited number of low-energy subbands contribute significantly to the low-energy eigenstates of the Hamiltonian. We therefore project the eigenstate (\ref{psi_expansion}) of the full Hamiltonian onto a low-energy subspace defined by subbands with $\varepsilon_{\alpha,o} < \varepsilon_{cut}$, where $\varepsilon_{cut}$ is a finite cutoff energy larger than any other relevant energy scale in the problem. Note that the accuracy of this low-energy projection can be tested by increasing $\varepsilon_{cut}$, i.e., including additional transverse modes into the low-energy basis. The basis is large-enough if further increasing it generates a negligible change of the final results.

Next, we point out that introducing a charge impurity breaks the translation invariance  along the $z$ axis, making the assumption of an infinite system rather inconvenient. To address this issue, we impose periodic boundary conditions with a supercell of length $\ell$ sufficiently large so that charge impurities in neighboring supercells have a negligible effect on one another.  In these conditions, the electrostatic potential within the large supercell will be practically identical to the potential of an infinitely long system within a region of length $\ell$ containing the impurity. We introduce the following Fourier transforms of the potential and charge density
\begin{align}
    \phi_i \left(\vec{r} \right) =& 
    \sum_q \widetilde{\phi}_{i,q}(x,y) e^{iG_q z}, \label{phiImp} \\
    \rho_i \left(\vec{r} \right) =&
    \sum_q \widetilde{\rho}_{i,q}(x,y) e^{iG_q z}, \label{rhoImp}
\end{align}
where $G_q = 2 \pi q/\ell$ is a reciprocal lattice vector, $q \in \mathbb{Z}$, and $i \in \{imp, red\}$ designates different components defined in Eqs. (\ref{chrDen}) and (\ref{phi1}). Plugging Eqs. (\ref{phiImp}) and (\ref{rhoImp}) into the Poisson equation (\ref{Pois}) yields 
\begin{equation}
    \left[
    \nabla_\perp \cdot 
    \left( \epsilon \nabla_\perp
    \right) 
    - \epsilon G_q^2
    \right] \widetilde{\phi}_{i,q}(x,y) = 
    -\widetilde{\rho}_{i,q}(x,y), \label{Helm}
\end{equation}
for all possible values of $q$, where $\nabla_\perp$ is the del operator in the $x{\text -}y$ plane. This reduces the original $3\text{-dimensional}$ Poisson equation to a set of independent $2\text{-dimensional}$ inhomogeneous Helmholtz equations with imaginary wavenumbers, $iG_q$. 
Note that the imaginary wavenumber in Eq. (\ref{Helm})  suppresses $\widetilde{\phi}_{i,q}$ away from the source, $\widetilde{\rho}_{i,q}$.
These $2D$ Helmholtz equations are significantly less costly numerically, as compared to the original $3D$ Poisson equation. As a result, we are able to efficiently perform high resolution calculations of the self-consistent potential near the impurity. By contrast, achieving similar results using a brute force approach to the $3D$ Poisson equation would require a dense discretization around the impurity, which would lead to significant costs in terms of both memory and computational time.

With periodic boundary conditions, the low-energy expansion of the eigenstates of the full Hamiltonian becomes
\begin{equation}
	|n, k_z\rangle =
	\sum_{\alpha = 1}^{\varepsilon_{\alpha,o} < \varepsilon_{cut}}
	\sum_{q} 
	|\alpha,q,k_z\rangle 
	A^{n,k_z}_{\alpha,q},
\end{equation}
where $k_z \in \left(-\pi/\ell,\pi/\ell\right]$ is the crystal momentum in the $z$-direction, $A^{n,k_z}_{\alpha,q} =  \langle\alpha,q,k_z | n, k_z\rangle \in \mathbb{C}$, and the basis state $|\alpha,q,k_{z}\rangle$ is given by
\begin{equation}
	\langle \vec{r}|\alpha,q,k_{z}\rangle =
 	\varphi_\alpha \left(x,y\right) 
	e^{i \left(G_q + k_z\right) z},
\end{equation}
with the bra-ket notation introduced for convenience.
Note that $k_z$ is a good quantum number due to the translation symmetry with period $\ell$. Calculation of the Hamiltonian matrix elements yields
\begin{align}
	\langle\alpha,q,k_z | H_o
    |\beta,q^\prime,k_z\rangle =& 
   ~\varepsilon_{\alpha,q}(k_z) 
   \delta_{\alpha,\beta} \delta_{q,q^\prime}~, \\
   \langle\alpha,q,k_z | H^\prime
    |\beta,q^\prime,k_z\rangle  =&
   ~\widetilde{V}_{imp,q^\prime-q}^{\alpha,\beta}
   + \widetilde{V}_{red,q^\prime-q}^{\alpha,\beta}~,
\end{align}
where 
\begin{align}
	\varepsilon_{\alpha,q}(k_z) =& 
	\varepsilon_{\alpha,o} 
	+ \frac{\hbar^2}{2 m^*} \left( G_q +k_z\right)^2,  \\
	\widetilde{V}_{i,q}^{\alpha,\beta} =&
    -e \int \varphi_\alpha^* 
    \widetilde{\phi}_{i,q} \varphi_\beta ~dxdy, \label{Vq}
\end{align} 
with $i \in \{imp, red\}$. Using this representation, the charge density can be expressed in the following compact form 
\begin{equation}
\begin{split}
    \widetilde{\rho}_{f,q}\left(x,y\right) =& 
    \frac{-e}{L}
    \sum_{n,k_z} \sum_{\alpha,\beta} \sum_{q^\prime}
    {A^{n,k_z}_{\alpha,q^\prime}}^* A^{n,k_z}_{\beta,q^\prime+q}
     \\
    & \times f\big[E_{n}(k_z),T\big] \varphi_{\alpha}^*(x,y) \varphi_{\beta}(x,y),
\end{split}
\end{equation}
where $E_{n}(k_z)$ is the eigenenergy of the $n^\text{th}$ eigenstate with crystal momentum $k_z$. 
Eq. (\ref{Vq}) shows that both $\phi_{imp}$ and $\phi_{red}$ generically have diagonal and off-diagonal matrix elements corresponding to intra- and inter-subband couplings. Consequently, the eigenstates of the full Hamiltonian will be linear combinations of basis states involving  several transverse modes. However, if the energy spacing between subbands is significantly larger than the perturbation terms, $\widetilde{V}_{imp,q}^{\alpha,\beta}$ and $\widetilde{V}_{red,q}^{\alpha,\beta}$, with $\alpha \neq \beta$, the inter-subband mixing is small and the subband index $\alpha$ becomes an ``almost  good'' quantum number. This motivates us to consider the \textit{independent subband approximation}, in which we neglect any Hamiltonian matrix element between different subbands, i.e. $\langle\alpha,q,k_z | H|\beta,q^\prime,k_z\rangle = 0$ for $\alpha \neq \beta$, when calculating the self-consistent potential. Within this approximation, the subband index becomes a good quantum number, and we can write the eigenstates as
\begin{equation}
	|\alpha,n, k_z\rangle =
	\sum_{q} 
	|\alpha,q,k_z\rangle 
	\mathbb{A}^{n,k_z}_{\alpha,q},
\end{equation}
where $\mathbb{A}^{n,k_z}_{\alpha,q} =  \langle\alpha,q,k_z |\alpha, n, k_z\rangle$.
The free charge density reduces to 
\begin{equation}
\begin{split}
    \widetilde{\rho}_{f,q}\left(x,y\right) =& 
    \frac{-e}{L}
    \sum_{n,k_z} \sum_{\alpha} \sum_{q^\prime}
    {\mathbb{A}^{n,k_z}_{\alpha,q^\prime}}^* \mathbb{A}^{n,k_z}_{\alpha,q^\prime+q}
     \\
    & \times f\big[E_{\alpha,n}(k_z),T\big] \left|\varphi_{\alpha}(x,y)\right|^2.
\end{split}
\end{equation}
Finally, we can write the matrix elements of $\phi_{red}$ in a compact form by introducing the \textit{subband Green's function}, $\widetilde{g}_{q,\alpha}$, defined as the solution of the Poisson equation,
\begin{equation}
    \left[
    \nabla_\perp \cdot 
    \left( \epsilon \nabla_\perp
    \right) 
    - \epsilon G_q^2
    \right] \widetilde{g}_{q,\alpha}(x,y) = 
    -e \left| \varphi_\alpha(x,y)\right|^2, \label{Greens}
\end{equation}
with trivial boundary conditions, and the \textit{Green's function tensor},
\begin{equation}
    \widetilde{g}_{q,\alpha}^{\beta,\gamma} = 
    \int \varphi_\beta^*
    \widetilde{g}_{q,\alpha}
    \varphi_\gamma 
    ~dx dy.
\end{equation}
With these notations, the relevant matrix elements become
\begin{equation}
    \widetilde{V}_{red,q}^{\beta,\gamma} = 
    \sum_{\alpha} \widetilde{g}_{q,\alpha}^{\beta,\gamma} 
    n_{\alpha,q}, \label{Vred}
\end{equation}
with
\begin{equation}
    n_{\alpha,q} = \frac{1}{L}
    \sum_{n,k_z} \sum_{q^\prime}
    {\mathbb{A}_{\alpha,q^\prime}^{(n,k_z)}}^* \mathbb{A}_{\alpha,q^\prime+q}^{(n,k_z)}
    f\big[E_{\alpha,n}(k_z),T\big], \label{nqalpha}
\end{equation}
for $q \neq 0$.
If $q = 0$, the structure of  Eq. (\ref{nqalpha}) remains the same, but the quantity $n_{\alpha,0}$ associated with the clean system must be subtracted, as it is already incorporated into $\phi_o$. 
Note that, while Eq. (\ref{Vred}) gives both diagonal and off-diagonal  matrix elements, within the independent subband approximation only the diagonal contributions containing  tensor elements of the form $\widetilde{g}_{q,\alpha}^{\beta,\beta}$ are relevant for the self-consistent calculation of the potential. Also, we point out that, once $\widetilde{g}_{q,\alpha}^{\beta,\gamma}$ and $\widetilde{V}_{imp,q}^{\alpha,\beta}$ have been calculated using the self-consistent wavefunctions of the clean system, the problem  reduces to finding $n_{\alpha,q}$ self-consistently. We perform this task using a simple iterative mixing scheme. An iteration is numerically inexpensive, since each  subband corresponds to an independent 1D Schr{\"o}dinger equation that determines the eigenstates $|\alpha, n, k_z\rangle$. In fact, the largest computation cost corresponds to calculating the elements $\widetilde{g}_{q,\alpha}^{\beta,\gamma}$ of the Green function tensor. Also note that, while the subbands are independent as far as solving the Schr{\"o}dinger equation is concerned, they still affect each other through Eq. (\ref{Vred}), since $n_{\alpha,q}$ enters the expression of  $\widetilde{V}_{red,q}^{\beta,\beta}$ for all $\alpha$ and $\beta$. Therefore our independent subband approximation still captures the main contribution due to inter-subband electrostatic screening. In addition, we have explicitly checked that neglecting inter-subband coupling  has a negligible effect on the spectrum of the full Hamiltonian. 

Once the self-consistent solution is found, we Fourier transform the matrix elements of the potential back to real space and  define the {effective impurity potential matrix} elements,
\begin{equation}
	V_{\alpha,\beta}\left(z\right) =
    \sum_q \left(
    \widetilde{V}_{imp,q}^{\alpha,\beta} 
    + \widetilde{V}_{red,q}^{\alpha,\beta}
    \right)
    e^{iG_q z}. \label{effPot}
\end{equation}
These quantities provide information regarding the amplitude and characteristic length scale of the potential inhomogeneity induced by the charge impurity.
Note that the diagonal element $V_{\alpha,\alpha}\left(z\right)$ can be interpreted as an effective 1D potential for the $\alpha$ subband. 
On the other hand, the  off diagonal element $V_{\alpha,\beta}\left(z\right)$ couples the subbands $\alpha$ and $\beta$  in a position dependent manner. 

For the numerical calculations we choose parameter values that roughly correspond to the currently existing InAs-Al and InSb-Al  nanowire-superconductor platforms, while being somewhat on the reasonably optimistic side.  We emphasize that our qualitative and semi-quantitative conclusions do not depend on the details of our parameter choice. Specifically, we have used the following parameter values: radius of the SM nanowire $R=35~$nm, dielectric thickness $d=10~$nm, superconductor thickness $W_{SC}=10~$nm, SM permittivity $\epsilon_{SM}=15.15$, dielectric permittivity $\epsilon_{d}=24$, effective mass $m_{eff}=0.023$, work function difference $V_{SC}=110~$meV, radius of the charge impurity $R_{imp}=2.5~$nm, supercell size $\ell = 500~$nm, energy cutoff for the transverse modes $\epsilon_{cut}=20~$meV, kinetic energy cutoff of plane waves along the $z$ direction $\epsilon_{cut}^{kin}=3~$eV, Fourier coefficients satisfying $|q| \leq 200$ are used for the electrostatic potential and charge density expansions, and transverse mesh spacing within the semiconductor for the Poisson, Helmholtz, and Schr{\"o}dinger equations $a_{SM}=1~$nm. 

\subsection{Results} \label{Results1} 

To understand the qualitative and quantitative characteristics of the effective potential generated by a charge impurity embedded inside the semiconductor wire, we start with a calculation of the impurity potential $\phi_{imp}$, which corresponds to the second term in the decomposition given by Eq. (\ref{phi1}). We note that $\phi_{imp}$ is the solution of the Poisson equation (\ref{Pois}) with a source term given by $\rho_{imp}$ from Eq. (\ref{Eq4}) and homogeneous Dirichlet boundary conditions on the surface of the superconductor and the metallic gate. Consequently, in addition to the bare $1/r$ potential of the charge impurity, $\phi_{imp}$ includes the screening effect due to the presence of the SC layer and metallic back gate. However, it does not include the screening effect due to the redistribution of the free charge within the wire, which corresponds to $\phi_{red}$ in Eq. (\ref{phi1}).

\begin{figure}[t]
\begin{center}
\includegraphics[width=0.4\textwidth]{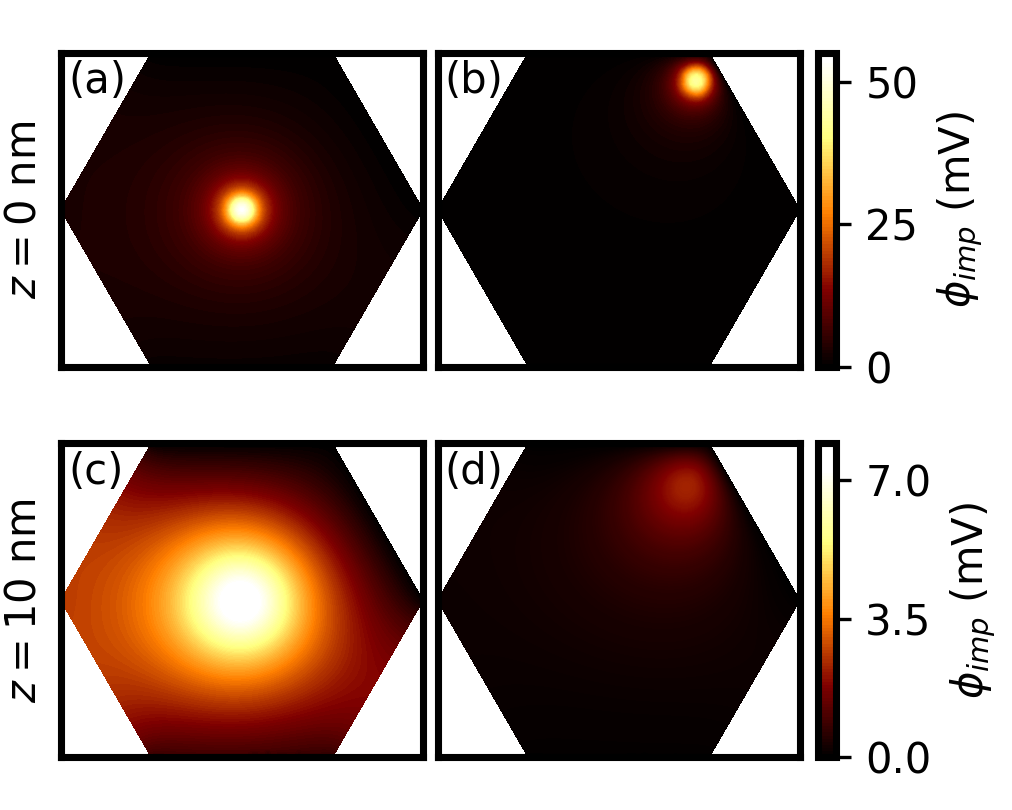}
\end{center}
\vspace{-7mm}
\caption{Impurity potential maps, $\phi_{imp}(x, y)$, within the semiconductor region for two impurity locations: middle of the wire, $(x_{imp},y_{imp})=(0, 0)$ [panels (a) and (c)]  and close to the SM-SC interface (top and upper right facets), $(x_{imp}, y_{imp})=(15, 25)~$nm,  [panels (b) and (d)].  The potential amplitudes at $z=0$, i.e., in plane containing the impurities, are comparable (top panels), while at $z=10~\text{nm}$ the potential of the central impurity is much stronger than the potential generated by the other impurity (lower panels) as a result of  weaker screening by the superconductor. Note the different energy scales for the upper and lower panels. The impurity charge $Q =e$ is used for both impurity locations.}
\label{FIG2}
\vspace{-1mm}
\end{figure}

Maps of the screened potential amplitudes at $z=0$ (i.e., in the plane containing the impurity) and $z=10~$nm for two different impurity locations are shown in Fig. \ref{FIG2}. The left panels correspond to an impurity located in the middle of the wire, while the right panels show the potential of an impurity located near the SM-SC interface. While at $z=0$ the potentials generated by the two impurities are comparable (see top panels in Fig. \ref{FIG2}), further away the potential of the central impurity is much stronger that the potential generated by the other impurity (lower panels). This indicates that the potential of the impurity located near the SM-SC interface has a significantly shorter decay length than the central impurity, which is the result of a stronger screening by the superconductor. We conclude that, while the characteristic length scale of the screened potential depends strongly on the location of the impurity relative to the SM-SC interface and the back gate, the maximum amplitude of $\phi_{imp}$ is on the order of tens of meV regardless of the location of the charge impurity. This is at least one order of magnitude larger than the characteristic energy scale associated with Majorana physics. Without additional screening, the presence of charge impurities inside the hybrid device would have catastrophic effects on the stability of topological superconductivity and Majorana zero modes.
This is a quantitative finding of extreme importance in the search for Majorana zero modes, as it clearly reveals the fragility of the quantum energy scale associated with Majorana physics (e.g., the topological gap $\sim 0.1~$meV or less), which can be easily overwhelmed by the huge (essentially classical) impurity energy scale ($\sim 10~$meV). This further emphasizes the critical need for clean samples and the role of screening in limiting the impurity-induced potential.

Next, we perform the full, self-consistent Schr{\"o}dinger-Poisson calculation and determine the effective impurity potential matrix elements defined by Eq. (\ref{effPot}). For concreteness, we focus on a system that, in the absence of the impurity, has the bottom of the fourth subband at the chemical potential, which is realized by properly tuning the gate potential $V_g$. Since Majorana physics is controlled by the top occupied subband, the relevant effective potential matrix elements $V_{\alpha,\beta}$ correspond, in this case, to $\alpha = 4$ and $\beta=3, 4, 5$, with the diagonal element $V_{4,4}(z)$ representing the intra-subband effective impurity potential and the off-diagonal elements $V_{4,3}$ and $V_{4,5}$ providing a measure of the impurity-induced inter-subband coupling. 
\begin{figure}[t]
\begin{center}
\includegraphics[width=0.45\textwidth]{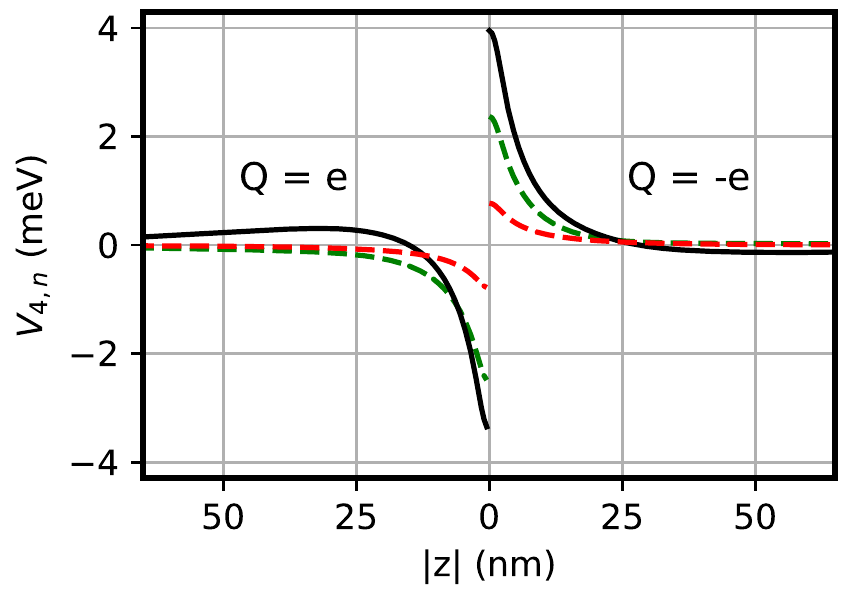}
\end{center}
\vspace{-7mm}
\caption{The dependence of the effective potential matrix elements on the distance $|z|$ from the plane containing the impurity for a system having the chemical potential near the bottom of the fourth subband. The left side corresponds to a positively charged impurity with $Q=+e$, while the right side corresponds to a negative charge, $Q=-e$. Both impurities are located at $(x_{imp}, y_{imp})=(23, 0)~$nm. 
The black solid lines correspond to the relevant intra-subband effective potential, $V_{4,4}$, while the green and red dashed lines represent the inter-subband matrix elements, $V_{4,3}$ and $V_{4,5}$, respectively.}
\label{FIG3}
\vspace{-1mm}
\end{figure}
The dependence of the effective potential matrix elements on the distance $|z|$ from the plane containing the impurity is shown in Fig. \ref{FIG3}. We consider two cases: positive charge impurity, $Q=+e$ (left side of Fig. \ref{FIG3}), and negative charge impurity, $Q=-e$ (right side). In both cases the location of the impurity in the transverse plane is given by $(x_{imp}, y_{imp})=(23, 0)~$nm. 
 First, we note that the off-diagonal contributions  are smaller than, but comparable to the diagonal term. If the inter-subband spacing is much larger than $\Delta E\sim 1~$meV, the impurity-induced inter-subband coupling is negligible and one can accurately describe the system within the independent subband approximation. If, on the other hand, the inter-subband spacing is comparable to (or lower than) $\Delta E$, inter-subband coupling becomes important and the system has to be treated explicitly as a multi-subband system. In this scenario, the system is expected to be prone to the formation of topologically-trivial low-energy states  due to impurity-induced inter-subband coupling \cite{Woods2019b}. On the other hand, in the independent-subband regime the system is expected to be less sensitive to  impurity-induced disorder. This study focuses on the more favorable scenario involving well separated subbands. We note that accessing this regime depends critically on ensuring low subband occupancy \cite{Woods2020a}. 
We emphasize that in systems characterized by small inter-subband energy separation, which is generically the case at high occupancy (e.g., for  $alpha > 10$), inter-subband coupling may prevent the realization of a robust topological phase even in the absence of disorder \cite{Woods2019b}. Here, we focus  on the situation corresponding to large inter-subband energy splittings and low subband occupancies, where the inter-subband coupling (induced  by, e.g., charge impurities) can be safely neglected. Note that, in principle,  the subband occupancy can be kept low by properly tuning the gate voltage, $V_g$.

The diagonal matrix elements (full black lines in Fig. \ref{FIG3}) are characterized by amplitudes of a few meV and decay lengths on the order of $10~$nm. In general, the amplitude of the potential generated by a negative charge is slightly larger that the amplitude of a positive charge potential corresponding to the same subband and impurity location. 
This is a screening effect arising from the free charge being made of electrons,  which are more effective in screening a positively charged impurity.
Note that the dependence of $V_{4,4}$ on $z$ is not monotonic, being characterized by a fast decay at short distances followed by a change of sign and a slow decay at long distances. Remarkably, this behavior, which turns out to be quite generic, is well captured by the following empirical function 
\begin{equation}
V_{\alpha,\alpha}(z) = B^{\alpha}_{imp} e^{-|z|/\lambda^\alpha_{imp}} - B^{\alpha}_{red} e^{-|z|/\lambda^\alpha_{red}},      \label{Vfit}
\end{equation}
where the four fitting parameters, $B^{\alpha}_{imp}$, $B^{\alpha}_{red}$, $\lambda^\alpha_{imp}$, and $\lambda^\alpha_{red}$, depend on the band index, $\alpha$, and also on the specific location of the impurity, $(x_{imp}, y_{imp})$. Details regarding the fitting procedure, its accuracy, and numerical fitting parameters are provided in Appendix \ref{Fit}. Note that the first and second terms in Eq. (\ref{Vfit}) account for the effective impurity and redistribution potentials, respectively. Moreover, while there are four fitting parameters in the Eq. (\ref{Vfit}), which, in principle, are independent, we show in Appendix \ref{Fit} that correlations between the fitting parameters imply that one only needs to input $B^{\alpha}_{imp}$ or $B^{\alpha}_{imp}$ and $\lambda^\alpha_{imp}$, i.e., two independent parameters, to obtain a realistic disorder potential.
This has two major implications. First, to understand the dependence of the effective impurity potential on the band index and the position of the impurity, it is enough to study the dependence of the amplitude and decay length on these parameters, which substantially simplifies our analysis. Second, the simple form of 
Eq. (\ref{Vfit}) provides an extremely useful phenomenological model for describing charge impurities embedded within SM-SC hybrid devices. Combined with our quantitative results described below, this enables the study of disorder generated by charge impurities without actually performing a full,  numerically-intensive Schr{\"o}dinger-Poisson calculation. The validation of the relatively simple empirical fitting of the impurity potential defined by Eq. (\ref{Vfit}) is an important result of our work.

Our next task is to determine the dependence of the amplitude and decay length characterizing the effective disorder potential $V_{\alpha,\alpha}(z)$ on the position of the impurity and the subband index. Here, we define the amplitude as $V_{\alpha,\alpha}(z=0)$, while the decay length $\xi_\alpha$  is obtained by finding $z$ such that $V_{\alpha,\alpha}(z) = V_{\alpha,\alpha}(0) \exp({-1})$. We emphasize that, within the independent subband regime, the only relevant matrix element is the diagonal element corresponding to the top occupied subband. In turn, the occupancy of the SM subbands is controlled by the applied gate potential $V_g$.  

\begin{figure}[t]
\begin{center}
\includegraphics[width=0.45\textwidth]{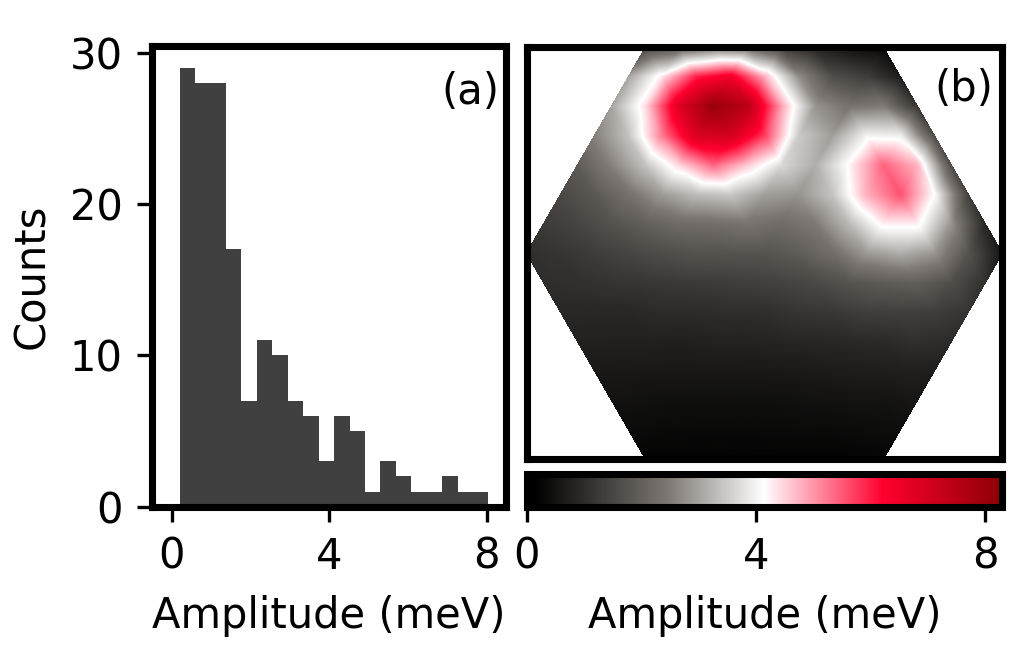}
\end{center}
\vspace{-7mm}
\caption{(a) Histogram of the effective potential amplitude $V_{2,2}(0)$. Data taken from 169 impurity locations sampled evenly over the hexagonal cross-section. (b) Effective potential amplitude $V_{2,2}(0)$ as a function of the impurity position, $(x_{imp}, y_{imp})$. Note that the largest amplitude corresponds to locations  where the second transverse mode has high spectral weight.}
\label{FIG4}
\vspace{-1mm}
\end{figure}

To acquire some intuition, we first consider a specific example involving a system having the bottom of the second subband near the chemical potential. A map showing the dependence of the amplitude $V_{2,2}(0)$ on the position of the impurity is provided in Fig. \ref{FIG4}(b). We note that the amplitude of the effective impurity potential depends strongly on the position of the impurity. The largest amplitude corresponds to locations  where the second transverse mode has high spectral weight. This is not surprising, considering that $V_{2,2}(0)$ is a matrix element of a short range quantity over the second subband. Also note that, as a result of having a finite work function difference, $V_{SC}$, the lowest energy modes tend to be localized in the vicinity of the SM-SC interface.  Higher energy modes, on the other hand, are more evenly spread over the cross section of the wire. The subband-dependent  amplitude of the effective impurity potential $V_{\alpha,\alpha}(0)$ exhibit a similar dependence on the position of the charge impurity. To describe quantitatively the distribution of potential amplitudes, we generate a histogram of the amplitude corresponding to 169 impurity locations sampled evenly over the hexagonal cross-section of the wire. The results are shown in Fig. \ref{FIG4}(a). Note that, as a result of the second subband being localized near the SM-SC interface, the distribution is  skewed toward lower amplitudes. For higher energy modes, the amplitude distributions are more uniform, as a consequence of the wider distribution of spectral weight associated with those modes. 

\begin{figure}[t]
\begin{center}
\includegraphics[width=0.4\textwidth]{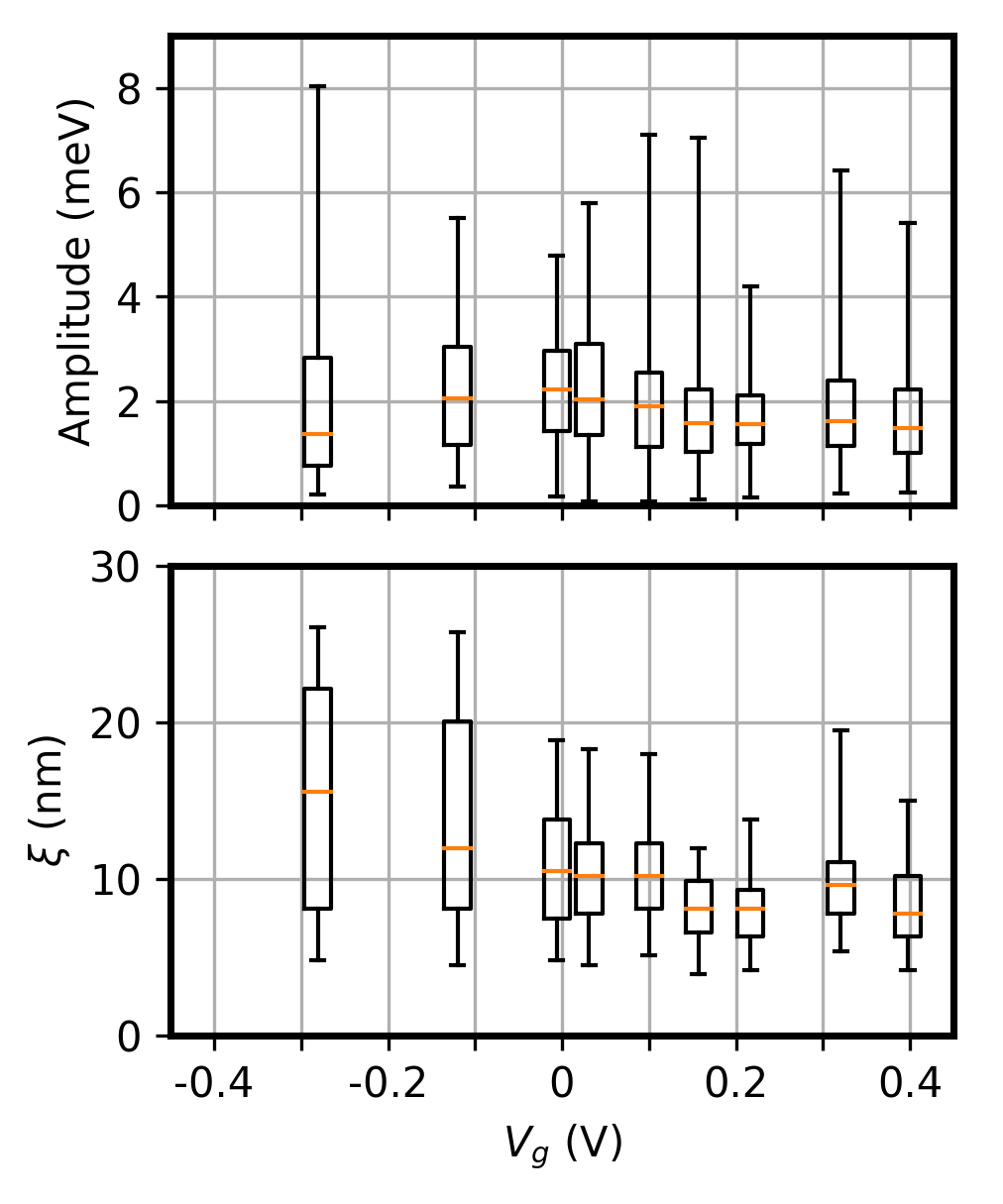}
\end{center}
\vspace{-7mm}
\caption{Distributions of the intra-subband effective potential amplitude, $V_{\alpha,\alpha}(0)$, (a) and decay length, $\xi_\alpha$, (b) for the subbands $\alpha=2-9$ with $Q = -e$. Note that the distributions corresponding to $\alpha=1$ are not shown. The bottom of each subband is  tuned to the Fermi level by adjusting the gate potential $V_g$. The orange lines indicate the median ($50$\%) of the distribution,  the boxes correspond to the $25-75$\% range, and the whiskers mark the upper and lower boundaries of the distribution. Each subband distribution is sampled over 169 impurity locations evenly distributed over the hexagonal cross-section of the semiconductor wire.}
\label{FIG5}
\vspace{-1mm}
\end{figure}

Our analysis of the position dependence of $V_{2,2}(0)$ suggests that, in general,  a compact characterization of the potential amplitude $V_{\alpha,\alpha}(0)$ can be obtained by simply focusing on the distribution obtained by sampling the hexagonal cross-section of the wire. Note that the effective potential $V_{\alpha,\alpha}$ is relevant when the bottom of the corresponding subband is in the vicinity of the Fermi level. 
We characterize the distributions by specifying the minimum and maximum values of the potential amplitude, as well as the values corresponding to the median ($50$\%), $25$\%, and $75$\%. 
A similar procedure can be used to compactly characterize the distribution of decay lengths. The results for subbands $2-9$ are shown in Fig. \ref{FIG5}.  The orange lines indicate the median ($50$\%),  the boxes correspond to the $25-75$\% range, and the whiskers mark the upper and lower boundaries of the distribution. We note that the distributions corresponding to a given subband $\alpha$ were obtained for a value of the applied gate potential $V_g$ that tunes the bottom of the subband near the chemical potential. As indicated in Fig. \ref{FIG5}, for $V_g=0$ the system has the fourth subband near the chemical potential. Accessing lower energy subbands requires depleting the wire, i.e., applying a negative gate potential. Higher energy bands, on the other hand, become relevant at positive $V_g$ values.  We note that the typical values of the effective potential amplitude are on the order of $2~$meV, significantly larger than the typical superconducting energy scales associated with Majorana physics. The typical decay lengths are in the range $8-12~$nm for $\alpha\geq 3$, while the lowest energy subbands are characterized by longer (typical) decay lengths and wider distributions due to the localization of the corresponding transverse modes near the SM-SC interface.

\begin{figure}[t]
\begin{center}
\includegraphics[width=0.4\textwidth]{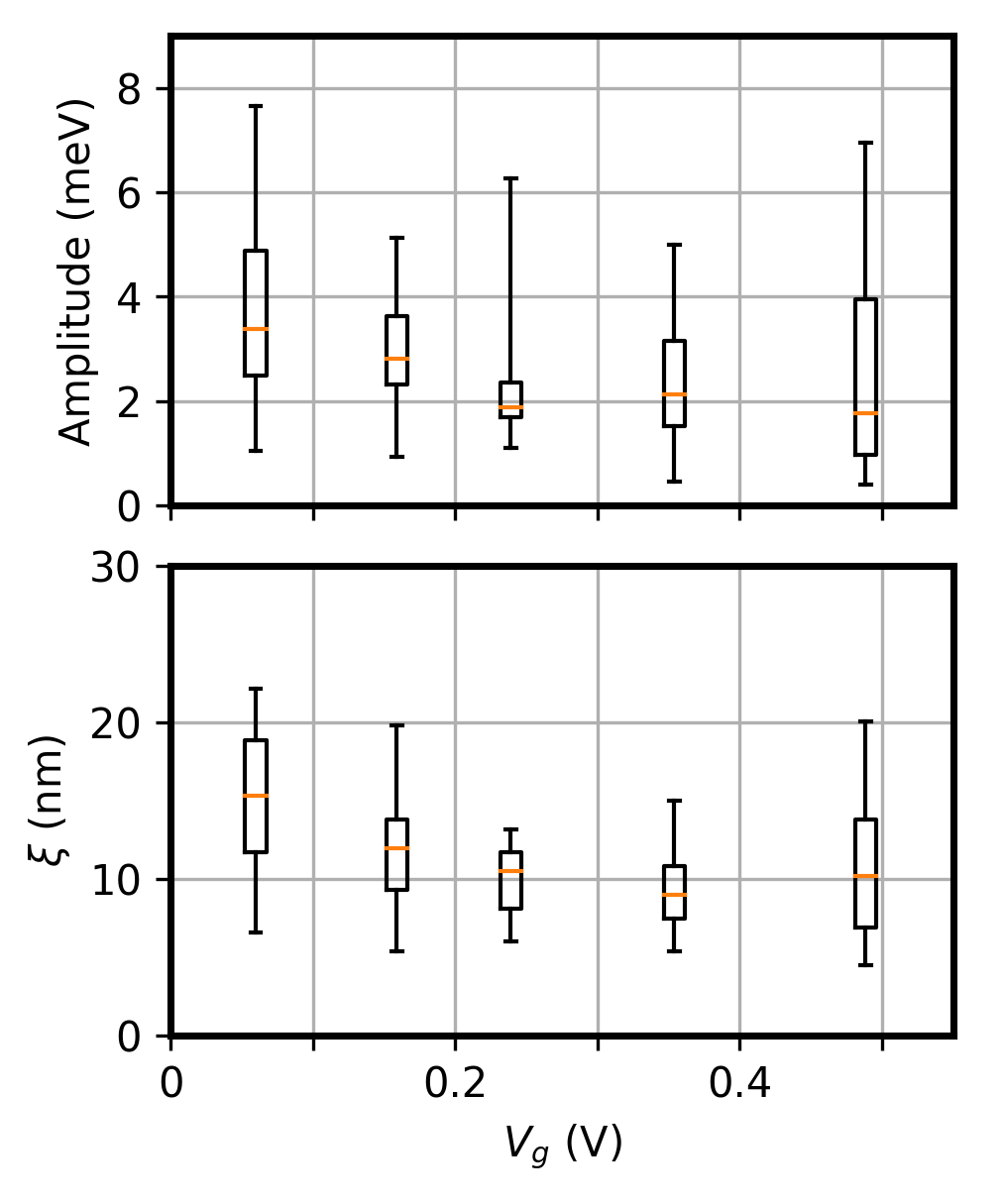}
\end{center}
\vspace{-7mm}
\caption{Same as Fig. \ref{FIG5}, but for a system without a superconductor layer. The distributions correspond to even subbands with index (from left to right) $2 \leq \alpha \leq 10$. Note that, as compared to the results shown in Fig. \ref{FIG5}, the typical values of the effective impurity potential amplitude are larger by at most a factor of two, while the typical decay lengths are only slightly larger, which indicates that screening by the superconductor has a rather limited effect.}
\label{FIG6}
\vspace{-1mm}
\end{figure}

An important question that can be raised at this point concerns  the role of the superconductor in screening the impurity potential. To address it, we consider a charge impurity embedded inside a semiconductor wire in the absence of the superconductor layer. Note that the only change with respect to the calculations described above is the elimination of the Dirichlet boundary condition $\phi=V_{SC}$ at the SC surface. The distributions of the intra-subband effective potential amplitude, $V_{\alpha,\alpha}(0)$, and decay length, $\xi_\alpha$, for the even subbands with $2 \leq \alpha \leq 10$ are shown in Fig. \ref{FIG6}. Note that in the absence of superconductor-induced band bending the values of $V_g$ associated with different subbands are different from the corresponding values in Fig. \ref{FIG5}. The key result of this calculation, which is revealed by the comparison of Figs. \ref{FIG5} and \ref{FIG6},  is that screening by the superconductor does not generate a dramatic effect, as it reduces the typical amplitude of the effective impurity potential by at most a factor of two and slightly shortens the typical decay length.  This behavior is mainly  due to the fact that the impurities inside the SM wire are typically located too far from the SM-SC interface for the superconductor  to drastically screen out the impurity potential.

\begin{figure}[t]
\begin{center}
\includegraphics[width=0.45\textwidth]{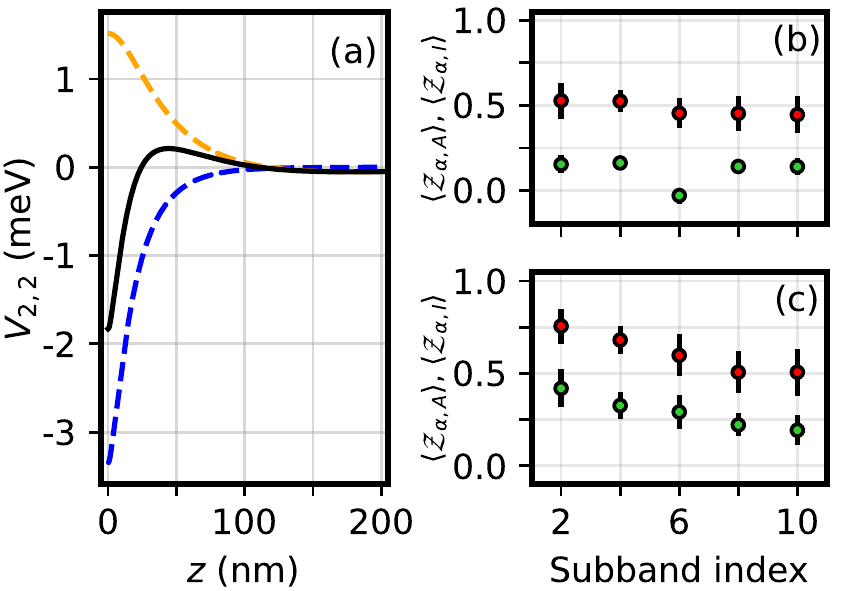}
\end{center}
\vspace{-4mm}
\caption{(a) Position dependence of the effective impurity potential $V_{2,2}$ (black line) and its impurity component,  $V_{2,2}^{imp}$ (blue dashed line), and free charge redistribution component, $V_{2,2}^{red}$ (orange dashed line), for a positively charged impurity placed at $x_{imp} = -18~\text{nm}$, $y_{imp} = 10~\text{nm}$ inside a wire having the second subband tuned to the Fermi level. (b) Average amplitude screening, $\mathcal{Z}_{\alpha}^{(A)}$ (red circles), and average integrated screening $\mathcal{Z}_{\alpha}^{(I)}$ (green circles), for a positively charged impurity, $Q=+e$, embedded  in a wire having the chemical potential tuned near the bottom of different subbands. The bars correspond to one standard deviation,  which ranges from 0.03 to 0.13. Each subband distribution is sampled over 169 impurity locations evenly distributed over the hexagonal cross-section of the semiconductor wire. (c) Same as (b) for a negatively charge impurity, $Q=-e$. Note that the screening by the free charge of negative impurities is significantly less effective than the screening of positive impurities.}
\label{FIG6bis}
\vspace{-1mm}
\end{figure}

Another important question regards the screening of the impurity potential due to the free charge redistribution in the wire. To characterize the renormalization of the band-dependent effective potential due to free charge redistribution, we introduce the amplitude screening factor, $\mathcal{Z}_{\alpha}^{(A)}$, and the integrated screening factor, $\mathcal{Z}_{\alpha}^{(I)}$, defined as follows
\begin{eqnarray}
\mathcal{Z}_{\alpha}^{(A)} &=&  \frac{V_{\alpha,\alpha}(z = 0)} {V^{imp}_{\alpha,\alpha}(z = 0)}, \label{ZA} \\
\mathcal{Z}_{\alpha}^{(I)} &=& \frac{ \int V_{\alpha,\alpha}(z) dz }{ \int V^{imp}_{\alpha,\alpha}(z) dz }, \label{ZI}
\end{eqnarray}
where $V_{\alpha,\alpha}^{imp}\left(z\right)$ is the real-space diagonal matrix element of the impurity potential,
\begin{equation}
V_{\alpha,\alpha}^{imp}\left(z\right) = \sum_q  \widetilde{V}_{imp,q}^{\alpha,\alpha} e^{iG_q z},
\end{equation}
with $\widetilde{V}_{imp,q}^{\alpha,\alpha}$ given by Eq. (\ref{Vq}). Note that, using Eq. (\ref{effPot}), the effective potential can be written as $V_{\alpha,\alpha}(z) = V_{\alpha,\alpha}^{imp}(z) + V_{\alpha,\alpha}^{red}(z)$, where  $V_{\alpha,\alpha}^{imp}$ includes the bare impurity potential contribution and the screening by the superconductor and the metallic gate, while $V_{\alpha,\alpha}^{red}$ is the contribution due to free charge redistribution. A specific example corresponding to a positive charge impurity embedded inside a system having the chemical potential near the bottom of the second subband is given in Fig. \ref{FIG6bis} (a). Note that $V_{\alpha,\alpha}^{red}$ has a larger decay length and a smaller amplitude than the impurity potential. This is a general property responsible for the sign change of the effective potential and the ``hump'' (``dip'' for negative impurities) feature starting near $z \approx 30~\text{nm}$.

The average screening factors averaged over different transverse impurity positions for a system with different occupancy levels are shown in Fig. \ref{FIG6bis} (b) and (c) for positively and negatively charged impurities, respectively. First, note that $\mathcal{Z}_{\alpha}^{(A)}$ is a measure of short-range screening, while $\mathcal{Z}_{\alpha}^{(I)}$ takes into account long-range contributions. Since in general $V_{\alpha,\alpha}^{red}$ has a longer decay length than $V_{\alpha,\alpha}^{imp}$, we have $\mathcal{Z}_{\alpha}^{(I)} < \mathcal{Z}_{\alpha}^{(A)}$. Second, the screening by the free charge of positive impurities is significantly more effective that the screening of negative impurities. In particular the integrated screening factor, $\mathcal{Z}_{\alpha}^{(I)}$, has values smaller than $0.2$ for all subbands, indicating that the contribution from the ``hump'' feature almost cancels the contribution from the central dip. In fact, in the case of the sixth subband the average integrated screening factor for $Q = e$ is actually negative, indicating over-screening by the free charge. In addition,  we note that the screening of negative impurities is more effective when the subband occupancy increases, while in the case of positive impurities the dependence on the subband index is weak. Our analysis demonstrates that screening due to free charge redistribution in the wire is a significant effect that has to be taken into account to obtain a quantitative description of the low-energy physics in the presence of charge impurities. This is physically reasonable, since the free charge inside the SM wire resides within the same spatial region as the impurity, making its screening effect quantitatively dominant.

We conclude this section with a comment on the relevance of the results obtained here to understanding Majorana physics in semiconductor-superconductor structures. On the one hand, the matrix elements of the effective impurity potential obtained numerically from the self-consistent solution of the Schr{\"o}dinger-Poisson problem can be used to investigate hybrid devices containing a finite number of randomly distributed charge impurities. The single impurity matrix elements should represent an excellent approximation, as long as the typical distance between neighboring impurities is much larger than the characteristic decay length $\xi$, so that each impurity can be considered as independent. In addition to the ``high energy'' ingredients described in Sec. \ref{Model1}, the model used in this type of investigation should include the key ingredients necessary for the emergence topological superconductivity, i.e., proximity-induced superconductivity, spin-orbit coupling, and Zeeman splitting. We pursue this path in the next section. On the other hand, the single impurity results described above can be used to construct phenomenological models with an effective impurity potential given by 
in Eq. (\ref{Vfit}) and relevant parameters, i.e., amplitude $A_\alpha$ and decay length $\xi_\alpha$, having distributions similar to those shown in Fig. \ref{FIG5} (see Appendix \ref{Fit} for more details regarding the construction of phenomenological models). This type of approach enables the efficient investigation of the disordered system over a large parameter space without the need to address a numerically demanding three-dimensional Schr{\"o}dinger-Poisson problem. Hence, in addition to the results discussed below, Majorana device modeling should indirectly benefit from our phenomenological characterization of the impurity potential given by Eq. (\ref{Vfit}).

\section{Multiple charge impurities} \label{Mult} 

In this section we consider a hybrid nanowire with multiple embedded charge impurities and investigate the effect of the impurity-induced  potential on the low energy physics, focusing on the fate of the Majorana zero energy modes that emerge in the clean system. Our analysis is based on two working assumptions. i) We consider systems with low/intermediate impurity concentrations, which are characterized by average distances between neighboring impurities that are much larger than the characteristic length of the effective (single) impurity potential. This allows us to work within the independent impurity approximation, in which each charge impurity generates an effective potential that is independent of the presence of other impurities and can be described using the approach discussed in the previous section. ii) We assume that the inter-subband spacing is much larger than all other relevant energy scales. This allows us to work within the independent band approximation, which neglects the effects of inter-subband coupling. Within this approximation, the low-energy physics can be accurately captured using an effective single band model. We note that  the independent band approximation is expected to break down in systems with high subband occupancy \cite{Woods2020a}. Also note that in systems with low inter-subband spacing the effects of impurity-induced disorder are expected to be significantly stronger than the effects described below, due to  additional contributions from impurity-induced inter-subband couplings \cite{Woods2019b}. So, the situation discussed here is, in some sense, the most optimistic scenario conducive to the emergence of topological Majorana modes; strong disorder, high subband occupancy, and, implicitly, small inter-subband spacing will simply make the situation worse, with topological physics being practically impossible to achieve in SM-SC hybrid platforms.
The effective single band model for a hybrid wire with multiple charge impurities is introduced in Sec. \ref{Model2}. The results of our numerical analysis are discussed in Sec. \ref{Results2}.

\subsection{Model} \label{Model2}

Within the independent subband approximation, the system can be described using an effective one-dimensional single-band model \cite{Lutchyn2010,Oreg2010} defined by the Bogliubov-de Gennes (BdG) Hamiltonian,
\begin{equation}
\begin{split}
	H =& 
	\left(
	-\frac{\hbar^2}{2 m^*} \partial_z^2 - \mu 
	-i \alpha_R \partial_z \sigma_y + \Gamma \sigma_z
	\right) \tau_z \\
	& - \Delta \sigma_y \tau_y + V_{imp}\left(z\right) \tau_z,
	\label{IsoHam}
\end{split}
\end{equation}
where $m^*$ is the effective mass, $\mu$ is the chemical potential, $\alpha_R$ is the Rashba spin-orbit coupling coefficient, $\Gamma$ is the Zeeman energy, $\Delta$ is the induced superconducting pairing, $V_{imp}$ is the effective potential generated by the presence of charge impurities, and $\sigma_i$ and $\tau_i$, with $ i = x,y,z$, are Pauli matrices in spin and particle-hole spaces, respectively. Note that all parameters in Eq. (\ref{IsoHam}) are assumed to be position independent, and we use the values $m^* = 0.023$, $\alpha_R = 20~\text{meV}\cdot\text{nm}$, and $\Delta = 0.3~\text{meV}$ unless stated otherwise. 

On the other hand, the impurity potential has the form 
\begin{equation}
	V_{imp}\left(z\right) =
	\sum_{m = 1}^{N_{imp}} 
	V_{\alpha, \alpha} \left(z - z_m; Q_m, x_m, y_m\right), \label{ImpPotential}
\end{equation}
where $N_{imp}$ is the total number of impurities embedded within the wire, 
$V_{\alpha,\alpha}$ is the effective potential generated by a single impurity, i.e., the intra-subband matrix element given by  Eq. (\ref{effPot}),  ${\bm r}_m=(x_m, y_m, z_m)$ describes the position of impurity $m$, and $Q_m$ indicates its charge. We assume charge neutrality and consider an equal number of positive ($Q = +e$) and negative ($Q = -e$) elementary charges distributed randomly throughout the wire. Each disorder realization corresponds to a specific set of $N_{imp}$ impurity positions $\{{\bm r}_m\}$ and a set of $N_{imp}$ charges $\{Q_m\}$.  Note that $\left(x_m,y_m\right)$ can take $169$ different values sampled evenly over the hexagonal cross-section of the nanowire, while $z_m$ can take any value corresponding to a lattice site of the discretized version of Eq. (\ref{IsoHam}) with $a_z = 4~\text{nm}$ being the lattice spacing.  For concreteness, we assume that chemical potential is tuned  near the bottom of the second subband, so that the relevant matrix elements  $V_{\alpha,\alpha}$ entering  Eq. (\ref{ImpPotential}) correspond to $\alpha=2$. These matrix elements are calculated self-consistently following the procedure described in Sec. \ref{Calc1}. The low-energy eigenvalues and the corresponding eigenstates of the Hamiltonian (\ref{IsoHam}) are then obtained using the Lanczos method \cite{ARPACK}.

To facilitate the connection with experimental tunneling spectroscopy, we also calculate the differential conductance for charge tunneling into the left or the right end of the wire. This is realized by connecting the proximitized wire to semi-infinite leads at both ends and using the Blonder-Tinkham-Klapwijk (BTK) formalism \cite{Blonder1982}. The normal leads are modeled by the Hamiltonians,
\begin{equation}
	H_{L(R)} = 
	\left(-\frac{\hbar^2}{2 m^*} \partial_z^2 - \mu_{l} 
	+ V_{L(R)}\left(z\right) \right) \tau_z, \label{HamLead}
\end{equation}
where the labels $L$ and $R$ designate the left and right leads, respectively, $\mu_l$ is the chemical potential of the leads, and $V_{L}$ and $V_{R}$ are tunnel barrier potentials at the left and right ends of the system, respectively. 
The tunnel barriers are square potential barriers of amplitude $V_{B}$ and length $L_{B} = 20~\text{nm}$  located at the ends of the corresponding leads directly adjacent to the proximitized wire. 
To evaluate the scattering matrix $S$, we consider the retarded Green's function,
\begin{equation}
	\mathcal{G}\left(\omega\right) = 
	\left[ 
	\omega - \bar{H} 
	- \Sigma_L\left(\omega\right) 
	- \Sigma_R\left(\omega\right)
	+i \eta
	\right]^{-1}, \label{Greens}
\end{equation}
where $\bar{H}$ is the (discretized) Hamiltonian containing the sites within the proximitized region, as well as the barrier sites, plus one additional site on each side of the system, immediately outside the corresponding  barrier region, $\Sigma_L$ and $\Sigma_R$ are the self-energies obtained by integrating out the degrees of freedom associated with the left and right leads \cite{Sancho1985}, respectively, and $\eta \in \mathbb{R}^{+}$ accounts for dissipative broadening \cite{Stenger2017,Liu2017}. The boundary elements of the Green's function (\ref{Greens}) are calculated using the recursive Green's function algorithm \cite{Wimmer2009}. In turn, these elements can be related to the scattering matrix, $S$, using the Fisher-Lee relations \cite{Fisher1981}. Finally, the scattering matrix elements are used to calculate the local conductance \cite{Blonder1982}, 
\begin{equation}
	G_{i} = \frac{e^2}{h}
	\left(2 - Tr({S^{ee}_{ii}}) + Tr({S^{eh}_{ii}})\right),
\end{equation}
where ${S^{ee}_{ii}}$ and ${S^{eh}_{ii}}$ describe the reflection of incoming electrons with energy $\omega$ into electrons and holes, respectively, and $i=\text{L, R}$.
The numerical values of the parameters used in the diferential conductance calculations are  $\mu_{l} = 20~\text{meV}$, 
 $V_{B} = 40~\text{meV}$, $L_{B} = 20~\text{nm}$, and  $\eta = 20~\mu\text{eV}$.

Before discussing the results, a few comments are warranted. By taking the effective impurity potential, $V_{imp}$, in Eq. (\ref{ImpPotential}) to be a sum of single impurity potentials, we are neglecting any change of the potential due to inter-impurity coupling. This is expected to be a good approximation, provided the typical spacing between charge impurities is larger than the single impurity potential decay length, i.e., in the low/intermediate impurity density regime.
The results shown in Fig. \ref{FIG5} indicate that the decay length is in the range $\xi \approx 5 - 25~\text{nm}$, which is significantly less than the typical impurity separation length for low/intermediate impurity densities. Note that for higher impurity densities we find that  Majorana physics is completely destroyed by disorder, a conclusion that is unlikely to be modified by including  inter-impurity coupling effects. The fact that strong disorder destroys the Majorana physics in nanowires and other superconducting systems is now well-accepted.
	
\begin{figure}[t]
\begin{center}
\includegraphics[width=0.45\textwidth]{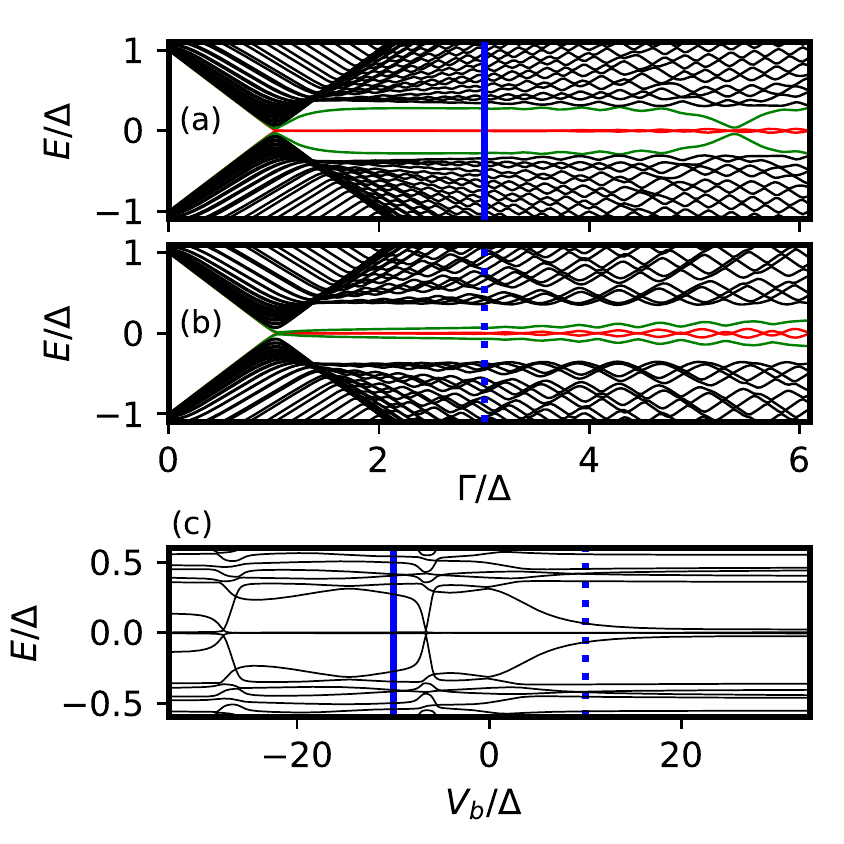}
\end{center}
\vspace{-7mm}
\caption{Low-energy spectrum as a function of Zeeman splitting for a wire of length $L = 4.2~\mu\text{m}$ having a square potential well (a) or barrier (b)  localized near its center. The width of the square potential is $L_b=50~\text{nm}$ and its height is (a) $V_b= -10\Delta$ and (b) $V_b = 10\Delta$.  Red and green lines correspond to the first and second lowest energy modes, respectively. (c) Spectrum as a function of $V_b/\Delta$ for a Zeeman field $\Gamma = 3\Delta$. Blue solid and dashed lines indicate matching parameters in panel (c) and panels (a) and (b), respectively.}
\label{FIGM1}
\vspace{-1mm}
\end{figure}
	
Finally, we note that the generalization of the single-band formalism discussed here to a multi-subband approach is straightforward. The generalized effective model is a one-dimensional  multi-subband model with inter-subband coupling induced by the off-diagonal matrix elements of the effective potential, 
 $V_{\alpha,\beta}$, with $\alpha \neq \beta$. As shown in Sec. \ref{Results1}, these elements are  typically smaller than, but comparable to the corresponding diagonal elements (see Fig. \ref{FIG3}). 
 The inter-subband coupling terms are expected to become relevant  when the inter-subband spacing $\Delta E$ between subbands close to the Fermi level is comparable to the magnitude of $V_{\alpha,\beta}$, which implies  $\Delta E\lesssim 1~$meV. For the case investigated here, which  corresponds to the second subband being tuned near the chemical potential, the inter-subband spacing is $\Delta E \sim10~\text{meV}$,  significantly larger than the amplitude of the effective potential matrix elements. Consequently, we can safely ignore the disorder-induced inter-subband coupling. High occupancy, on the other hand, is associated with a reduction of the inter-subband spacing \cite{Woods2020a} and a multi-subband approach becomes necessary. We emphasize that in the multi-subband regime the system is less robust against disorder \cite{Bagrets2012,Woods2019b}. Therefore, our independent-subband treatment provides upper bounds for impurity concentrations consistent with various aspects of Majorana physics.   
 In other words, we are considering the most favorable scenario in order to predict the upper bound on the allowed disorder that would still enable topological Majorana physics to emerge in realistic SM-SC structures.

\subsection{Results} \label{Results2}

\begin{figure}[t]
\begin{center}
\includegraphics[width=0.45\textwidth]{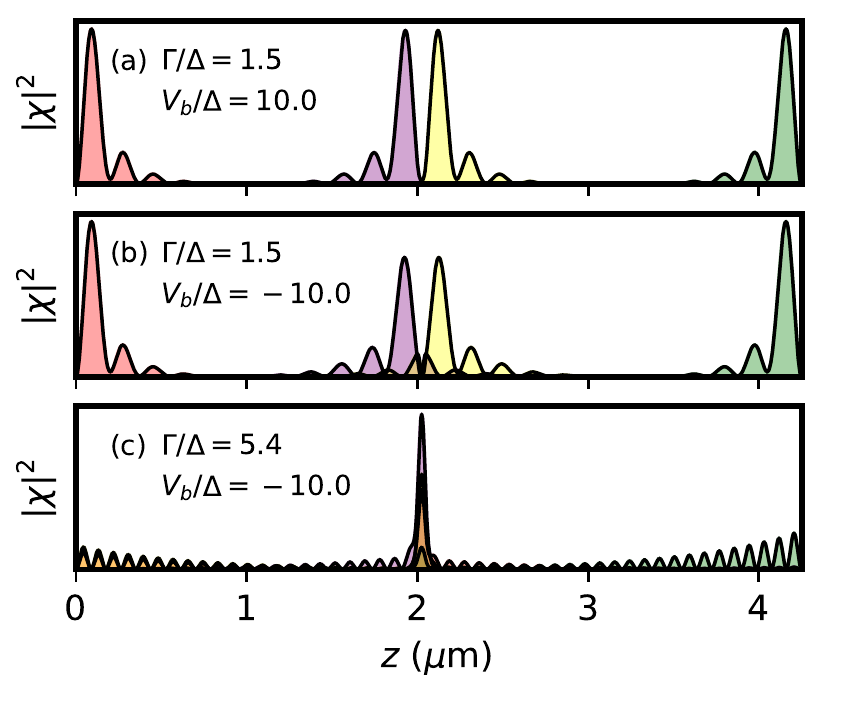}
\end{center}
\vspace{-7mm}
\caption{Position dependence of the amplitude of the Majorana wave functions, $|\chi_n^A|^2$ and $|\chi_n^B|^2$, corresponding to the lowest energy states ($n=1,2$) in Fig. \ref{FIGM1}. The values of the potential height $V_b$ and Zeeman field $\Gamma$ are indicated inside each subplot. Note that the lowest energy states  ($n=1$, red lines in Fig. \ref{FIGM1}) correspond to a pair of Majorana modes localized near the two ends of the nanowire (red and green modes), while the potential-induced in-gap states ($n=2$, green lines in Fig. \ref{FIGM1}) correspond to a pair of (partially) overlapping Majorana modes localized near the middle of the wire (purple and yellow modes). At the Andreev crossing corresponding to $\Gamma \approx 5.4\Delta$ in Fig. \ref{FIGM1}(a) the two Majorana modes completely overlap [panel (c)].}
\label{FIGM2}
\vspace{-1mm}
\end{figure}

The numerical results discussed in this section correspond to a charge neutral system containing an equal number of positively and negatively charged impurities with charges $Q=+e$ and $Q=-e$, respectively. Positive charges create local potential wells, while negatively charged impurities generate effective potential barriers. To gain some intuition regarding the effects induced by the two types of potential perturbations (i.e., ``well'' and ``barrier''), we first consider a wire of length $L = 4.2~\mu\text{m}$ having an ``artificial'' potential perturbation localized near the middle of the wire and consisting of a square potential well (barrier) of width $L_b = 50~\text{nm}$ and height $V_b= -10\Delta$ ($V_b= +10\Delta$), where $\Delta=0.3~$meV is the induced pair potential.  The dependence of the corresponding low-energy spectra on the applied Zeeman field is shown in Fig. \ref{FIGM1}, panels (a) and (b). Note that the short-range potential perturbation induces sub-gap states (green lines in Fig. \ref{FIGM1}) when the system is in the topological regime \cite{Sau2013,Moore2018}, which can act as a source of quasiparticle poisoning in Majorana qubits \cite{Karzig2021}. Also note that the characteristic energy of the in-gap mode generated by the potential barrier is much lower than the energy of the in-gap mode generated by the potential well, except for an isolated zero energy crossing at Zeeman field $\Gamma \approx 5.4\Delta$. 

The difference between the in-gap mode induced by the potential well and that generated by the potential barrier is further illustrated by the dependence of these modes on the amplitude of the square potential. This dependence is shown in fig. \ref{FIGM1}(c) for a fixed value of the Zeeman field, $\Gamma = 3\Delta$. Note that the potential well generates an in-gap mode with energy comparable to the topological gap, except a few isolated Andreev crossings. By contrast, the mode generated by the potential barrier collapses toward zero energy with increasing $V_b$. This is a specific example of a near-zero energy subgap mode induced by an inhomogeneous potential, a scenario extensively discussed in the literature.

To identify the nature of the in-gap modes, we calculate the corresponding wave functions in the Majorana representation. More specifically, let $\psi_{\pm E_n}(z)$, with $0\leq E_1 \leq E_2$, be the lowest energy eigenstates of the BdG Hamiltonian. We define the following Majorana components associated with the low-energy BdG states \cite{Chiu2016}
\begin{eqnarray}
\chi_n^A(z) &=& \frac{1}{\sqrt{2}}\left[\psi_{E_n}(z)+\psi_{- E_n}(z) \right], \nonumber \\
\chi_n^B(z) &=& \frac{i}{\sqrt{2}}\left[\psi_{E_n}(z)-\psi_{- E_n}(z) \right]. \label{chi_n}
\end{eqnarray}
Note that $\chi_n^A$ and $\chi_n^B$ are not eigenstates of the BdG Hamiltonian, except for $E_n=0$, and we have $\langle \chi_n^A|H|\chi_n^A\rangle =\langle \chi_n^B|H|\chi_n^B\rangle= 0$  and $\langle \chi_n^A|H|\chi_n^B\rangle= i E_n$. The position dependence of the amplitude  of the Majorana wave functions corresponding to the in-gap states from Fig. \ref{FIGM1} are shown in Fig. \ref{FIGM2}. The lowest energy states, $n=1$ (red lines in Fig. \ref{FIGM1}), correspond to a pair of Majorana modes localized near the two ends of the nanowire (red and green modes in Fig. \ref{FIGM2}). On the other hand, the in-gap states induced by the square potential perturbation, n=2 (green  lines in Fig. \ref{FIGM1}), correspond to a pair of (partially) overlapping Majorana modes localized near the middle of the wire (purple and yellow modes  in Fig. \ref{FIGM2}). Note that the Majorana modes generated by the potential well  [Fig. \ref{FIGM2}(b)] have a significantly stronger overlap than the Majorana modes generated by the potential barrier  [Fig. \ref{FIGM2}(a)]. Furthermore, at the Andreev crossings, the two Majorana modes $\chi_2^A$ and  $\chi_2^B$ completely overlap, generating a ``regular'' Andreev bound state localized in the potential well. In general, however, the in-gap modes generated by the local potential perturbation can be viewed as a pair of partially overlapping quasi-Majorana modes \cite{Vuik2019} or, alternatively, as a partially separated Andreev bound state (ps-ABS) \cite{Stanescu2019a}. As shown below, partially overlapping/separated Majorana modes emerge generically in proximitized wires in the presence of positively/negatively charged impurities. 

\begin{figure}[t]
\begin{center}
\includegraphics[width=0.45\textwidth]{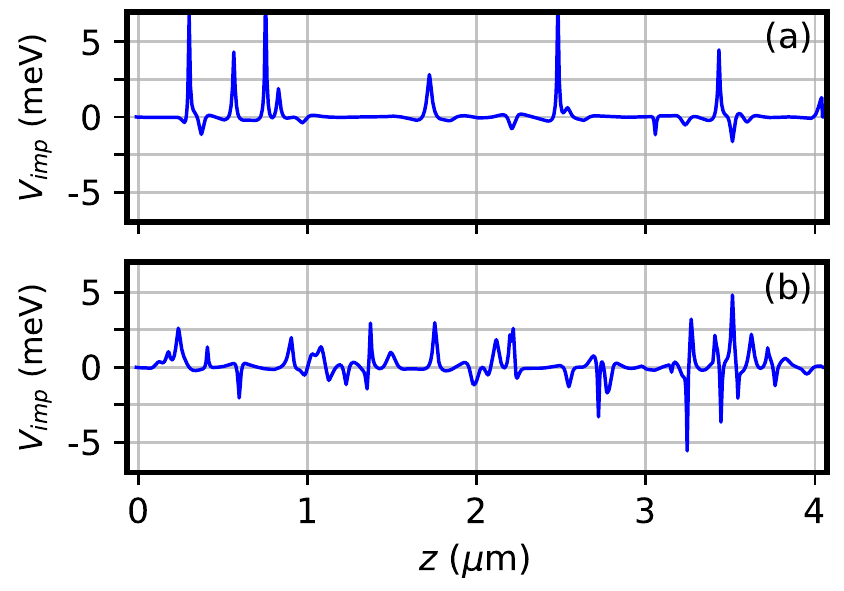}
\end{center}
\vspace{-4mm}
\caption{Position dependence of the effective impurity potential for two specific disorder realizations corresponding to impurity densities (a)  $n_{imp} = 1.6 \cdot 10^{15}~$cm$^{-3}$ (linear density $\lambda_{imp}=5~\mu$m$^{-1}$) and (b)  $n_{imp} = 4.7 \cdot 10^{15}~$cm$^{-3}$ (linear density $\lambda_{imp}=15~\mu$m$^{-1}$). The chemical potential of the wire is tuned near the bottom of the second subband. These impurity potentials are used in the calculations discussed in subsections \ref{ss1} and \ref{ss2}.}
\label{FIGM2bis}
\vspace{-1mm}
\end{figure}

Next, we characterize the effective potential generated by charge impurities embedded within the wire by providing some specific examples and  
calculating the correlation function $\langle V_{imp}(z)V_{imp}(z^\prime)\rangle$. The position dependence of the effective impurity potential $V_{imp}(z)$ given by Eq. (\ref{ImpPotential}) for two disorder realizations with impurity densities $n_{imp} = 1.6 \cdot 10^{15}~$cm$^{-3}$ and $n_{imp} = 4.7 \cdot 10^{15}~$cm$^{-3}$, respectively, are shown in Fig. \ref{FIGM2bis}. The first example corresponds to a low impurity density of about $5$ impurities per micron, while the second example corresponds to an intermediate regime with $15$ impurities per micron. These are relatively low impurity concentrations for semiconductor materials, but within the current technological capability.
Note that the amplitude of the strongest potential peaks exceeds $5~$meV, which corresponds to about $17\Delta$, a significant perturbation (more than an order of magnitude larger than the SC gap) even taking into account its relatively short range. The properties of the system in the presence of the effective potential shown in Fig. \ref{FIGM2bis} (a) are discussed in Sec. \ref{ss1}, while the intermediate impurity density regime corresponding to $V_{imp}$ given in Fig. \ref{FIGM2bis} (b) is investigated in Sec. \ref{ss2}.

\begin{figure}[t]
\begin{center}
\includegraphics[width=0.45\textwidth]{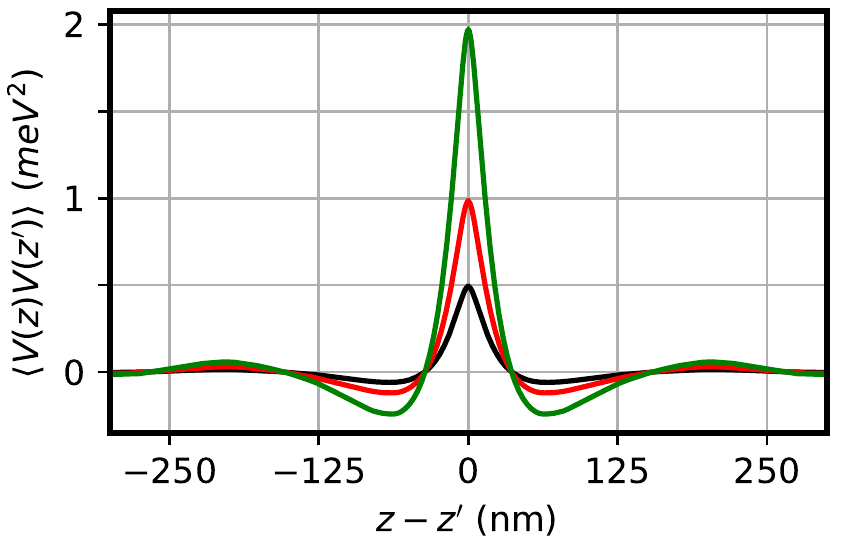}
\end{center}
\vspace{-4mm}
\caption{Correlation of the impurity potential for a system with impurity densities $n_{imp} = 0.25 \cdot 10^{16}~\text{cm}^{-3}$ (black), $n_{imp} = 0.5 \cdot 10^{16}~\text{cm}^{-3}$ (red), $n_{imp} = 1 \cdot 10^{16}~\text{cm}^{-3}$ (green), which correspond to linear densities $\lambda_{imp} = 7.9~\mu\text{m}^{-1}$, $\lambda_{imp} = 15.9~\mu\text{m}^{-1}$, and $\lambda_{imp} = 31.8~\mu\text{m}^{-1}$, respectively. The system is charge neutral (i.e., contains an equal number of $Q=+e$ and $Q=-e$ impurities, and has the chemical potential near the bottom of the second subband. Each correlation function was obtained by averaging over $5 \cdot 10^5$ disorder realizations. Note that the potential correlation scales with the impurity density.}
\label{FIGM3}
\vspace{-1mm}
\end{figure}

To obtain a more generic characterization of the effective impurity potential, we consider many disorder realizations consistent with given values of the impurity density and calculate the correlation function $\langle V_{imp}(z)V_{imp}(z^\prime)\rangle$.
The results for a system with impurity densities $n_{imp} = 0.25 \cdot 10^{16},  0.5 \cdot 10^{16}, 1 \cdot 10^{16}~\text{cm}^{-3}$, which correspond to linear densities $\lambda_{imp} = 7.9, 15.9, 31.8~\mu\text{m}^{-1}$, respectively, are shown in Fig. \ref{FIGM3}. Each curve was obtained by averaging over $5 \cdot 10^5$ disorder realizations.  Note that the potential correlation function scales with the impurity density. For the intermediate density, $n_{imp} =0.5 \cdot 10^{16}~\text{cm}^{-3}$, the correlation function is characterized by a central peak of height $\propto 1~$meV$^2$ and width at half maximum of about $40~$nm. 

Based on previous studies of disorder effects in Majorana nanowires \cite{Kells2012,Prada2012,Liu2012,Sau2013a,Cole2016,Liu2017a,Stanescu2019a,Pan2020,Pan2021a}, we know that the presence of disorder generally  induces low-energy sub-gap states. Also, the simple example illustrated in Figs. \ref{FIGM1} and \ref{FIGM2} suggests that, at least under certain conditions, these sub-gap states consist of partially overlapping Majorana modes (or ps-ABSs)  localized throughout the wire, in general away from the ends of the system. Note, however, that the presence of such non-topological (often called ``trivial'') ABSs does not necessarily affect the ``genuine'' topological Majorana zero modes (MZMs) that emerge in the topological regime at the ends  of the system, as shown in Fig. \ref{FIGM2}. Therefore, it is of crucial importance to characterize quantitatively the spatial separation between Majorana modes and the edge-to-edge correlation associated with the presence of MZMs at the ends of the wire and investigate the effect of charge impurity-induced disorder on these quantities. To this end, we introduce the  \textit{Majorana separation length}, $\ell_{sep}$, defined as follows. Let 
$\psi_{E_n}$, with $E_n \geq 0$, be a positive energy eigenstate of the BdG Hamiltonian and $\chi_n^{(L/R)}$ be its left/right Majorana components. The corresponding Majorana separation length is defined as
\begin{equation}
\ell_{sep}^{\left(n\right)} = \langle z_{n,R}\rangle - \langle z_{n,L} \rangle, \label{lsepn}
\end{equation}
where $\langle z_{n,L(R)}\rangle$ is the expectation value of the position along the wire corresponding to the left (right) Majorana component. 
Explicitly, we have
\begin{equation}
    \langle z_{n,J} \rangle = 
    \sum_{\nu}
    \sum_{i = 1}^{N_z}
    \left| \chi_{n}^{\left(J\right)}(z_i,\nu) \right|^2  z_i,
\end{equation}
where $J \in\{ L,R\}$,  $N_z$ is total number of sites, $z_i$ is the (discretized) $z$-coordinate corresponding to site $i$, and we sum over the spin and particle-hole degrees of freedom indexed by $\nu$. Finally, we have 
\begin{equation}
\ell_{sep} = {\rm Max}_n \left[\ell_{sep}^{\left(n\right)}~\!\mathcal{F}\left(E_n,\mathcal{U},\Omega\right)\right],  \label{lsep}
\end{equation}
where $\mathcal{F}$ is a function that filters out the states outside a small energy window centered at $E=0$. The details of the filtering are not important, as this simply corresponds to the energy resolution defining ``zero energy'' or ``zero bias'' in the experiment.
We choose the filter function to have the form
\begin{equation}
	\mathcal{F}\left(E,\mathcal{U},\Omega\right) = 
	\frac{1}{2} \left[
	\tanh{\left(\frac{E - \mathcal{U}}{\Omega}\right)} - 
	\tanh{\left(\frac{E + \mathcal{U}}{\Omega}\right)}
	\right].    \label{Ff}
\end{equation}
\begin{figure}[t]
\begin{center}
\includegraphics[width=0.45\textwidth]{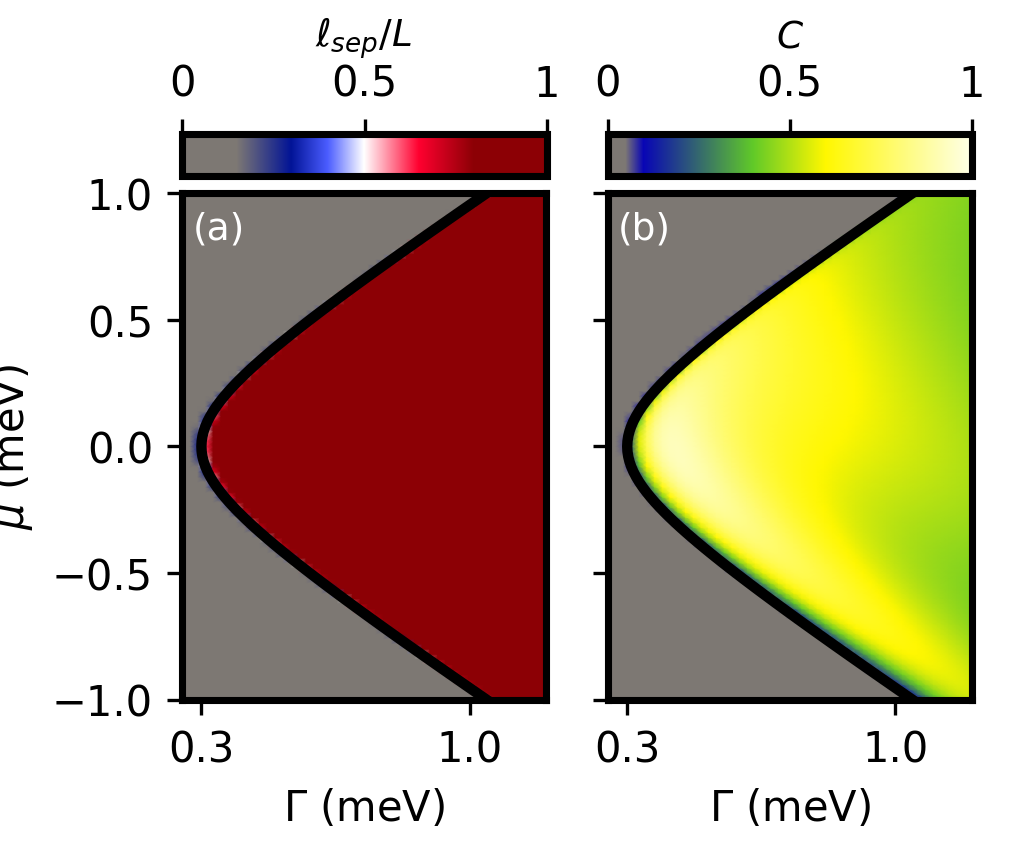}
\end{center}
\vspace{-7mm}
\caption{(a) Majorana separation, $\ell_{sep}$, and (b) edge-to-edge correlation, $C$, maps for a clean system of length $L = 4~\mu\text{m}$. The black lines indicate the (bulk) topological quantum phase transition corresponding to $\Gamma = \sqrt{\mu^2 + |\Delta|^2}$. The edge length used in the definition of $C$ [see Eq. (\ref{Wn})] is $\ell_e=200~$nm.}
\label{FIGM4}
\vspace{-1mm}
\end{figure}
Note that $\mathcal{F} \approx 0$ for $|E| \gg \mathcal{U}$ and $\mathcal{F} \approx 1$ for $E = 0$, while it smoothly interpolates between these values near $|E| \approx \mathcal{U}$ over an energy scale $\Omega$. Throughout the rest of this work we set $\mathcal{U} = 0.2 \Delta$ and $\Omega = 0.1 \Delta$. These are, most likely, fairly generous  estimates for defining the zero-energy modes.
Hence, according to Eq. (\ref{lsep}),  $\ell_{sep}$ measures the largest separation length between the left and right Majorana components of BdG states having a sufficiently low energy, so as to be operationally considered a zero-energy state.
Next, we define the edge-to-edge correlation associated with the BdG eigenstate $\psi_{E_n}$ as
\begin{equation}
	C_{n} = 
	\sqrt{W_{n}^{(L)} W_n^{(R)}}~\!
	\mathcal{F}\left(E_n,\mathcal{U},\Omega\right), \label{Cn}
\end{equation}
with $W_n^{(L/R)}$ being the spectral weight at the left/right end of the system. Explicitly, we have
\begin{equation} 
    W_n^{(J)} = \sum_{\nu} \sum_{i}^{(\ell_e)_J} 
    \left| \chi_{n}^{\left(J\right)}(z_i,\nu) \right|^2, \label{Wn}
\end{equation} 
where $J \in\{ L,R\}$ and the summation over $i$ is restricted to sites that are within a  distance $l_e$ of the corresponding edge. Let $n_0$ be the state characterized by the largest Majorana separation, i.e., $\ell_{sep}^{\left(n_0\right)} = \ell_{sep}$. Typically, $n_0=1$, i.e., the largest Majorana separation corresponds to the lowest energy mode, unless there is a ``regular'' (i.e., non-separated) Andreev bound state.  We define the {\em edge-to-edge correlation} as $C=C_{n_0}$. Note that  $0 \leq C \leq 1$, with $C\approx 1$ corresponding to a low energy BdG state having its Majorana components localized at the ends of the system, each within a distance $\ell_e$ of the corresponding edge. 

\begin{figure}[t]
\begin{center}
\includegraphics[width=0.45\textwidth]{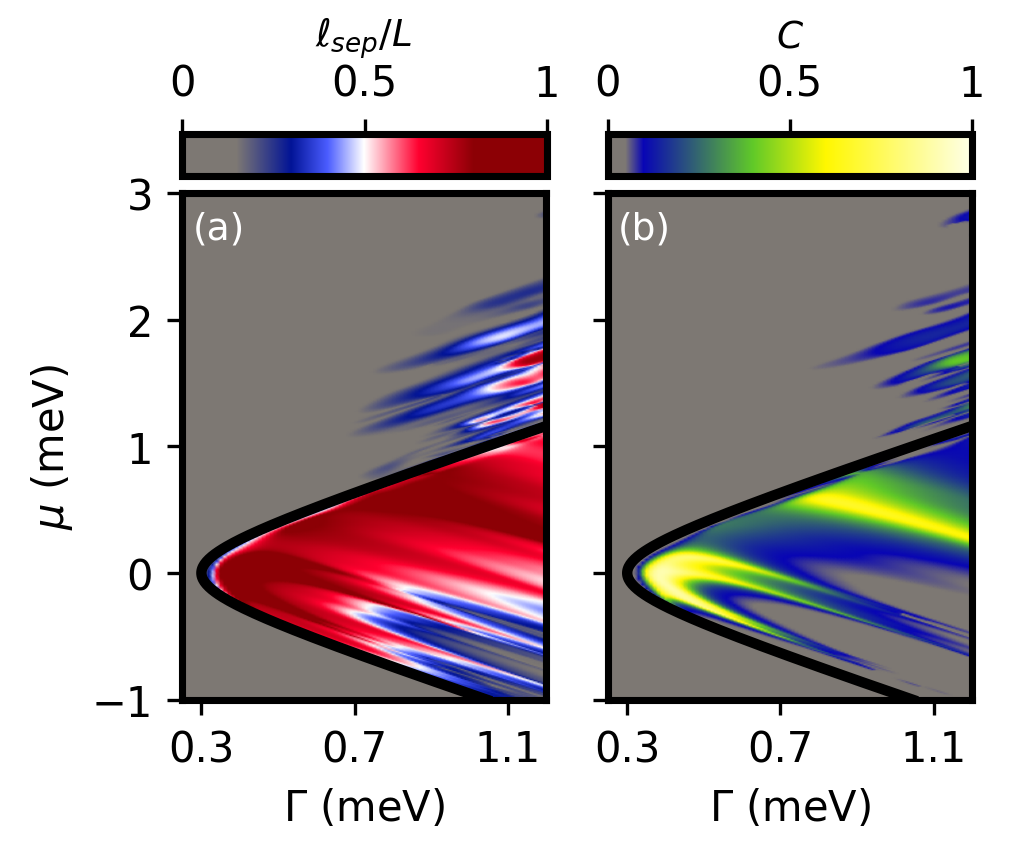}
\end{center}
\vspace{-7mm}
\caption{(a) Majorana separation, $\ell_{sep}$, and (b) edge-to-edge correlation, $C$, maps for a  disordered system of length $L = 4~\mu\text{m}$
with impurity density $n_{imp} = 1.6 \cdot 10^{15}~\text{cm}^{-3}$ ($\lambda_{imp} = 5~\mu\text{m}^{-1}$). The black lines indicate the topological quantum phase transition for a clean system. The edge length used in the definition of $C$ [see Eq. (\ref{Wn})] is $\ell_e=200~$nm.  Note that non-negligible values of $\ell_{sep}$ and $C$ occur outside the nominally topological region, while these quantities are significantly suppressed in some areas within this region.}
\label{FIGM5}
\vspace{-1mm}
\end{figure}

To benchmark these quantities, we start with a clean system of length $L=4~\mu$m and calculate the dependence of the Majorana separation, $\ell_{sep}$, and edge-to-edge correlation, $C$, on the Zeeman field and chemical potential. The corresponding ``phase diagrams'' are shown in Fig. \ref{FIGM4}. The black lines mark the theoretically known phase boundary \cite{Stanescu2011} associated with the topological quantum phase transition. Remarkably, the area characterized by large values of the Majorana separation, $\ell_{sep}\lesssim L$, and large edge-to-edge correlations, $C>0.5$, practically coincides with the topological phase. This indicates that the two quantities capture meaningful information about the Majorana zero modes and the topological quantum phase transition. Note, for example, that $C$ decreases with increasing Zeeman field as result of increasing the Majorana localization length, $\xi$, which transfers some of the spectral weight outside the edge regions defined by the length scale $\ell_e$ in Eq. (\ref{Wn}). We emphasize that generating two dimensional maps of the relevant quantities as functions of various control parameters, such as the Zeeman splitting and the chemical potential (or applied back gate potential), provides significantly more information than focusing on specific parameter values. As shown below, such maps are mandatory for properly understanding the effects of disorder and should represent the standard in both theoretical and experimental investigations of hybrid systems. We urge experimentalists to always characterize the presence of (near) zero-energy modes by providing two-dimensional ``phase diagram'' maps  in the magnetic field-gate voltage parameter space.

\begin{figure}[t]
\begin{center}
\includegraphics[width=0.45\textwidth]{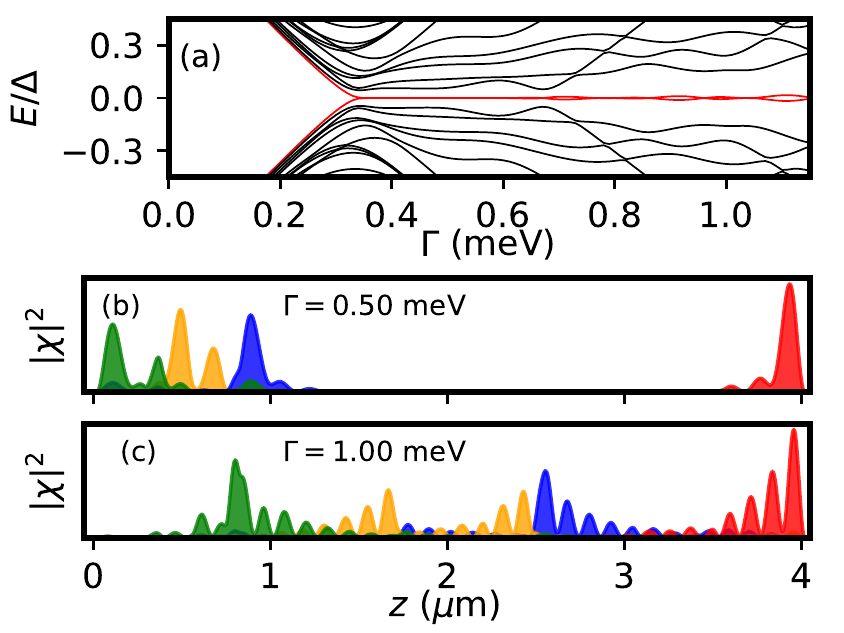}
\end{center}
\vspace{-7mm}
\caption{(a) Low-energy spectrum as a function of the Zeeman field for a system with the same parameters as in Fig. \ref{FIGM5} and $\mu = 0$. Red lines denote the lowest energy mode. (b) and (c) Spatial profiles of the Majorana components corresponding to the lowest BdG eigenstate (red and green) and second lowest energy eigenstate (blue and yellow) for $\Gamma = 0.5~\text{meV}$ and $\Gamma = 1~\text{meV}$, respectively. Note that in (c) the left Majorana component of the lowest energy state (green) is localized away from the corresponding edge,  which causes the collapse of the edge-to-edge correlation $C$ in Fig. \ref{FIGM5} in the area around $\mu = 0, \Gamma = 1~${meV}.}
\label{FIGM6}
\vspace{-1mm}
\end{figure}

\subsubsection{Low impurity density regime} \label{ss1}

We are now ready to consider a system with randomly distributed charge impurities and investigate the effects of this type of disorder using the quantities introduced above. We start with a specific disorder realization corresponding to a relatively low impurity density, $n_{imp} = 1.6 \cdot 10^{15}~\text{cm}^{-3}$, which means $\lambda_{imp} = 5$ impurities per micron. The position dependence of the impurity potential $V_{imp}(z)$ for this disorder realization is shown in Fig. \ref{FIGM2bis} (a).
 The maps of the Majorana separation and edge-to-edge correlation as functions of Zeeman field and chemical potential are shown in Fig. \ref{FIGM5}. A comparison of these maps with the corresponding ``phase diagrams'' in Fig. \ref{FIGM4} reveals two distinctive features: the emergence of areas with significant values of $\ell_{sep}$ and $C$ outside the nominally topological region and the substantial suppression of these quantities in certain areas within the topological region. We emphasize that, although the quantitative details of the phase diagram in Fig. \ref{FIGM5} depend on the specific disorder realization and on the corresponding impurity potential (see Fig. \ref{FIGM2bis}) used in the calculation, these two distinctive qualitative features are generic.

\begin{figure}[t]
\begin{center}
\includegraphics[width=0.45\textwidth]{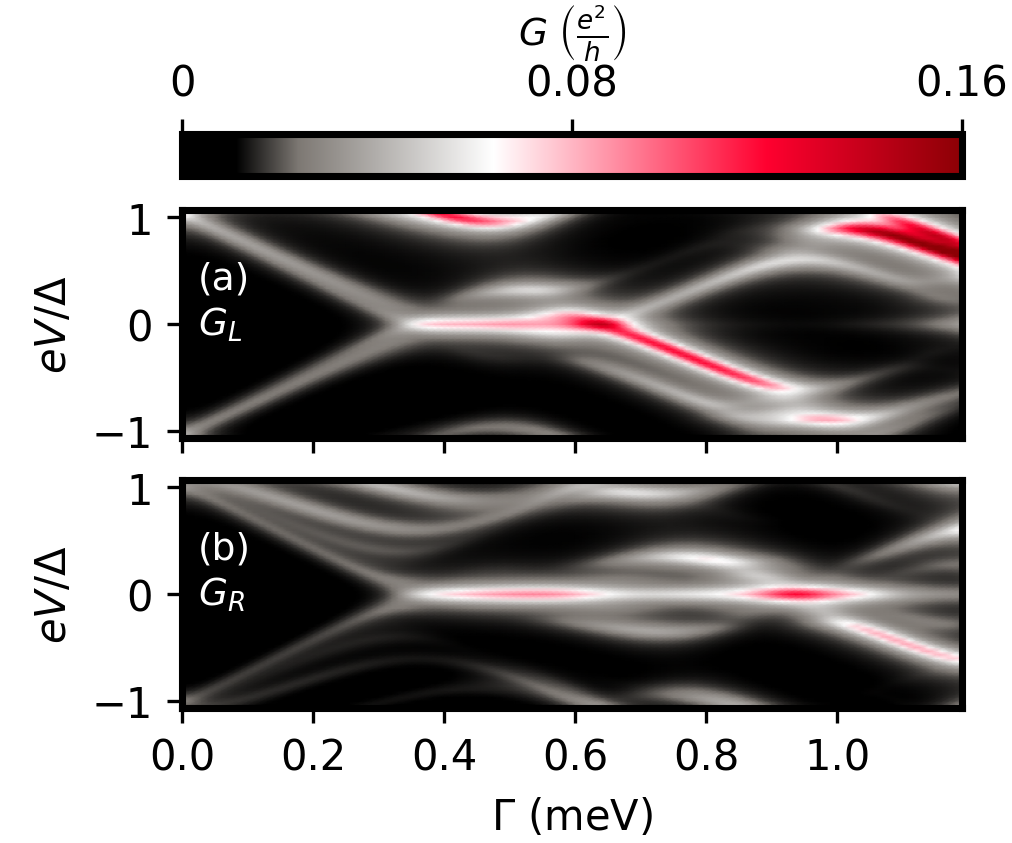}
\end{center}
\vspace{-4mm}
\caption{Local differential conductance at the left (a) and right (b) ends of the wire for a system with the same parameters as in Fig. \ref{FIGM6}. Note that the zero-bias conductance peak characterizing $G_L$  is suppressed between $0.7 \lesssim \Gamma \lesssim 1.1~\text{meV}$ as a result of the left Majorana mode being pushed away from the  edge,  as shown in Fig. \ref{FIGM6} (c).}
\label{FIGM7}
\vspace{-1mm}
\end{figure}

To better understand the significance of these features, we calculate the low-energy spectrum as a function of the Zeeman field for a fixed value of the chemical potential, as well as the spatial profile of the Majorana components corresponding to certain representative low-energy modes. The results for $\mu=0$ are shown in Fig. \ref{FIGM6}. The low-energy spectrum in Fig. \ref{FIGM6}(a) shows the emergence of a near-zero energy mode for Zeeman fields $\Gamma \gtrsim 0.3~$meV (red lines). The lowest energy mode is separated from other finite energy states by a small gap that increases significantly for $\Gamma \gtrsim 0.75~$meV. This behavior may be surprising if judged based on the information in Fig. \ref{FIGM5}, which, for $\mu=0$, shows a strong suppression of $C$ at higher values of the Zeeman field. However, the spatial profiles of the Majorana components shown in Fig. \ref{FIGM6}(b) and (c) clarify the physics. Indeed, for $\Gamma=0.5~$meV  the lowest energy state consists of two well separated Majorana modes localized near the ends of the system (green and red). The left (green) Majorana has some overlap with a ps-ABS localized nearby (yellow and blue), which represents the second lowest BdG state, but is weakly affected by the presence of this bound state. Consequently, $\ell_{sep}$ is comparable to the length $L$ of the wire and the edge-to-edge correlation $C$ is large. By contrast, at $\Gamma=1~$meV  the left (green) Majorana mode is ``pushed'' away from the end of the system, which results in a reduction of the Majorana separation length and the collapse of the  edge-to-edge correlation. 

The example discussed above shows that a hybrid system with a low concentration of charged impurities is consistent with the emergence of well separated, near-zero energy Majorana modes. However, the presence of disorder may ``push'' these modes away from the ends of the system, which results in low values of the edge-to-edge correlation. In other words, the system can host ``genuine'' MZMs, but they may be ``invisible'' to local probes coupled to the ends of the wire. This severely limits the relevance of tunnel spectroscopy as a tool for detecting the emergence of Majorana zero modes in the presence of disorder, even in the weakly disordered situation.
To make further connection with experiment, we calculate the local differential conductance for charge tunneling into the left and right end of the system. The results corresponding to a system with the same parameters as in Fig. \ref{FIGM6} are shown in Fig. \ref{FIGM7}. One can clearly notice two low-energy modes coalescing toward zero energy and generating robust zero-bias conductance peaks (ZBCPs) at both ends of the system. At the left end, the ZBCP persists from $\Gamma = 0.3~$meV to $\Gamma\approx 0.7~$meV, then it appears to split. However, as revealed by the data in Fig. \ref{FIGM6}, the apparent splitting is due to a ps-ABS localized near the left end, while the ``actual'' Majorana mode (i.e., the ``green'' Majorana) does not become gapped, becoming instead ``invisible'' to local measurements at the edge, as it gets pushed away from the end of the wire.  Within the range $0.7 \lesssim \Gamma \lesssim 1.1~\text{meV}$ there is a robust ZBCP at the right end of the wire, but no ZBCP at the left end. This example clearly illustrates the difficulty of correctly interpreting tunneling conductance results in the presence of disorder. First, apparent splittings of the ZBCP can be misleading, as they are not necessarily associated with the mode that generates the ZBCP. Second, the absence of edge-to-edge correlation does not necessarily imply the absence of robust, well-separated Majorana modes; it may simply mean that (at least) one of these modes is localized away from the end of the wire. We note that the conductance calculations shown in  Fig. \ref{FIGM7} were done in the tunneling limit, i.e., for high values of the potential barrier amplitude. In addition, we considered some finite dissipation,  $\eta = 20~\mu\text{eV}$. As a result, the height of the ZBCP is much smaller than the quantized value and there is some particle-hole asymmetry \cite{DSarma2016,Stenger2017,Liu2017}.  These issues are well-understood and do not in any way affect our key qualitative conclusion of disorder possibly pushing the zero mode away from the end and making it invisible in standard tunneling spectroscopy.  In some sense, this invisibility of the topological Majorana in the tunneling measurement (a false negative) is the ironic counterpart of the ps-ABS misleadingly producing non-topological zero bias conductance peaks mimicking Majorana zero modes (a false positive)!

\begin{figure}[t]
\begin{center}
\includegraphics[width=0.45\textwidth]{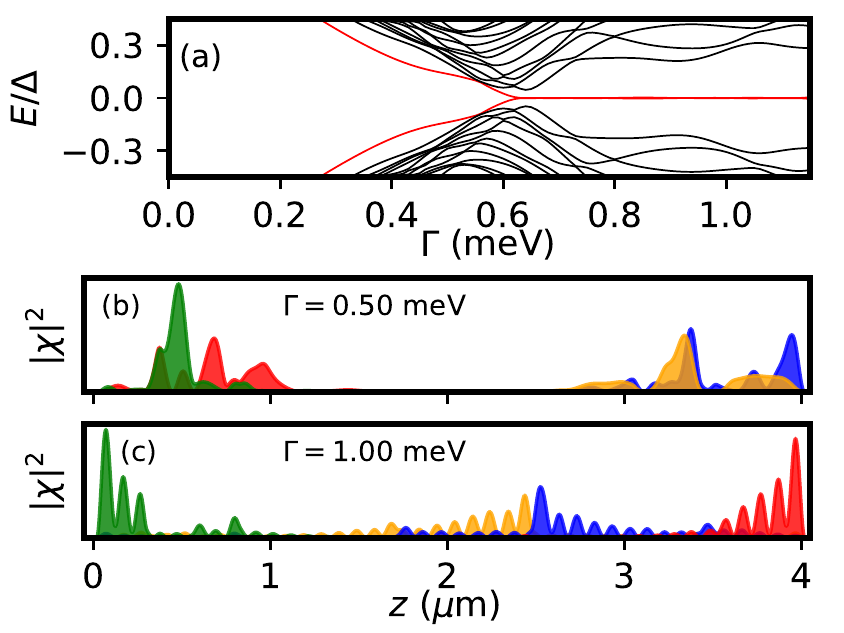}
\end{center}
\vspace{-4mm}
\caption{(a) Low-energy spectrum as a function of the Zeeman field for a system with the same parameters as in Fig. \ref{FIGM5} and $\mu = 0.5~$meV. Red lines denote the lowest energy mode. (b) and (c) Spatial profiles of the Majorana components corresponding to the lowest BdG eigenstate (red and green) and second lowest energy eigenstate (blue and yellow) for $\Gamma = 0.5~\text{meV}$ and $\Gamma = 1~\text{meV}$, respectively. 
Note that in (b), which corresponds to the trivial regime, the Majorana modes strongly overlap, generating two ABSs localized near the ends of the system.}
\label{FIGM8}
\vspace{-1mm}
\end{figure}

Next, we consider another horizontal cut through the phase diagram  in Fig. \ref{FIGM5} corresponding to $\mu=0.5~$meV. For this value of the chemical 
potential, the system is characterized by large Majorana separations and edge-to-edge correlations in the topological regime, i.e., for 
$\Gamma \gtrsim 0.6~$meV. Indeed, the spectrum shown in Fig. \ref{FIGM8} (a) is characterized by a robust zero energy mode (red line for 
$\Gamma  \gtrsim 0.6~$meV) and a sizable topological gap. Note the presence of finite energy in-gap states in the topologically trivial regime (e.g., red lines for $\Gamma \lesssim 0.6~$meV). These topologically trivial in-gap modes consist of Andreev bound states with strongly overlapping Majorana components 
localized near the ends of the wire, as shown in Fig. \ref{FIGM8} (b). In the topological regime, on the other hand, the system is characterized by well separated Majorana modes localized at the ends of the system, as shown in Fig. \ref{FIGM8} (c) (the green and red Majoranas) and is consistent with the large values of $\ell_{sep}$ and $C$ in Fig. \ref{FIGM5}. 

\begin{figure}[t]
\begin{center}
\includegraphics[width=0.45\textwidth]{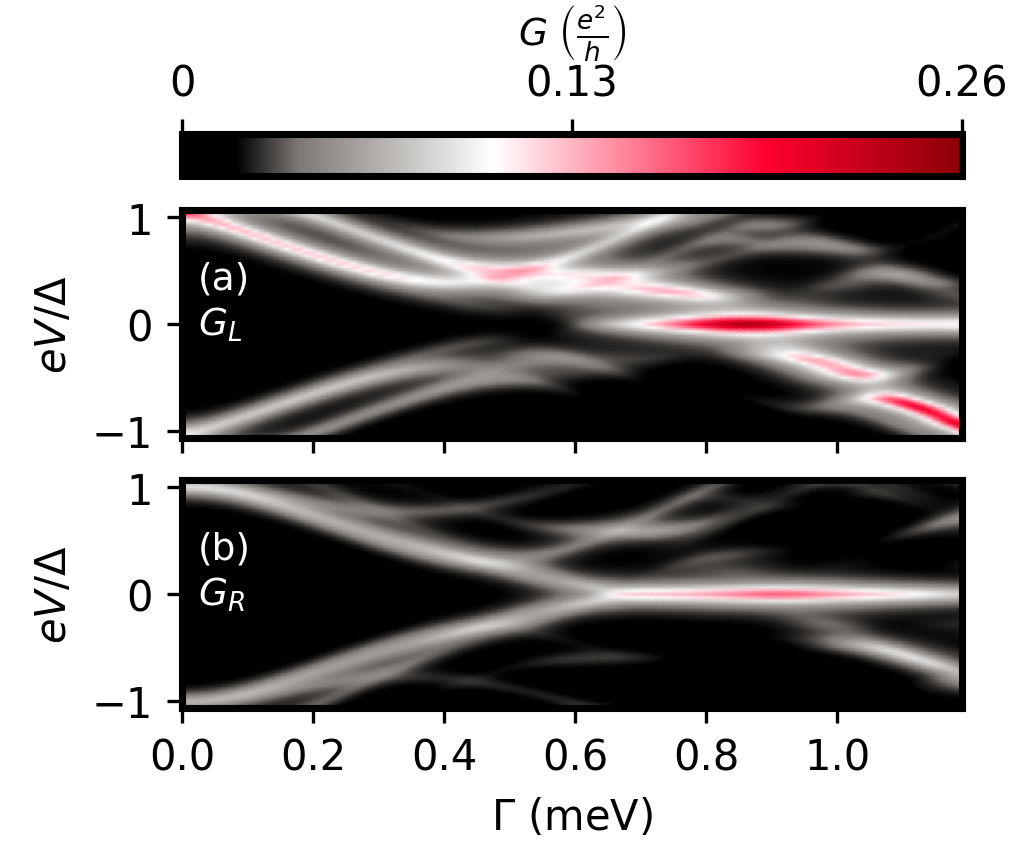}
\end{center}
\vspace{-4mm}
\caption{Local differential conductance at the left (a) and right (b) ends of the wire for a system with the same parameters as in Fig. \ref{FIGM8}. Correlated zero-bias conductance peaks occur at the two end of the system for $\Gamma \gtrsim 0.65$,  consistent with the large $C$ values for $\mu = 0.5~$meV and $\Gamma \gtrsim 0.65\text{meV}$ in Fig. \ref{FIGM5}. Note the significant enhancement of the ZBCP in (a) due to the Majorana mode hybridizing with a bound state localized within the barrier region, which crosses zero energy at $\Gamma\approx 0.8~$meV.}
\label{FIGM9}
\vspace{-1mm}
\end{figure}

The conductance traces corresponding to the $\mu=0.5~$meV cut are shown in Fig. \ref{FIGM9}. The presence of the MZMs is revealed by the emergence of robust ZBCPs at both ends of the system. Note, however, that the emergence of the ZBCP looks rather different at the two ends, with two low-energy modes coalescing toward zero energy clearly visible at the right end and no apparent gap closing at the left end.  This behavior  is due to the fact that the right Majorana mode is adiabatically connected to the ABS localized at the right end of the system, while the left Majorana is connected to a trivial mode that has low spectral weight at the left end of the system and couples weakly to the corresponding probe, thus remaining ``invisible.'' Another significant feature that is clearly manifested in Fig. \ref{FIGM9} (a) is the enhancement of the ZBCP weight/height due to the Majorana mode hybridizing with a bound state localized in the barrier region. Indeed, in Fig. \ref{FIGM9} (a)  one can clearly notice an ABS crossing zero energy at $\Gamma\approx 0.8~$meV. This mode is absent from the low-energy spectrum shown in Fig. \ref{FIGM8} (a), a clear indication that it is generated by the very presence of the barrier region that couples the system to the normal lead, as this is not included in the calculation of the spectrum. This type of enhancement of the ZBCP due to coupling to an ABS localized at the end of the system is also visible in Fig. \ref{FIGM7}. 
The results presented in Fig. \ref{FIGM9} and discussed above indicate a serious problem regarding tunnel conductance measurements: the end-to-end conductance correlations, which are often thought to be the decisive signature for the existence of topological MZMs, may very well be quite imprecise (or even absent) in the presence of (even weak) disorder.  The absence of such correlations can be quite generic in disordered systems and may imply either that one of the MZMs cannot be accessed through tunnel spectroscopy at the wire end (because it was pushed away), or that the observed zero mode is simply trivial. Comparing the conductance traces at the two ends of the system cannot discriminate between these possibilities. However, generating two-dimensional conductance maps over large parameter regions may provide additional information, as discussed below.

\begin{figure}[t]
\begin{center}
\includegraphics[width=0.45\textwidth]{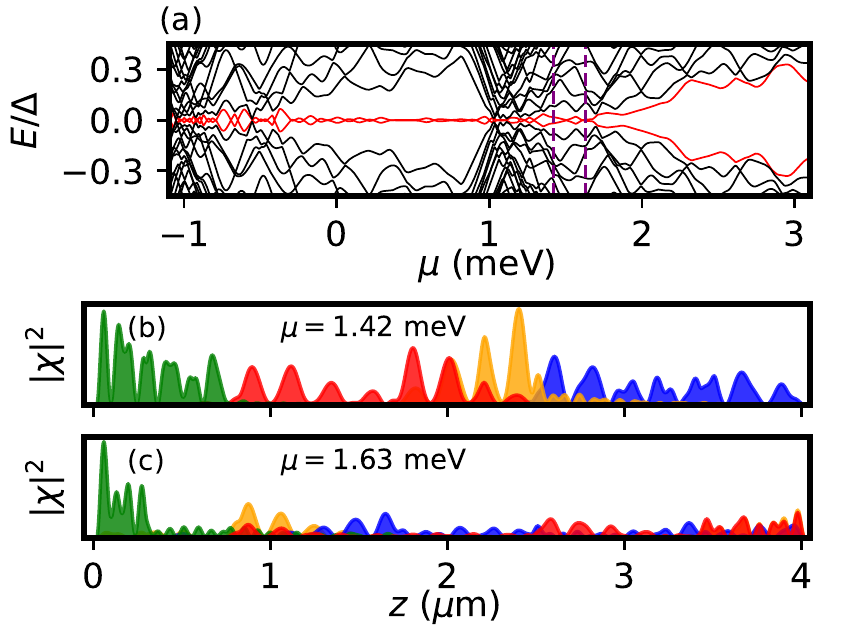}
\end{center}
\vspace{-4mm}
\caption{(a) Low-energy spectrum as a function of the chemical potential for a system with the same parameters as in Fig. \ref{FIGM5} and $\Gamma = 1.1~\text{meV}$. Red lines denote the lowest energy mode. (b) and (c) Spatial profiles of the Majorana components corresponding to the lowest BdG eigenstate (red and green) and second lowest energy eigenstate (blue and yellow) for two values of the chemical potential marked by dashed purple lines in (a). In (c) the Majorana components of the lowest energy mode (green and red) have nonzero spectral weights at the ends of the system, which results in a finite edge-to-edge correlation $C$.}
\label{FIGM10}
\vspace{-1mm}
\end{figure}

Having clarified the features that characterize the nominally topological region of the phase diagrams in Fig. \ref{FIGM5}, the natural question  concerns the nature of the low-energy states responsible for the emergence of high Majorana separations and significant edge-to-edge correlations in the trivial region (of the pristine system) with $\mu > 1~$meV. To address this question, we consider a vertical cut at fixed Zeeman field $\Gamma = 1.1~\text{meV}$. The dependence of the low-energy spectrum on the chemical potential along this cut is shown in Fig. \ref{FIGM10} (a). For $-1 \lesssim \mu \lesssim 1~$meV the system is in the nominally topological regime and one can clearly notice the a near-zero energy mode (red lines) protected by a finite gap over most of this interval. The gap collapses for $\mu \lesssim -0.25~$meV. Most interestingly, low-energy modes are also present for $1 \lesssim \mu \lesssim 1.75~$meV, i.e., in the nominally trivial regime. To clarify the nature of these states, we calculate their Majorana components for two values of the chemical potential.The results are shown in  Fig. \ref{FIGM10} (b) and (c). The low-energy states can be viewed as superposition of several partially-overlapping Majorana modes. Accidentally,  Majorana components associated with the lowest energy state can have significant weights at the ends of the system, which generates a finite edge-to-edge correlation, as shown in Fig. \ref{FIGM5} (b). Such zero  modes accidentally arising from the disorder-induced overlap of several Majorana modes cannot be construed as being topological.

\begin{figure}[t]
\begin{center}
\includegraphics[width=0.45\textwidth]{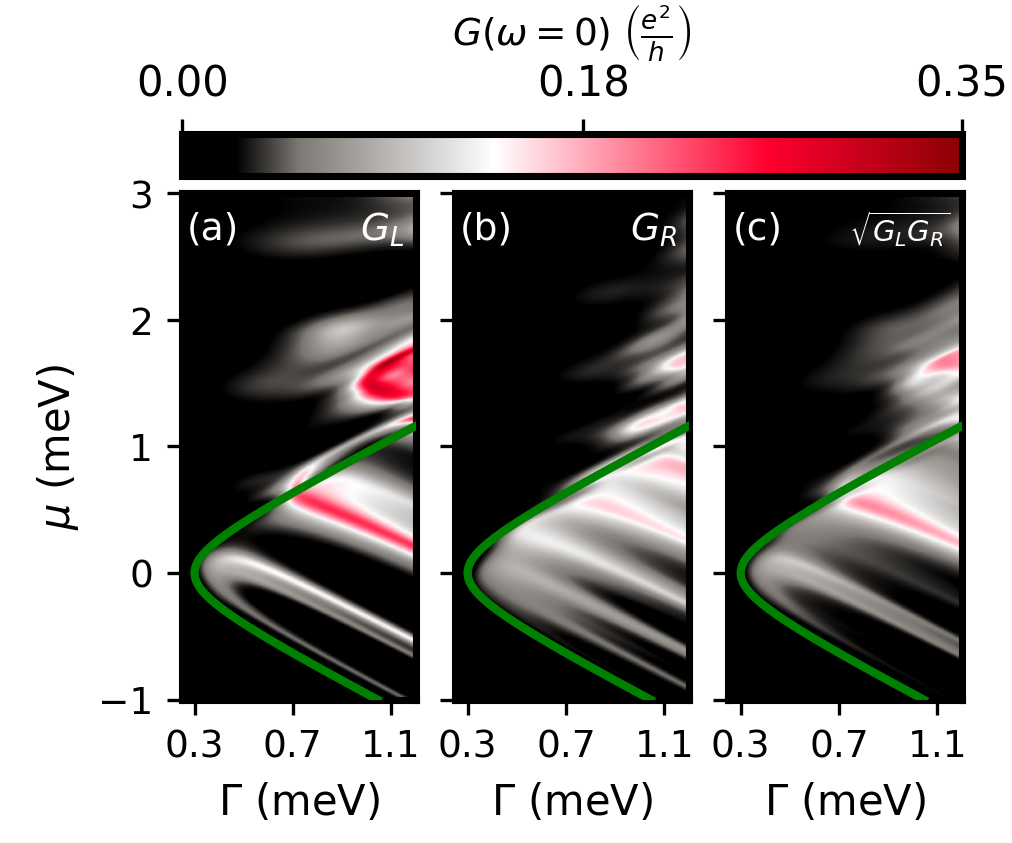}
\end{center}
\vspace{-4mm}
\caption{Zero-bias differential conductance maps for a system with the same parameters as in Fig. \ref{FIGM5}, but having normal leads and tunnel barriers attached at both ends. The conductance at the left  ($G_L$) and right ($G_R$) ends of a system and the geometric average ($C_G = \sqrt{G_L G_R}$) 
are shown in (a), (b), and (c), respectively. Note that the $C_G$ map closely resembles the edge-to-edge correlation map, $C$, in Fig. \ref{FIGM5}.}
\label{FIGM11}
\vspace{-1mm}
\end{figure}

We have already pointed out the importance of generating two dimensional maps of the relevant quantities as functions of various control parameters. To further emphasize this point, we calculate the zero-bias differential conductance maps corresponding to charge tunneling into the left ($G_L$) and right ($G_R$) ends of a system having the same parameters as in Fig. \ref{FIGM5}. In addition,  we define the geometric average of the left and right conductivities as a practical measure of the edge-to-edge correlation. Specifically, we define
\begin{equation}
C_G = \sqrt{G_L G_R}.   \label{CG}
\end{equation}
The results are shown in Fig. \ref{FIGM11}. We note that the $C_G$ map in Fig. \ref{FIGM11} (c) closely resembles the edge-to-edge correlation map, $C$, in Fig. \ref{FIGM5}. This observation has two important implications. First, $C_G$ provides a good measure of the edge-to-edge correlation that can be easily determined experimentally.  Second, for large scale calculations (e.g., when doing statistics involving many disorder realizations -- see below), one can focus on the numerically-less-expensive quantity $C$, instead of the more experimentally-relevant quantity $C_G$, since we find the two to be representing equivalent physics, even in the presence of disorder.
In addition, we note that for low-impurity concentrations, the (zero energy) conductance maps provide a reasonably good correspondence with the phase diagram of the clean system, particularly in the low-field regime. However, as shown below, this correspondence fades away upon increasing the impurity concentration.  This suggests that the systematic mapping of the zero-bias conductance at both ends of the system and of the corresponding correlation $C_G$ can provide a powerful experimental tool for assessing the strength of the effective disorder potential. Finally, we note that $C_G$ has the highly desirable practical property that it does not require identical tunnel barriers at the two ends. As long as a differences between the two barriers amounts to an overall enhancement/suppression of $G_L$ relative to $G_R$, the corresponding factor is irrelevant when calculating the correlation $C_G$. 
We note that our calculated conductance shown in Figs. \ref{FIGM9} and \ref{FIGM11} is characterized by zero bias values ($G_L$, $G_R$, and $C_G$) smaller than the so-called Majorana quantization value of $2e^2/h$, as we consider relatively high tunnel barriers and include a dissipation term. We emphasize that, in the presence of disorder, fine-tuning the parameters to obtain quantized values of the zero-bias conductance does not provide additional information regarding the nature of the underlying low-energy mode. Instead,  producing detailed conductance maps over extended ranges of tuning parameters, similar to those in Fig. \ref{FIGM11}, can provide additional information, including estimates of the disorder strength. We think that generating such comprehensive maps is  what experiments should focus on, rather than fine-tuning  parameters to achieve Majorana quantization.

\subsubsection{Intermediate impurity density regime} \label{ss2}

How does the phenomenology discussed above depend on the concentration of charge impurities, i.e., on the disorder strength? To address this question, we consider another specific disorder realization corresponding to an intermediate impurity density, $n_{imp} = 4.7 \cdot 10^{15}~\text{cm}^{-3}$, which means $\lambda_{imp} = 15$ impurities per micron. This is still relatively low disorder in terms of the bulk doping magnitude, but it is three times larger than the low-disorder case ($\lambda_{imp} =5~\mu$m$^{-1}$) considered above.
The position dependence of the impurity potential $V_{imp}(z)$ for this disorder realization is shown in Fig. \ref{FIGM2bis} (b). We carry out the same calculations as above and construct  the maps corresponding to the Majorana separation, $\ell_{sep}$, and edge-to-edge correlation, $C$,  as functions of Zeeman field and chemical potential. The results are shown in Fig. \ref{FIGM5}. 

\begin{figure}[t]
\begin{center}
\includegraphics[width=0.45\textwidth]{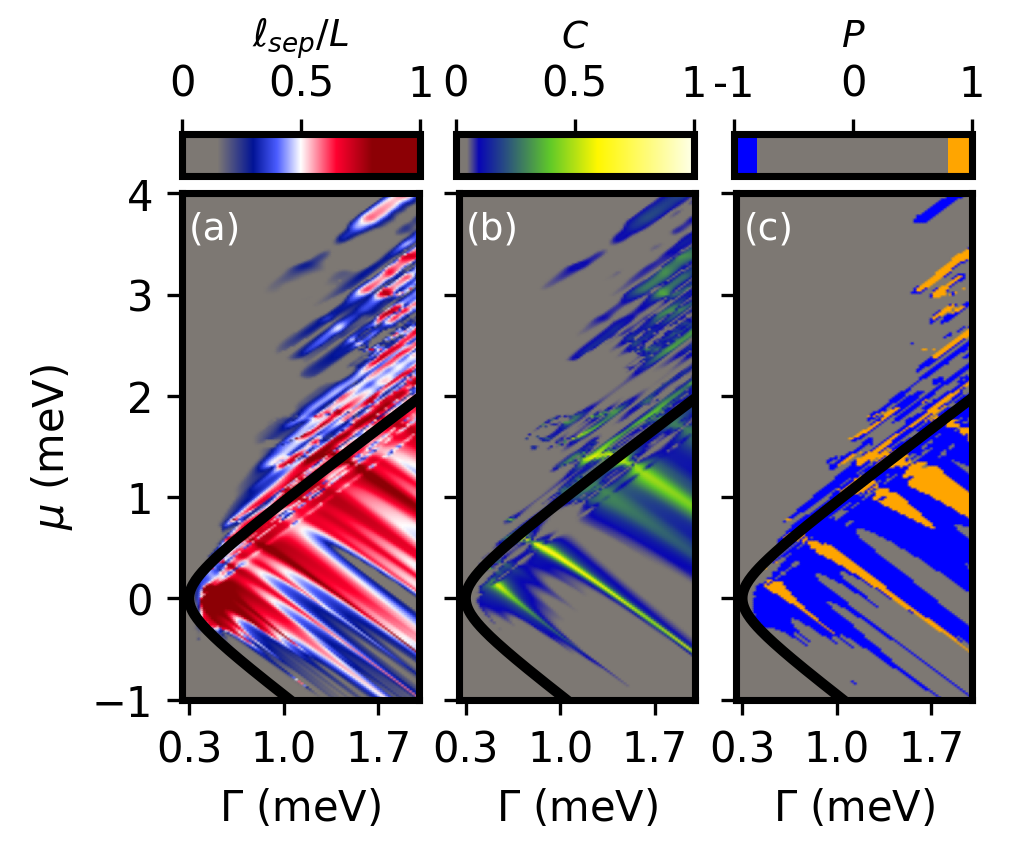}
\end{center}
\vspace{-4mm}
\caption{(a) Majorana separation, $\ell_{sep}$, (b) edge-to-edge correlation, $C$, and (c) projection, $P$, maps for a  disordered system of length $L = 4~\mu\text{m}$ with impurity density $n_{imp} = 4.7 \cdot 10^{15}~\text{cm}^{-3}$ ($\lambda_{imp} = 15~\mu\text{m}^{-1}$). The impurity potential $V_{imp}(z)$ for this disorder realization is shown in Fig. \ref{FIGM2bis} (b).  The black lines indicate the topological quantum phase transition for a clean system. The projection map in (c) corresponds to $\ell_{min} = 0.5L$ and $C_{min} = 0.25$.}
\label{FIGM12}
\vspace{-1mm}
\end{figure}

In addition, we introduce a ``projection map'' based on the following quantity:
\begin{equation}
P(\ell_{sep}, C) =
\begin{cases}
   ~~~~\!0, & \ell_{sep} \leq \ell_{min} \\
  -1, & \ell_{sep} > \ell_{min} \text{ and } C < C_{min} \\
    ~~~~\!1, & \ell_{sep} > \ell_{min} \text{ and } C > C_{min}  
\end{cases}.
\end{equation}
In essence, $P=0$ corresponds to low Majorana separation lengths (according to a criterion determined by $\ell_{min}$), $P=-1$ signals well separated Majoranas that do not generate a substantial edge-to-edge correlation (e.g., because one of the Majorana modes is pushed away from the end of the system by the disorder potential), while $P=1$ corresponds to the desired scenario involving well separated Majoranas and substantial edge-to-edge correlation. The projection map corresponding to $\ell_{min} = 0.5L$ and $C_{min} = 0.25$ is shown in Fig. \ref{FIGM12} (c). As compared to the corresponding maps in Fig. \ref{FIGM5}, the suppression of the Majorana separation and edge-to-edge correlation inside the nominally topological region is  
significantly stronger. When comparing the two figures, note that $\Gamma$ extends to higher values in Fig. \ref{FIGM12} than Fig. \ref{FIGM5}.  Nonetheless, there is a substantial area -- blue region in panel (c) -- corresponding to large values of the Majorana separation ($\ell_{sep} > 2~\mu$m), but weak edge-to-edge correlation. This suggests that, even at this level of impurity concentration, there are segments of the wire that can be viewed as effectively topological, but their presence cannot be revealed by local measurements at the ends of the wire. By contrast, the areas corresponding to large values of $C$  are reduced to a few small islands. The underlying disorder-induced nonperturbative rearrangement of the Majorana spatial locations and the corresponding signatures revealed by the Majorana phase diagrams are important findings of our work.

\begin{figure}[t]
\begin{center}
\includegraphics[width=0.45\textwidth]{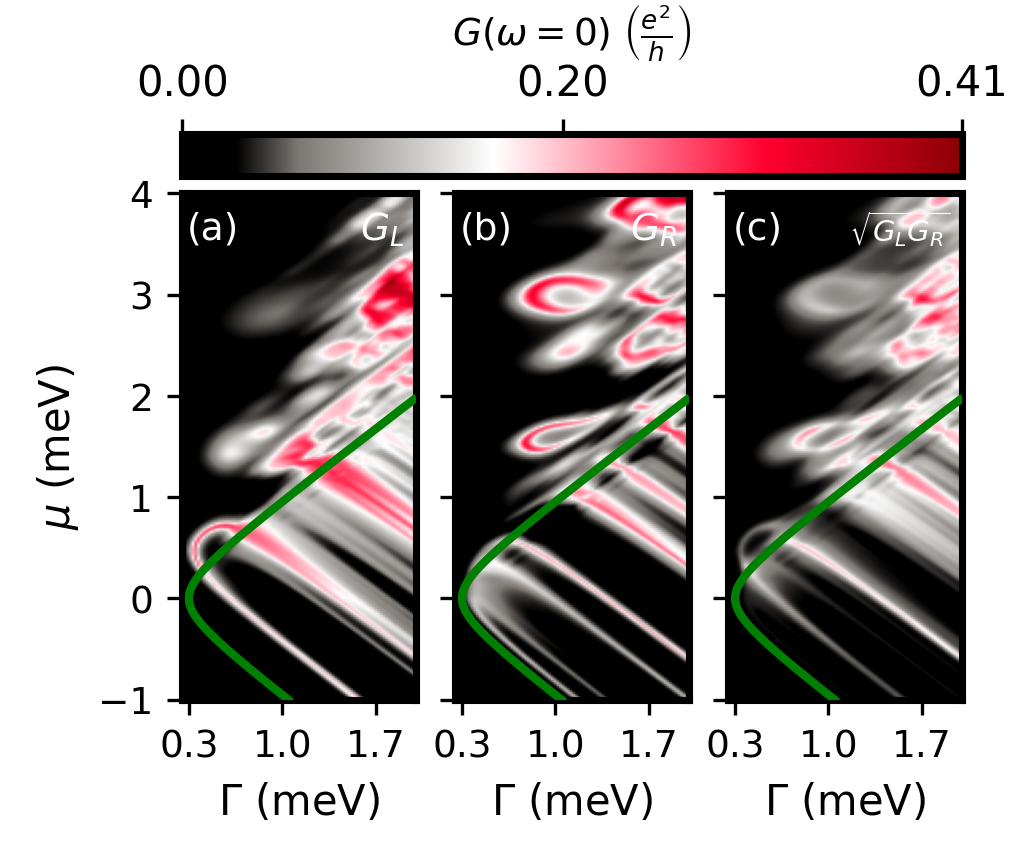}
\end{center}
\vspace{-4mm}
\caption{Zero-bias differential conductance maps for a system with the same parameters as in Fig. \ref{FIGM12}, but having normal leads and tunnel barriers attached at both ends. The conductance at the left  ($G_L$) and right ($G_R$) ends of a system and the geometric average ($C_G = \sqrt{G_L G_R}$)  are shown in (a), (b), and (c), respectively. Note that the correspondence between the conductance maps and the topological phase boundary for the clean system (green line) is weak.}
\label{FIGM14}
\vspace{-1mm}
\end{figure}

To help connect these features to experimentally measurable quantities, we generate the corresponding zero bias conductance maps, as well as the geometric correlation $C_G$, for the intermediate disorder case. The results are shown in Fig. \ref{FIGM14}.  First, we note the close resemblance between the $C$ map in Fig. \ref{FIGM12} (b) and the $C_G$ map in Fig. \ref{FIGM14} (c), with the exception of a few additional, loop-like features present in the $C_G$ map that will be discussed below. Second, we point out that, unlike the low impurity density case shown in Fig. \ref{FIGM11}, the areas of high zero-bias conductance are almost equally distributed inside and outside the nominally topological region. This suggest a shift of the chemical potential associated with the emergence of low-energy modes toward higher values as the impurity density increases, which is consistent with previous studies \cite{Adagideli2014,Woods2020b}. Note that this is not due to an actual shift of the impurity-induced effective potential, as the average value of $V_{imp}$ is close to zero regardless of the impurity concentration (see Fig. \ref{FIGM2bis}). 

\begin{figure}[t]
\begin{center}
\includegraphics[width=0.45\textwidth]{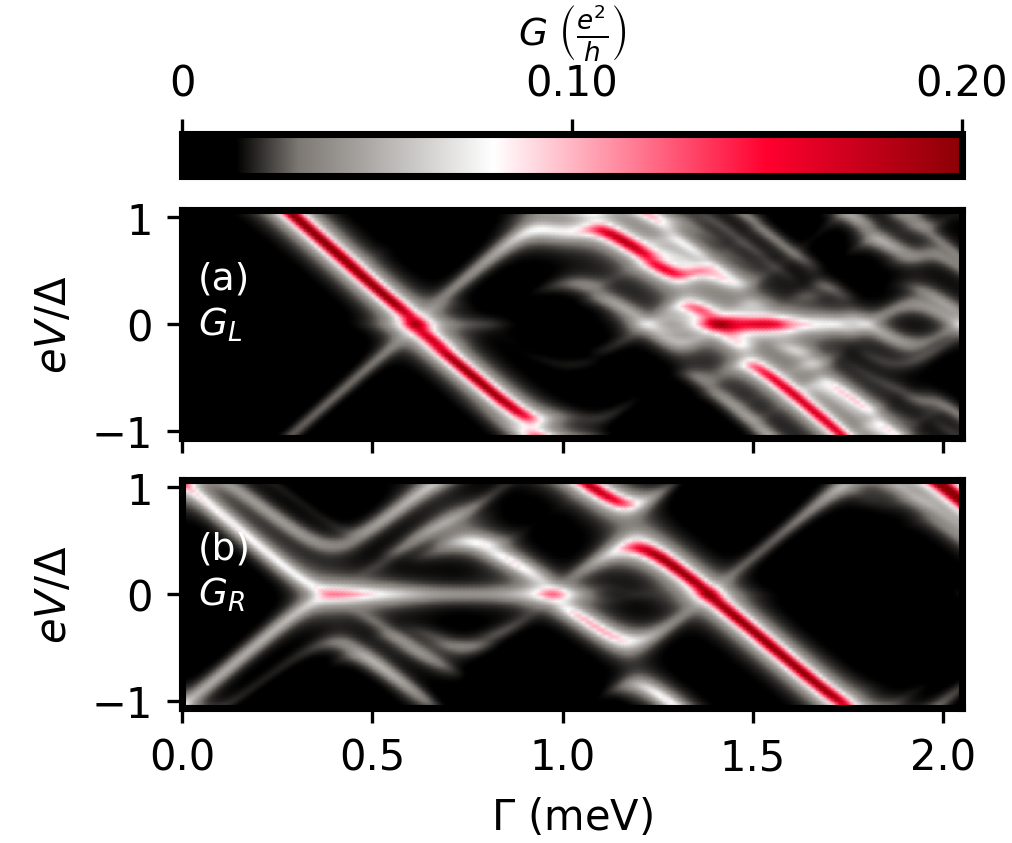}
\end{center}
\vspace{-4mm}
\caption{Local differential conductance at the left (a) and right (b) ends of the wire for a system with the same parameters as in Fig. \ref{FIGM12} and chemical potential $\mu = 0$. Note the strong features associated with Andreev bound states that cross zero energy at different values of the Zeeman field. In (a) the hybridization of these states with the Majorana mode leads to an enhancement of the ZBCP (extremely faint near $\Gamma\approx 0.6~$meV and clearly visible above $\Gamma \approx 1.4~$meV).}
\label{FIGM13}
\vspace{-1mm}
\end{figure}

To shed further light on the nature of various streaky and loopy high-conductance features in Fig. \ref{FIGM14}, we consider the differential conductance as function of the applied Zeeman field and potential bias for two specific values of the chemical potential, $\mu=0$ and $\mu=3~$meV, respectively. The first trace cuts through several narrow, uniformly dispersing high-conductance features that are characteristic to the nominally topological region (see Fig. \ref{FIGM14}). As revealed by the results shown in Fig. \ref{FIGM13}, these features are associated with Andreev bound states crossing zero energy at different values of the Zeeman field. Note that robust ZBCPs signaling the presence of well separated Majorana modes are clearly visible at both ends of the system, but within different intervals of Zeeman fields. The presence of these ZBCPs at $\mu=0$ is consistent with the large values of the Majorana separation in Fig. \ref{FIGM12} (a), while their emergence within different $\Gamma$ intervals is consistent with the low values of $C$ in Fig. \ref{FIGM12} (b). Also note that, as mentioned before, the ZBCP is strongly enhanced as a result of the Majorana modes hybridizing with the ABSs localized near the ends of the wire. Particularly interesting is the faint ZBCP near $\Gamma\approx 0.6~$meV, which is ``revealed'' by the strong ABS mode that crosses zero energy at that value of the Zeeman field. 

\begin{figure}[t]
\begin{center}
\includegraphics[width=0.45\textwidth]{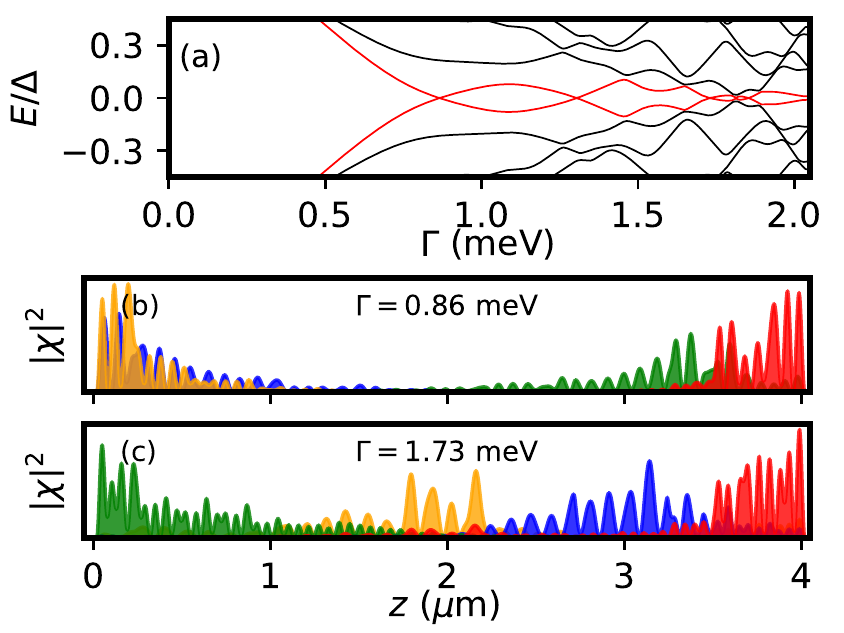}
\end{center}
\vspace{-4mm}
\caption{(a) Low-energy spectrum as a function of the Zeeman field for a system with the same parameters as in Fig. \ref{FIGM12} and $\mu = 3~\text{meV}$. Red lines denote the lowest energy mode. (b) and (c) Spatial profiles of the Majorana components corresponding to the lowest BdG eigenstate (red and green) and second lowest energy eigenstate (blue and yellow) for two values of the Zeeman field, $\Gamma = 0.86~\text{and}~1.73~\text{meV}$, respectively. Note that the lowest energy state in (b) has partially separated Majorana components (i.e., quasi-Majorana modes)  localized near the right edge, while the lowest energy state in (c) has well-separated Majorana components.}
\label{FIGM15}
\vspace{-1mm}
\end{figure}

Next, we focus on the $\mu=3~$meV trace, which cuts through a loop-like feature in Fig. \ref{FIGM14} (b) that has no equivalent in Fig. \ref{FIGM12}.  The corresponding low-energy spectrum is shown in Fig. \ref{FIGM15} (a). Note that, with increasing Zeeman field, several low-energy modes accumulate near zero energy, with the first one crossing zero at $\Gamma \approx 0.8~$meV (red lines). To understand the nature of the low-energy states, we calculate their component Majorana modes. As shown in Fig. \ref{FIGM15} (b), for $\Gamma = 0.86~$meV the lowest energy BdG state consists of a partially separated ABS (ps-ABS) localized near the right end of the wire (red and green Majorana components in  Fig. \ref{FIGM15} (b)). On the other hand, the second lowest energy state is a ``regular'' ABS consisting of two nearly overlapping Majorana components (orange and blue) localized at the left end of the system. As a consequence, both the Majorana separation and the edge-to-edge correlation have small values in the area around $\mu=3~$meV, $\Gamma= 0.86~$meV (see Fig. \ref{FIGM12}).
At a higher Zeeman field,   $\Gamma = 1.73~$meV, the Majorana components of the lowest energy mode -- green and red in  Fig. \ref{FIGM15} (c) -- are well separated and localized near the ends of the wire. This explains the large Majorana separation and the finite edge-to-edge correlation characterizing the corresponding region of the ``phase diagrams'' in Fig. \ref{FIGM12}. Note, however, that these well separated  Majorana modes  have a significant overlap with the Majorana components of higher energy states, with which they can easily hybridize in the absence of an energy gap that would protect them. Consequently, $\ell_{sep}$ and $C$ are highly sensitive to variations of the control parameters, which explains the ``small islands'' structure of the corresponding region of the  phase diagram in Fig. \ref{FIGM12}.    

\begin{figure}[t]
\begin{center}
\includegraphics[width=0.45\textwidth]{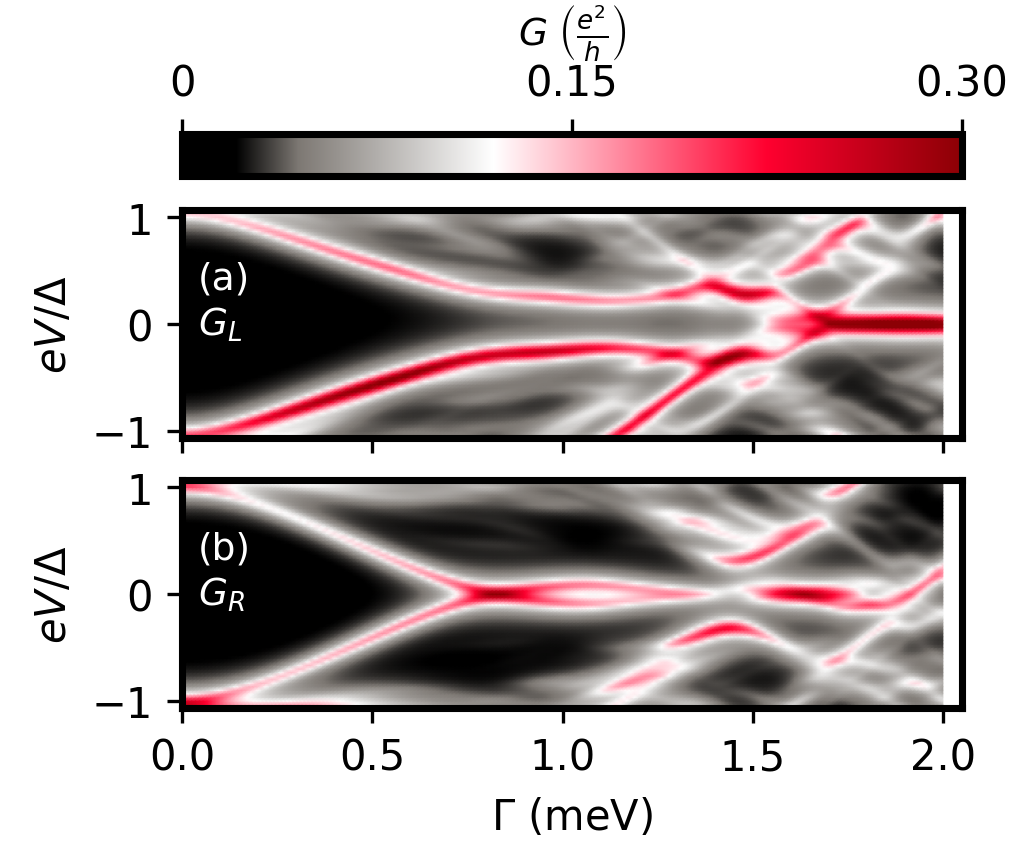}
\end{center}
\vspace{-4mm}
\caption{Local differential conductance at the left (a) and right (b) ends of the wire for a system with the same parameters as in Fig. \ref{FIGM15}, but having normal leads and tunnel barriers attached at both ends. The left conductance has no ZBCP for $\Gamma \lesssim 1.5~\text{meV}$, while the right conductance is characterized by a strong nearly-zero energy feature associated with the loop-like feature in Fig. \ref{FIGM14} (b) and generated by the quasi-Majorana mode shown in Fig. \ref{FIGM15} (b). At larger Zeeman fields, the differential conductance is characterized by ZBCP at both ends of the system, which is consistent with a finite edge-to-edge correlation.}
\label{FIGM16}
\vspace{-1mm}
\end{figure}

Our analysis of the low energy spectrum corresponding to $\mu=3~$meV suggests that the loop-like feature visible in Fig. \ref{FIGM14} around that value of the chemical potential is associated with the quasi-Majorana mode (or ps-ABS) emerging at the right edge of the system [see Fig. \ref{FIGM15} (b)]. To confirm this finding, we calculate the differential conductance at the left and right ends of the system along the same constant $\mu$ cut as the spectrum in 
Fig. \ref{FIGM15} (b). The result in Fig. \ref{FIGM16} (b) clearly shows the emergence of a nearly-zero bias conductance peak at the right edge of the system that practically traces the lowest energy mode [red lines in  Fig. \ref{FIGM15} (a)] for $\Gamma \lesssim 1.4~$meV. A maximum of the zero-bias conductance occurs at $\Gamma \approx 0.8~$meV, where the quasi-Majorana mode crosses zero energy and the $\mu=3~$meV cut intersects the loop-like feature [see Figs. \ref{FIGM14} (b) and \ref{FIGM15} (a)]. We conclude that the loop-like features that characterize the zero-bias conductance maps in Fig. \ref{FIGM14} outside the nominally topological region are generated by quasi-Majorana modes (or ps-ABSs) localized near the ends of the system. 

Turning now our attention to the left end of the system, we notice  [see Fig. \ref{FIGM16} (a)]  the presence of strong finite bias conductance peaks for $\Gamma \lesssim 1.4~$meV. These peaks are generated by the ABS localized at the left end of the system and representing the second-lowest BdG state (see Fig. \ref{FIGM15}). We note that, as a result of finite broadening, the contribution of this state to the zero-bias conductance $G_L$ is finite, although small. However,  when combined with the large quasi-Majorana contribution to $G_R$, it generates a non-zero contribution to the correlation $C_G$, which can be clearly seen as ``shadow'' loop-like feature in Fig. \ref{FIGM14} (c). This spurious correlation feature can be eliminated by considering the finite bias conductance and suppressing $C_G$ if the left and right contribution are not associated with conductance peaks located within the same energy window $(E-\delta E, E+\delta E)$, where $\delta E$ is determined by the energy resolution. Nonetheless, the zero-bias conductance maps, including the $C_G$ map,  can play a crucial role as a first step in characterizing the system and evaluating the effects of disorder. We suggest that this type of comprehensive maps, rather than fine-tuned and post-selected ``good looking'' traces, including traces with conductance $\sim O(2e^2/h)$, 
 should be the standard  protocol for the experimental characterization of hybrid semiconductor-superconductor devices. Finally, we note that for $\Gamma \gtrsim 1.5~$meV the conductance is characterized by ZBCPs at both the left and right ends, as shown in Fig. \ref{FIGM16}. This is consistent with the finite edge-to-edge correlation expected in this regime based on the ``phase diagrams'' shown in Fig. \ref{FIGM12}. 

We conclude this section with a few additional remarks on the ``phase diagrams'' shown in Figs. \ref{FIGM12} and \ref{FIGM14}. First, we note that within the nominally topological regime all ``phase diagrams'' are characterized by stripy features that disperse downwards in $\mu$ with increasing Zeeman field. We have shown that in the case of the conductance maps these features are associated with Andreev bound states localized near the ends of the system that cross zero energy. In certain cases the presence of these ABSs may enhance an otherwise ``invisible'' ZBCP generated by well separated Majorana modes, which results in a finite edge-to-edge correlation. Second, we note that the features located outside the nominally topological region have qualitatively different characteristics. The  conductance maps show several rounded, loop-like features that we identified as being associated with partially separated Majorana modes (or quasi-Majoranas). As discussed above, these features can be eliminated from the correlation map using additional finite bias information. The remaining features have a stripy character and are present in all ``phase diagrams.'' However, unlike the stripy features emerging in the topological region, these ``trivial stripes'' disperse upward in $\mu$  with increasing Zeeman field. Note that a qualitatively similar behavior can be observed even at lower impurity concentrations, as revealed by the ``phase diagrams'' in Figs. \ref{FIGM5} and \ref{FIGM11}. These observations suggest that detailed zero-bias conductance maps could help identify nominally topological regions even when the presence of disorder suppresses the ``standard'' Majorana phenomenology expected in a clean system. Note however, that these results are not expected to hold if the system is characterized by a small inter-subband spacing (i.e., it is not in the in the independent subband regime) or if the disorder strength exceeds a certain threshold (i.e., the system is in the strong disorder regime).  For small inter-subband spacings, even weak disorder will make the system behave as a random disordered class D system because of the essentially random nature of the resultant inter-subband couplings that become comparable to the intra-subband terms.

\subsubsection{Charge impurity statistics}

We have investigated the effects of impurity-induced disorder for two specific disorder realizations corresponding to two different impurity concentrations. The natural questions are: i) What is the generic behavior of the system for arbitrary disorder realizations corresponding to a given impurity concentration?  ii) What is the dependence of the results on the impurity concentration? To effectively address these questions, we need to define some quantities that provide a ``global'' description of the two-dimensional maps discussed in the previous section. To this end, we first define the ``filter function'' $\chi(\mu, \Gamma; \ell_{min}, C_{min}, E_{min})$ that selects control parameter values consistent with certain minimum requirements associated with the presence of well separated Majorana modes capable of generating edge-to-edge correlations. Specifically, we have 
\begin{equation}
\small{
\chi\left(\mu,\Gamma\right) \!=\! \Theta\left(\ell_{sep}\!-\!\ell_{min}\right)\Theta\left(C\!-\!C_{min}\right)\Theta\left(E_{g}\!-\!E_{min}\right),
} 
\end{equation}
where $\Theta(x)$ is the step function,  $\Theta(x>0)=1, \Theta(x<0)=0$, and $E_g = E_2 - E_1$, with $E_1$ and $E_2$ being lowest and second lowest positive eigenenergies, respectively, is the quasiparticle gap separating the lowest energy state from the rest of the spectrum. Note that $\chi=1$ if the Majorana separation length is larger than $\ell_{min}$, the edge-to-edge correlation larger than $C_{min}$ and the quasiparticle gap larger than $E_{min}$, while  $\chi=0$ otherwise. Next, we introduce the quantity $M(\Gamma)$ defined as the total chemical potential range that satisfies the ``good Majorana'' criterion, $\chi(\mu, \Gamma)=1$, for a given value of the Zeeman field. Specifically, we have  
\begin{equation}
M\left(\Gamma\right) = \int \chi\left(\mu,\Gamma\right) d\mu. \label{MDef}
\end{equation}
\begin{figure}[t]
\begin{center}
\includegraphics[width=0.45\textwidth]{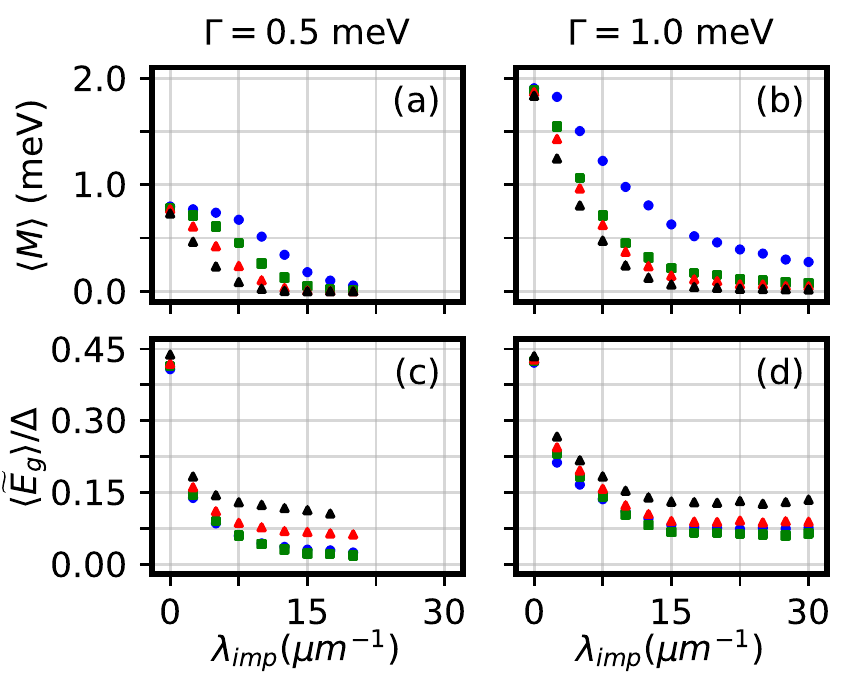}
\end{center}
\vspace{-4mm}
\caption{Disorder-averaged chemical potential range, $\langle M\rangle$, and quasiparticle gap, $\langle \widetilde{E}_g \rangle$, as functions of the impurity density for a system of length $L=4~\mu$m.  The first (a,c) and second (b,d) columns correspond to  $\Gamma = 0.5~\text{meV}$ and $\Gamma = 1~\text{meV}$, respectively. The results corresponding to different sets of filter function parameters, $(\ell_{min}/L, C_{min}, E_{min}/\Delta)$, are color coded: $(0.5, 0, 0)$ -- blue, $(0.5, 0.2, 0)$ -- green, $(0.5, 0.2, 0.05)$ -- red, $(0.5, 0.2, 0.1)$ -- black. Note that the maximum impurity density, $\lambda_{imp} = 30~\mu\text{m}^{-1}$, corresponds to $n_{imp} = 9.4 \cdot 10^{15}~\text{cm}^{-3}$.}
\label{FIGM17}
\vspace{-1mm}
\end{figure}
Note that for a clean system and ``reasonable'' values of $\ell_{min}$, $C_{min}$, and $E_{min}$ we have $M\left(\Gamma\right) =0$ for  $\Gamma < \Delta$, i.e., in the topologically trivial regime, and  $M\left(\Gamma\right) = 2\sqrt{\Gamma^2 - \Delta^2}$ for $\Gamma > \Delta$. In other words, for a clean system  $M\left(\Gamma\right)$ is a measure of the ``thickness'' of the topological region along the $\mu$ direction at a given value of the Zeeman field. For example,  $\Gamma\rightarrow \Delta$ (from above) implies $M\rightarrow 0$, precisely giving the lowest Zemman field associated with the pristine TQPT.
In addition, we define the average quasiparticle gap within the region satisfying the ``good Majorana'' condition as
\begin{equation}
\widetilde{E}_{g}\left(\Gamma\right) = \frac{1}{M\left(\Gamma\right)} \int E_g\left(\mu,\Gamma\right)\chi\left(\mu,\Gamma\right) d\mu. \label{EgDef}
\end{equation}

\begin{figure}[t]
\begin{center}
\includegraphics[width=0.45\textwidth]{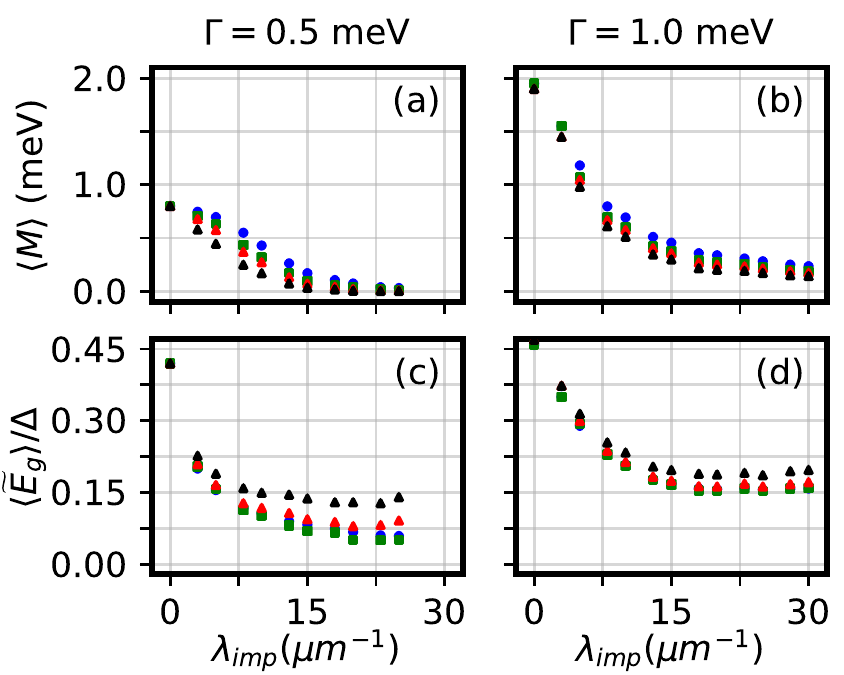}
\end{center}
\vspace{-4mm}
\caption{Same as Fig. \ref{FIGM17} for a wire of length  $L = 2~\mu\text{m}$.}
\label{FIGM18}
\vspace{-1mm}
\end{figure}

To test the relevance of these quantities, we calculate the disorder averages $\langle M\rangle$ and $\langle \widetilde{E}_g \rangle$ as functions of the impurity concentration for two values of the Zeeman field and different sets of filter function parameters, $(\ell_{min}/L, C_{min}, E_{min}/\Delta)$. 
The results for a wire of length $L=4~\mu$m are shown in Fig. \ref{FIGM17}, while the results corresponding to a shorter wire with $L=2~\mu$m are presented in Fig. \ref{FIGM18}. The averages corresponding to each value of the impurity density, $\lambda_{imp}$, were calculated using $500$ different disorder realizations.
Note that if a given impurity realization is characterized by $M = 0$, $\widetilde{E}_g$ is undefined, and we do not include it in the calculation of $\langle \widetilde{E}_g \rangle$.
First, we observe that $\langle M\rangle$ collapses with increasing impurity density reaching negligible values for impurity densities of the order $10-20$ impurities per micron. This means that for higher impurity concentrations there are practically no ``good Majoranas'' in the system. We point out that for the Majorana separation criterion we used a rather generous value, $\l_{min}=0.5L$, which does not guarantee the localization of the well-separated Majorana modes near the ends of the wire. This is particularly significant in Fig. \ref{FIGM17} (b), where introducing the edge-to-edge correlation requirement $C_{min}=0.2$ strongly reduces  $\langle M\rangle$ as compared to the case $C_{min}=0$ (blue dots). On the other hand, the fact that the blue dots in Fig. \ref{FIGM17} (b) correspond to finite values of $\langle M\rangle$ over the entire range of impurity densities reveals that, even in the presence of relatively strong disorder, the system contains well-separated Majoranas. However, these Majoranas do not generate
edge-to-edge correlations. In other words, some segments of a long wire are likely to be in the topological superconducting phase, but these segments have a concentration-dependent typical length (which is unknown experimentally) that is less than the length $L$ of the wire. Therefore, their presence cannot be established  based on the edge-to-edge correlation, which is negligible. This observation is consistent with the specific examples discussed in sections \ref{ss1} and \ref{ss2}.  Note that for the shorter system (see Fig. \ref{FIGM18}) imposing the additional filter $C > 0.2$ does not reduce $\langle M\rangle$ drastically. This is due to the fact  that Majorana modes with $\ell_{sep} > L/2$ are significantly more likely to generate edge-to-edge correlations in a shorter wire, as compared to a longer wire.  Finally, regarding the average quasiparticle gap, $\langle \widetilde{E}_g \rangle$, we notice a sharp drop at low impurity density, followed by a slower decline toward a density-independent plateau, which starts at $\lambda_{imp}\approx 15~\mu$m$^{-1}$. The height of the plateau is determined by the average inter-state spacing, which depends on the length of the wire being proportional to $1/L$. 

\begin{figure}[t]
\begin{center}
\includegraphics[width=0.45\textwidth]{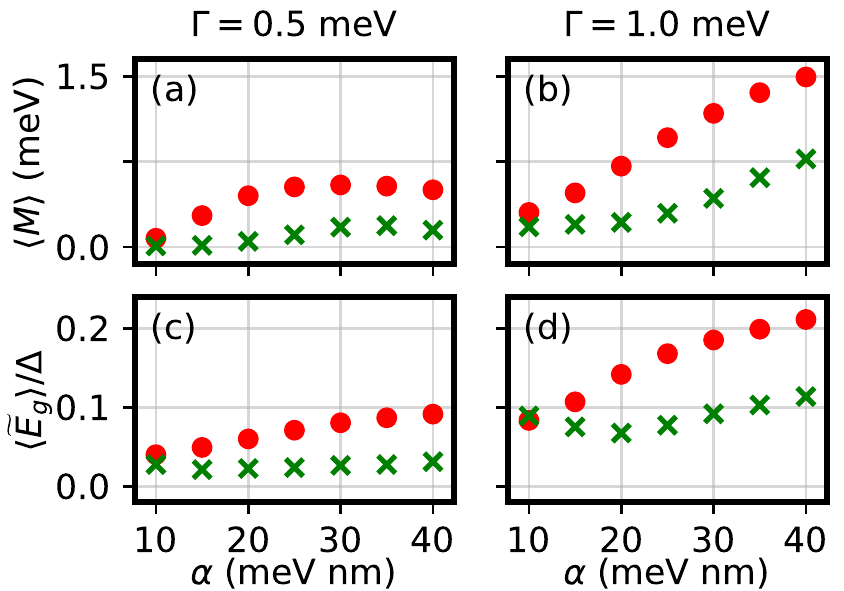}
\end{center}
\vspace{-3mm}
\caption{Disorder-averaged chemical potential range, $\langle M\rangle$, and quasiparticle gap, $\langle \widetilde{E}_g \rangle$, as functions of the
spin-orbit coupling strength, $\alpha$, for a wire of length $L = 4~\mu\text{m}$. The red circles and green crosses correspond to $\lambda_{imp} = 7.5~\mu\text{m}$, and $\lambda_{imp} = 15~\mu\text{m}$, respectively. The filter function parameters are $\ell_{min}/L = 0.5$, $C_{min} = 0.2$, and $E_{min} = 0$.}
\label{FIGM19}
\vspace{-1mm}
\end{figure}

The ``global'' quantities introduced above provide useful tools for studying the effects of disorder on the Majorana physics. Most importantly, they reveal the strong dependence of the Majorana physics on the impurity concentration. In particular, observing edge-to-edge correlations at relatively low values of the Zeeman field requires reducing the impurity density below a certain threshold of about $15-20$ impurities per micron. This type of analysis can be also useful for optimizing the system parameters. As an example, we consider the dependence on the spin-orbit coupling strength. Fig. \ref{FIGM19} shows the  
dependence of $\langle M\rangle$ and $\langle \widetilde{E}_g \rangle$ on spin-orbit coupling strength $\alpha$ for a wire of length $L = 4~\mu\text{m}$ for two impurity densities and two values of the Zeeman field. Typically, increasing the spin-orbit coupling strength enhances both $\langle M\rangle$ and $\langle \widetilde{E}_g \rangle$. However, for $\lambda_{imp} = 15~\mu\text{m}^{-1}$ (green crosses) the dependence of the average energy gap on $\alpha$ is weak, while  $\langle M\rangle$ shows a significant enhancement only at larger values of the Zeeman field and for $\alpha \gtrsim 25~$meV$\cdot$nm. Finally, we point out that throughout this work the value of the spin-orbit coupling strength was $\alpha = 20~$meV$\cdot$nm, which we consider as relatively optimistic. While for large enough Zeeman fields $\langle M\rangle$ and $\langle \widetilde{E}_g \rangle$ can be enhanced by having a stronger spin-orbit coupling, there is not much room for optimizing the low-field regime. Note, however, that at large field values the topological gap itself may be rather small and, again, optimization becomes a challenge even in this regime.

\section{Conclusions} \label{Conclusion}

We have carried out a comprehensive microscopic theoretical study of disorder effects arising from the inevitable presence of charge impurities in superconductor-semiconductor nanowire hybrid structures, focusing on the fate of the Majorana zero modes expected to emerge in these systems.  The work consists of four closely connected, but distinct, theoretical components: (1) developing a fully self-consistent realistic  Schr{\"o}dinger-Poisson scheme to calculate the effective impurity potential arising from the presence of charge impurities, which takes into account electrostatic and screening effects due to the  superconductor and potential back gate, as well as the screening by  the free charge in the wire; (2) carrying out full solutions of the BdG equations in the presence of disorder by incorporating the effective impurity potential calculated self-consistently for a multi-band system, as well as the superconducting proximity effect, spin-orbit coupling, and applied Zeeman field; (3) obtaining, based on the solutions of the BdG equations, effective ``phase diagrams'' as functions of the control parameters (i.e., Zeeman field and chemical potential) in the presence of disorder and investigating their dependence on the disorder strength; (4) calculating the tunnel conductance at both ends of the system and generating the corresponding ``phase diagrams'',   which provides insight into the existing tunnel spectroscopy experiments on Majorana nanowires and suggests new directions for enhancing the relevance of such measurements. 
Since the work involves multiple aspects, we have specific conclusions regarding each component of the theory already included in the corresponding section of this article.  Instead of  repeating what is already described and discussed in depth in sections \ref{Single} and \ref{Mult}, we summarize our most important conclusions regarding the role of charge impurity-induced disorder from the perspective of the ongoing search for non-Abelian Majorana modes in superconductor-semiconductor nanowire hybrid structures. 

We show that the superconductor plays a rather limited role in screening the impurity potential, while substantial screening arises from the free charges in the nanowire.  We provide a simple two-parameter empirical fitting formula for the effective screened potential, which should be useful for future  simulations of Majorana devices. Quantitatively, we find that the effective impurity potential has typical amplitudes of the order of $1.5-2~$meV and typical decay lengths of about $8-12$nm.

We find that disorder produces zero energy states outside the pristine topological phase boundary and we analyze in depth the nature of these states and their possible experimental signatures.  We also find that, within the nominally topological regime, the system can host well separated Majorana modes even in the presence of significant disorder levels, but typically the presence of these modes is not associated with a significant edge-to-edge correlation. 
A key finding in this context is that disorder may often push  Majorana zero modes away from the wire ends, thus making them invisible to local (end-of-wire) tunnel spectroscopy.  Thus, it is entirely possible (and likely) to miss the presence of Majorana zero modes in a disordered nanowire when using  tunneling spectroscopy simply because this is a local probe sensitive only to states localized at the wire ends.  Hence, in the presence of disorder,  long segments within the bulk of the wire may be topologically nontrivial, with Majorana modes emerging at their ends, but the wire ends themselves may contain no Majorana modes, which dramatically reduces the probability of observing edge-to-edge correlations. 

We establish that detailed two-dimensional maps of the zero-bias conductance  as a function of Zeeman splitting (i.e. magnetic field in the laboratory) and chemical potential (i.e. gate voltage in the laboratory) may be the most effective operational way to search for the ``hidden'' topological superconductivity and the associated  Majorana modes. The current experimental focus on looking for large zero bias peaks with conductance $\sim 2e^2/h$ by fine-tuning the control parameters is unlikely to solve the outstanding questions regarding the nature of the low-energy states responsible for these peaks.
First, a large zero bias peak obtained through careful fine-tuning and post selection may have nothing to do with topological Majorana modes, and second, this procedure is likely to lead to strong confirmation bias in the experiment.  Instead, creating zero bias conductance maps  in the extensive parameter space of gate voltage and magnetic field using the cleanest possible samples and comparing these maps to our theoretical results may be a much more systematic way of searching for Majorana physics, without suffering from any confirmation bias. In addition, this would provide much needed estimates of the disorder strength characterizing actual superconductor-semiconductor hybrid devices and an effective way of testing future materials improvements that aim at reducing disorder.

We find that for reasonably realistic (but still somewhat optimistic) parameter choices, genuine, well-separated topological Majorana modes should exist in nanowires for impurity densities up to $5 \cdot 10^{15}~$cm$^{-3}$, which corresponds to around 15 impurities per micron. This would mean that a $2-4$ micron long nanowire can contain up to $30-60$ charge impurities, but cleaner samples, with charge impurity density below $10^{15}~$cm$^{-3}$, may be necessary in practice, since we ignored any disorder arising from possible interface defects or imperfections.  Such a low intrinsic doping of less than $10^{15}~$cm$^{-3}$ is a challenge, but is by no means out of reach in semiconductor materials growth, as impurity contents below $10^{13}~$cm$^{-3}$ have been achieved in MBE-grown GaAs structures \cite{Pfeiffer1989}.

\renewcommand{\thefigure}{A\arabic{figure}}

\setcounter{figure}{0}

Our final conclusion is that charge impurities cause serious problems, but by no means destroy the topology in hybrid nanowires, as long as their concentration is maintained below a certain threshold.  Future experiments  should provide estimates of the disorder levels that characterize existing hybrid systems, while a systematic effort should be dedicated to the production of much cleaner wires, with significantly lower impurity content, where Majorana zero modes could emerge easily and manifest the full range of their expected phenomenology.

Implications of our work for the realization of Majorana zero modes and topological qubits are obvious and far-reaching.  Since semiconductor nanowire-superconductor hybrid platforms are by far the leading TQC candidates, by virtue of the tunability of the system through electrical gating and varying magnetic field, and because semiconductor growth enables the realization of very pure materials, our detailed macroscopic, quantitative analysis of all relevant aspects of Majorana physics in the SM-SC platform in the presence of charge impurity disorder provides the community with clear and quantitative guidelines on how to make progress: Obtain samples with $10^{15}$ per cm$^3$ or less impurity content,  produce two-dimensional parameter maps of the zero bias differential conductance over extended parameter regions, beware of the possibility that impurities may push the Majorana bound states away from the edges, so that topology may be hidden in tunnel spectroscopy at the ends, do not focus on trying to find Majorana quantization (which are often spurious), instead focus on the totality of the parameter space rather than fine-tuning, carry out conductance correlations the way proposed in the current work as a function of magnetic field and gate voltage by doing tunneling from both ends, and try to obtain nonlocal correlations not just from the two ends, but along the wire.  Our work establishes the existence of topological Majorana modes in the system in the presence of some amount of charge impurities even when the disorder potential is significantly larger than the SC gap, provided that the impurity concentration is not too high.  This is a highly encouraging result which should inspire new efforts toward creating Majorana qubits.

\begin{acknowledgments}
This  work  is  supported  by  NSF Grant No. 10026278 and by the Laboratory for Physical Sciences.
\end{acknowledgments}

\appendix

\begin{figure}[t]
\begin{center}
\includegraphics[width=0.45\textwidth]{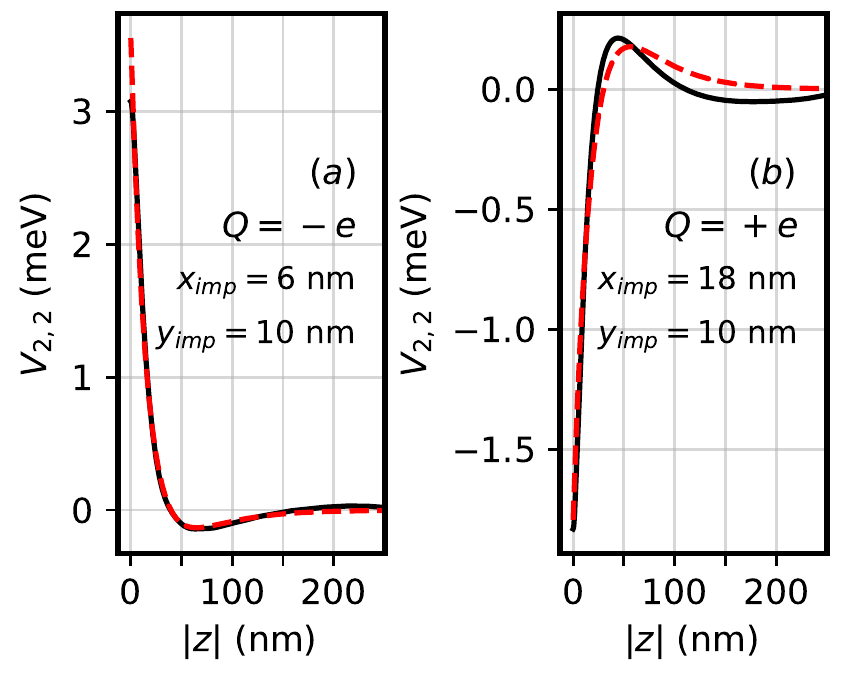}
\end{center}
\vspace{-3mm}
\caption{Examples of effective potential (black solid lines) and the fitted effective potential (red dashed lines) from impurities of charge (a) $Q = -e$ and (b) $Q = e$ using the fitting Eqs. (\ref{A1}-\ref{A3}). The transverse positions of the impurities are indicated in the panels. The average absolute error between the exact and fitted potentials for $|z| \leq 100~\text{nm}$ is (a) $0.03~\text{meV}$ and (b) $0.07~\text{meV}$, respectively.  }
\label{FIGA1}
\vspace{-1mm}
\end{figure}

\section{Fitting effective impurity potentials to an analytic function} \label{Fit}

\begin{figure}[t]
\begin{center}
\includegraphics[width=0.45\textwidth]{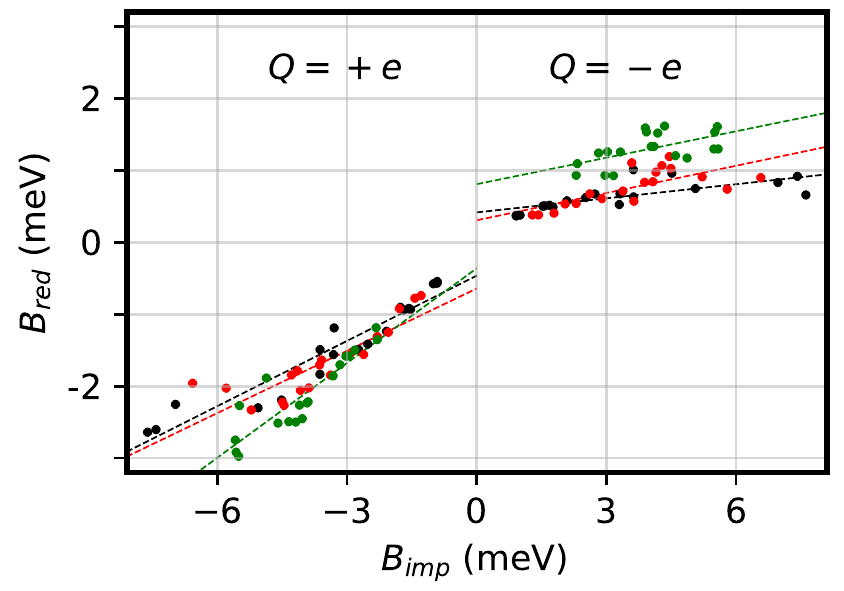}
\end{center}
\vspace{-8mm}
\caption{Effective redistribution amplitude $B^\alpha_{red}$ vs effective impurity amplitude $B^\alpha_{imp}$ when the $\alpha = 2 \text{ (black dots), } 3 \text{ (red), or } 4 \text{ (green)}$ subband is tuned to the Fermi level. Amplitudes are extracted from fitting the effective potential of $19$ evenly spaced impurity locations within the transverse profile of the nanowire. Dashed lines are linear regression fits to the matching color data. See Table \ref{table1} for fitting parameters. }
\label{FIGA2}
\vspace{-1mm}
\end{figure}

As alluded to in the main text [see also Eq. (29)], the effective potential of a single charge impurity located at $z = 0$ can be well captured by fitting both the effective impurity and redistribution potentials to exponential functions,
\begin{eqnarray}
	V_{\alpha,\alpha}\left(z\right) = 
	V_{\alpha,\alpha}^{imp}\left(z\right) + V_{\alpha,\alpha}^{red}\left(z\right), \label{A1} \\ 
	V_{\alpha,\alpha}^{imp}\left(z\right) =  
	B^\alpha_{imp}e^{-|z|/\lambda^\alpha_{imp}}, \label{A2} \\
	V_{\alpha,\alpha}^{red}\left(z\right) = 
	-B^\alpha_{red}e^{-|z|/\lambda^\alpha_{red}}, \label{A3} 
\end{eqnarray}
where $B^\alpha_{imp}$ and $B^\alpha_{red}$ are the amplitudes of the effective impurity and redistribution potentials, respectively, $\lambda^\alpha_{imp}$ and $\lambda^\alpha_{red}$ are the corresponding decay lengths, and $\alpha$ is the subband index. We place a minus sign in front of $B^\alpha_{red}$ to emphasize that the redistribution potential (partially) suppresses the impurity charge potential. Two examples of this fitting are shown in Fig. \ref{FIGA1}, for a negative and positive elementary impurity charge with the $\alpha = 2$ subband tuned to the Fermi level. We find excellent agreement between the actual and fitted potential in both cases. Indeed, the average absolute error, $|V_{2,2}^{exact} - V_{2,2}^{fit}|$, within $100~\text{nm}$ of the impurity is only (a) $0.03~\text{meV}$ and (b) $0.07~\text{meV}$, respectively. Note that we found the average absolute error to be of this order for all impurity locations sampled. Generically, we find that cases with a negative charge impurity fit slightly better to Eqs. (\ref{A1}-\ref{A3}) than positive charge impurity cases. This is due to a more prominent ``hump'' feature after first crossing $V_{\alpha,\alpha} = 0$ for positively charged impurities compared to negatively charged impurities. Nevertheless, the fitting is exceptional for both impurity charge signs.

\begin{figure}[t]
\begin{center}
\includegraphics[width=0.45\textwidth]{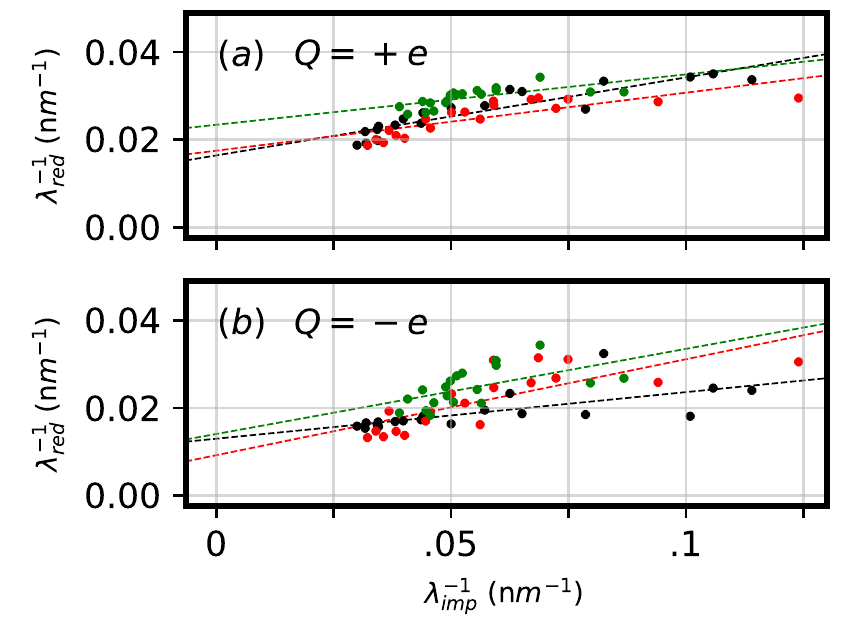}
\end{center}
\vspace{-8mm}
\caption{Comparison of impurity and redistribution (inverse) decay lengths, $(\lambda^\alpha_{imp})^{-1}$ and $(\lambda^\alpha_{red})^{-1}$, when the $\alpha = 2 \text{ (black dots), } 3 \text{ (red), or } 4 \text{ (green)}$ subband is tuned to the Fermi level. Decay lengths are extracted from fitting the effective potential of $19$ evenly spaced impurity locations within the transverse profile of the nanowire. (a) Top and (b) bottom panels correspond to $Q = +e$ and $Q = -e$, respectively. Dashed lines are linear regression fits to the matching color data. See Table \ref{table2} for fitting parameters. }
\label{FIGA3}
\vspace{-1mm}
\end{figure}

\begin{figure}[t]
\begin{center}
\includegraphics[width=0.45\textwidth]{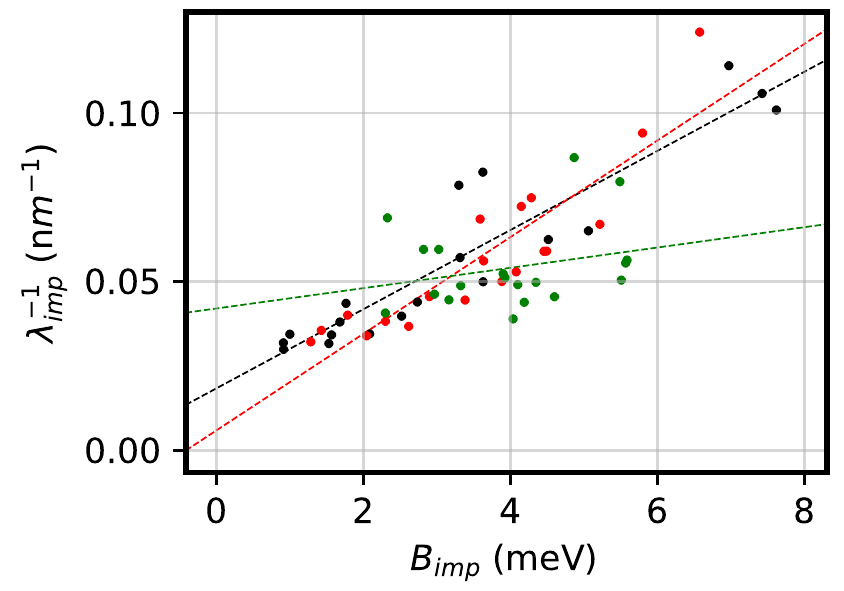}
\end{center}
\vspace{-8mm}
\caption{Comparison of impurity amplitude $B^\alpha_{imp}$ and (inverse) decay length $(\lambda^\alpha_{imp})^{-1}$ when the $\alpha = 2 \text{ (black dots), } 3 \text{ (red), or } 4 \text{ (green)}$ subband is tuned to the Fermi level. Amplitude and decay lengths are extracted from fitting the effective potential of $19$ evenly spaced impurity locations within the transverse profile of the nanowire. Impurity charge $Q = -e$. Dashed lines are linear regression fits to the matching color data. See Table \ref{table3} for fitting parameters. }
\label{FIGA4}
\vspace{-1mm}
\end{figure}

While Eqs. (\ref{A1}-\ref{A3}) represent an excellent approximation for the effective potential, it requires $4$ fitting parameters, which may be cumbersome if one wants to construct a phenomenological model of charge impurity disorder without explicitly performing numerically expensive Schr{\"o}dinger-Poisson calculations. This motivates us to investigate whether the various fitting parameters display correlations to reduce the number of necessary input parameters. We indeed find this to be the case. The resulting correlations are shown in Figs. \ref{FIGA2}, \ref{FIGA3}, and \ref{FIGA4} and discussed below.

The effective redistribution amplitude $B^\alpha_{red}$ as a function of the effective impurity amplitude $B^\alpha_{imp}$ is shown in Fig. \ref{FIGA2} for several different subbands tuned to the Fermi level. The amplitudes come from fitting the effective impurity and redistribution potentials to Eqs. (\ref{A2}, \ref{A3}) with each data point  corresponding to a different impurity location in the transverse profile of the wire. For this data set, 19 evenly spaced positions in the nanowire's cross section were sampled. We observe a general linear trend between the two amplitudes for all three subbands in which the magnitude of the redistribution amplitude increases with increasing magnitude of the impurity amplitude, as seen in the linear fit lines (dashed lines). The positive sign of the slope makes physical sense since increasing the magnitude of the impurity amplitude should increase the redistribution of free charge around the impurity to (partially) counteract the perturbation of the electrostatic environment.  What's not obvious, however, is that a linear relationship should capture the dependency rather well. After all, the Schr{\"o}dinger-Poisson equations should be expected to behave non-linearly due to the interplay between the various occupied subbands. To quantify how well the linear fit captures the relationship, we have gathered the fitting parameters into Table \ref{table1}. In particular, we wish to bring attention to the coefficient of determination, $r^2$, which indicates how much of the variance of the data is explained by the linear model. For all except the ($\alpha = 4, Q = -e$) case, the linear fit explains over half of the variance ($r^2 > 0.5$). Moreover, $Q = e$ cases display particularly high $r^2$ value. We also notice that the $r^2$ coefficient diminishes on average with increasing subband index, $\alpha$, suggesting the Schr{\"o}dinger-Poisson equations are behaving with increasing non-linearity as occupation is increased. 

Similar to the effective potential amplitudes, we compare the (inverse) decay lengths of the effective redistribution and impurity potentials in Fig. \ref{FIG3} with the subband $\alpha = 2, 3, \text{ or } 4$ tuned to the Fermi level. Again we observe general linear trends and fit the data from each subband to a line. The fitting parameters are gathered in Table \ref{table2}. The $r^2$ coefficients are of similar size to what was found in studying the relationship between the potential amplitudes (Table \ref{table1}), but are slightly smaller, indicating the decay lengths behave in a slightly more non-linear manner.

Finally, we study the correlation between the effective impurity potential's amplitude $B^\alpha_{imp}$ and (inverse) decay length $(\lambda_{imp}^\alpha)^{-1}$ in Fig. \ref{FIGA4}. In contrast to Figs. \ref{FIGA2} and \ref{FIGA3}, we only consider $Q = -e$, since flipping the sign to $Q = e$ only changes the sign of the amplitude, $B^\alpha_{imp}$. The fitting parameters are gathered in Table \ref{table3}. On the one hand, we observe large coefficients of determination, $r^2 = 0.87, 0.80$, for $\alpha = 2,3$, respectively. On the other hand, $r^2 = 0.07$ for $\alpha = 3$. These indicate that the relationship between the effective impurity amplitude and inverse decay length is well captured by a linear fit for $\alpha = 2,3$, but not $\alpha = 4$. Evidently, as the wavefunction moves away from the SM-SC interface with increasing $\alpha$ the electrostatics become more subtle and the relationship between the amplitude and (inverse) decay length becomes more complicated.

We're now in a strong position to create realistic phenomenological models of charge impurity disorder in SM-SC hybrids nanowires using only 1 or 2 parameters from which we need to sample. We accomplish this by leveraging the information we've just laid out regarding the linear relationships between the various fitting parameters. In the case of low-occupancy ($\alpha \leq 3$ in this case) the relationships between the 4 fitting parameters, $B^\alpha_{imp}$, $B^\alpha_{red}$, $(\lambda^\alpha_{imp})^{-1}$, and $(\lambda^\alpha_{red})^{-1}$, are well described by all three linear relationships studied in this appendix. Therefore, one only needs to sample the $B^\alpha_{imp}$ distribution to create a realistic model of disorder. Given a $B^\alpha_{imp}$ value, we only have to plug it into the linear equations given in Tables \ref{table1}-\ref{table3} and the corresponding parameters, $m$ and $b$ (also in the tables), to obtain the other three fitting parameters. In the case of higher occupancy ($\alpha \geq 4$ in this case) the relationship between $B^\alpha_{imp}$ and $(\lambda^\alpha_{imp})^{-1}$ is not represented well by a linear fit. Therefore, we need to sample from both $B^\alpha_{imp}$ and $(\lambda^\alpha_{imp})^{-1}$ to create a realistic disorder model. The other two fitting parameters, $B^\alpha_{red}$ and $(\lambda^\alpha_{red})^{-1}$, however, can still be obtained using the linear equations and corresponding parameters in Tables \ref{table1} and \ref{table2}. We then have a convenient and accurate way of producing realistic disorder potential profiles due to charge impurities in Majorana SM-SC hybrid nanowires.


\begin{table}[h!]
\caption{\label{table1}Fitting parameters of dashed lines in Fig. \ref{FIGA2} corresponding to the fitting equation, $B^\alpha_{red} = m B^\alpha_{imp} + b$. Coefficient of determination $r^2$ for each linear fit is given in the final column, where $r^2 =1$ indicates a perfect fit.}
\begin{ruledtabular}
\begin{tabular}{l|ccc}
$(\alpha,Q)$&$m$&$b~\text{(meV)}$&$r^2$\\
\hline
$(2,+e)$ & 0.30 & -0.46 & 0.91\\
$(2,-e)$ & 0.07 & 0.41 & 0.53\\
$(3,+e)$ & 0.29 & -0.64 & 0.75\\
$(3,-e)$ & 0.13 & 0.30 & 0.52\\
$(4,+e)$ & 0.43 & -0.36 & 0.79\\
$(4,-e)$ & 0.12 & 0.81 & 0.34\\
\end{tabular}
\end{ruledtabular}
\end{table}
 
\begin{table}[h!]
\caption{\label{table2}Fitting parameters of dashed lines in Fig. \ref{FIGA3} corresponding to the fitting equation, $(\lambda^\alpha_{red})^{-1} = m (\lambda^\alpha_{imp})^{-1} + b$. Coefficient of determination $r^2$ for each linear fit is given in the final column, where $r^2 =1$ indicates a perfect fit.}
\begin{ruledtabular}
\begin{tabular}{l|ccc}
$(\alpha,Q)$&$m$&$b~(\text{nm}^{-1})$&$r^2$\\
\hline
$(2,+e)$ & 0.18 & 0.016 & 0.83\\
$(2,-e)$ & 0.11 & 0.013 & 0.46\\
$(3,+e)$ & 0.13 & 0.017 & 0.63\\
$(3,-e)$ & 0.22 & 0.009 & 0.59\\
$(4,+e)$ & 0.12 & 0.023 & 0.45\\
$(4,-e)$ & 0.19 & 0.014 & 0.32\\
\end{tabular}
\end{ruledtabular}
\end{table}

\begin{table}[h!]
\caption{\label{table3}Fitting parameters of dashed lines in Fig. \ref{FIGA4} corresponding to the fitting equation, $(\lambda^\alpha_{imp})^{-1} = m B^\alpha_{imp}+ b$. Coefficient of determination $r^2$ for each linear fit is given in the final column, where $r^2 =1$ indicates a perfect fit.}
\begin{ruledtabular}
\begin{tabular}{l|ccc}
$(\alpha,Q)$&$m~(\text{meV}^{-1} \text{ nm}^{-1})$&$b~(\text{nm}^{-1})$&$r^2$\\
\hline
$(2,-e)$ & 0.012 & 0.018 & 0.87\\
$(3,-e)$ & 0.014 & 0.005 & 0.80\\
$(4,-e)$ & 0.003 & 0.042 & 0.07\\
\end{tabular}
\end{ruledtabular}
\end{table}

\phantom{a}\par
\phantom{a}\par
\phantom{a}\par

\FloatBarrier


%

\end{document}